\documentclass{article}

\usepackage{arxiv}

\usepackage[utf8]{inputenc} % allow utf-8 input
\usepackage[T1]{fontenc}    % use 8-bit T1 fonts
\usepackage{hyperref}       % hyperlinks
\usepackage{url}            % simple URL typesetting
\usepackage{booktabs}       % professional-quality tables
\usepackage{amsfonts}       % blackboard math symbols
\usepackage{nicefrac}       % compact symbols for 1/2, etc.
\usepackage{microtype}      % microtypography
\usepackage{amsmath}
\usepackage{cleveref}       % smart cross-referencing
\usepackage{lipsum}         % Can be removed after putting your text content
\usepackage{graphicx}
\usepackage{natbib}
\usepackage{doi}

\usepackage{graphicx}
\usepackage{lmodern}

\usepackage{array}
\usepackage{esint}
\usepackage{amsfonts}
\usepackage{siunitx}
\usepackage[draft]{ed}

\usepackage{tabularx}

\usepackage{footmisc}
\usepackage{hyperref}
\usepackage{subfig}

\usepackage{fancyvrb}

\usepackage{cprotect}

\usepackage{latexml}
\usepackage{soul}

\usepackage{caption} 
\definecolor{Mycolor2}{HTML}{0f4c5c}
\sethlcolor{black}
\renewcommand{\_}{{\fontfamily{ptm}\selectfont\textunderscore}}

\makeatletter
\newcommand\newtag[2]{#1\def\@currentlabel{#1}\label{#2}}
\makeatother

\title{Discovery and Recognition of Formula Concepts using Machine Learning}

\author{ {\hspace{1mm}Philipp Scharpf} \\
	University of Konstanz \\
        Germany \\
	\texttt{philipp.scharpf@uni-konstanz.de} \\
	\And
	{\hspace{1mm}Moritz Schubotz} \\
	University of Wuppertal and FIZ Karlsruhe \\
        Germany \\
	\texttt{moritz.schubotz@fiz-karlsruhe.de} \\
        \And
	{\hspace{1mm}Howard S. Cohl} \\
	Applied and Computational Mathematics Division \\
	National Institute of Standards and Technology \\
	Gaithersburg, Maryland, USA \\
	\texttt{howard.cohl@nist.gov} \\
        \And
	{\hspace{1mm}Corinna Breitinger} \\
	Institute of Computer Science \\
	University of Göttingen \\
	Germany \\
	\texttt{corinna.breitinger@uni-goettingen.de} \\
        \And
	{\hspace{1mm}Bela Gipp} \\
	Institute of Computer Science \\
	University of Göttingen \\
	Germany \\
	\texttt{gipp@uni-goettingen.de} \\
}

\renewcommand{\headeright}{Scharpf et al.}
\renewcommand{\undertitle}{arXiv Preprint}
\renewcommand{\shorttitle}{Formula Concept Retrieval using Machine Learning}
\newcommand{\wdl}[2]{%\href{https://wikidata.org/wiki/#2}
{#1 (#2)}}
\begin{document}
\maketitle

\begin{abstract}
Citation-based Information Retrieval (IR) methods for scientific documents have proven effective for IR applications, such as Plagiarism Detection or Literature Recommender Systems in academic disciplines that use many references. In science, technology, engineering, and mathematics, researchers often employ mathematical concepts through formula notation to refer to prior knowledge. Our long-term goal is to generalize citation-based IR methods and apply this generalized method to both classical references and mathematical concepts. In this paper, we suggest how mathematical formulas could be cited and define a Formula Concept Retrieval task with two subtasks:~Formula Concept Discovery (FCD) and Formula Concept Recognition (FCR). While FCD aims at the definition and exploration of a `Formula Concept' that names bundled equivalent representations of a formula, FCR is designed to match a given formula to a prior assigned unique mathematical concept identifier. We present machine learning-based approaches to address the FCD and FCR tasks. We then evaluate these approaches on a standardized test collection (NTCIR arXiv dataset). Our FCD approach yields a precision of 68\% for retrieving equivalent representations of frequent formulas and a recall of 72\% for extracting the formula name from the surrounding text.
FCD and FCR enable the citation of formulas within mathematical documents and facilitate semantic search and question answering as well as document similarity assessments for plagiarism detection or recommender systems.

\end{abstract}

% keywords can be removed
\keywords{Mathematical Information Retrieval \and Search \and Machine Learning \and Classification \and Clustering \and Wikidata}

%=====================================================================================
% Introduction
%=====================================================================================

\section{Introduction}\label{sec:intro}

\begin{figure}[ht]
\centering
\includegraphics[width=\textwidth]{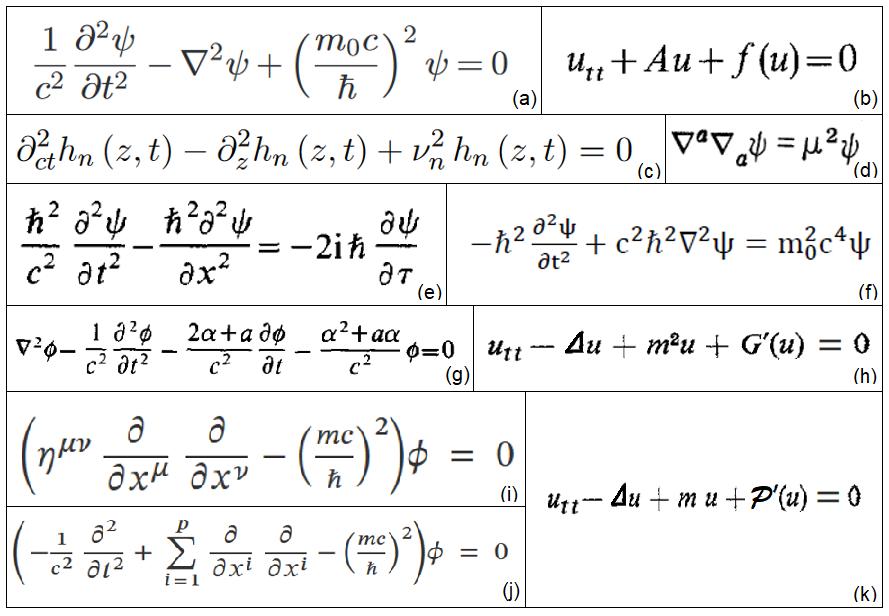}
\caption{Representations of the Klein--Gordon equation extracted from physics papers - 
(a): \cite{arbab2010derivation}, (b): \cite{pecher1984nonlinear}, (c): \cite{tretyakov2010derivation}, (d): \cite{detweiler1980klein}, (e): \cite{kaloyerou1989evolution}, (f): \cite{haroun2017derivation}, (g): \cite{tiwari1988derivation}, (h): \cite{strauss1978numerical}, (i): \cite{nlab:homepage}
, (j): \cite{nlab:homepage}
, (k): \cite{morawetz1968time}.
Some of the representations are written in a general, potentially nonlinear form. With constraints given for the parameters in the respective publications, the equations become the linear Klein--Gordon equation.}
\label{fig:Klein-Gordon-equation}
\end{figure}

Documents from Science, Technology, Engineering, and Mathematics (STEM) often contain a significant amount of mathematical formulas~\cite{DBLP:conf/lwa/HambasanK15}.
Formulas are a vital non-textual component to understand the content of STEM documents. Systems, such as semantic search engines, question answering systems,
and document recommender systems, should also be capable of processing formulas and their connections with the surrounding text and mathematical expressions.
In information science and technology, the semantics of natural language is typically grasped via conceptualization~\cite{yucong2011formalizing}.
According to~\cite{gruber1993translation}, the term conceptualization refers to the process of simplifying the representation of objects of discourse and specifying a semantic vocabulary in an ontology (knowledge system).
Analogously, to capture the semantics of mathematical language in formulas, we argue for the introduction of a mathematical \textit{Formula Concept}, which we define as a collection of equivalent formulas with different representations
(see also Section \ref{sec:fcd} below). This extends the definition of the \textit{formula content} comprising constituents, relations, and semantics of a formula, which was introduced in ~\cite{DBLP:conf/sigir/ScharpfSG18}.
We select the Klein--Gordon equation as an example for mathematical conceptualization. Figure \ref{fig:Klein-Gordon-equation} shows different representations of the Klein--Gordon equation\footnote{\url{https://en.wikipedia.org/wiki/Klein-Gordon_equation}} from quantum mechanics (also referred to as a relativistic wave equation).
These representations of the Klein--Gordon equation in the academic literature appear to be diverse, but they all represent the same mathematical concept. Employing additional mathematical Formula Concept examples, we illustrate and discuss differences and explain the resulting challenges of this conceptualization process in detail. We introduce two tasks:~\textit{Formula Concept Discovery (FCD)} and \textit{Formula Concept Recognition (FCR)} to (1) identify Formula Concepts and (2) find formulas which are instances of particular Formula Concepts.
We present implementations to automatically perform the FCD and FCR tasks using machine learning techniques.

\paragraph{Novelty of Contribution.}

This paper extends our previous publication~\cite{DBLP:conf/sigir/ScharpfSCG19}, in which we introduced the first FCD retrieval method implementation.
We extend our study of Formula Concepts by two additional FCD retrieval methods, three additional tasks, and the entire section on FCR experiments.
A strong focus of this work is placed on the in-depth analytical examination of example Formula Concepts. We discuss 36 different representations of the Klein--Gordon equation, Einstein's field equations, and Maxwell's equations.
Analyzing their differences, we identify 13 challenges for FCD to derive requirements for the practical implementation of an FCD framework.
Furthermore, we investigate the Formula Concept vector space of our examples in four different formula encodings (vector representations).
Additionally, we examine the separability or delineation of different Formula Concepts by computing classification accuracy (SVM classifier) and cluster purity ($k$-means clusterer). We also generate formula similarity maps in different encoding measures to illustrate FC class coherence.
Finally, we present and discuss several of our FCR implementations, including search rankings and additional machine learning methods.

%=====================================================================================
% Related Work
%=====================================================================================

\section{Related Work} \label{sec:rel.work}

This section reviews and explains some background knowledge necessary to understand this research project.
This includes our own preliminary work and achievements to tackle FCD, related methods of Mathematical Entity Linking, formula knowledge bases, STEM document dataset sources, and mathematical information system applications.

We recently introduced a first machine learning approach for Formula Concept Discovery~\cite{Scharpf2019b}. Using {\tt Doc2Vec}~\cite{DBLP:conf/icml/LeM14} encodings and $k$-means clustering, equivalent representations of formulas were retrieved and evaluated. The experiment was carried out on a selection of astrophysics papers from the NTCIR arXiv dataset~\cite{DBLP:conf/ntcir/AizawaKOS14}. We took formulas that occurred most often in the corpus (duplicates) as a cluster seed. Furthermore, for the major part of the test selection candidates, a valid Formula Concept name could be retrieved from the surrounding text. For almost all of the retrieved name candidates, a Wikidata QID was available.\footnote{%
The mention of specific products, trademarks, or brand names is for purposes of
identification only. Such mention is not to be interpreted in any way as an endorsement
or certification of such products or brands by the National Institute of Standards and
Technology, nor does it imply that the products so identified are necessarily the best
available for the purpose. All trademarks mentioned herein belong to their respective
owners.
}

In this paper, we extend our Formula Concept Discovery method by novel Formula Concept Recognition methods. Both approaches involve two steps: knowledge-base population and content referencing. These can both be described in terms of Mathematical Entity Linking (MathEL)~\cite{Scharpf2021,Scharpf2021b}. MathEL approaches link mathematical formulas to unique URLs in a semantic knowledge base. If the URLs are part of Wiki web resources, MathEL can be regarded as the `Wikification' of mathematical content~\cite{DBLP:conf/icadl/KristiantoTA16}.

In Natural Language Processing, Entity Linking entities are typically linked to Wikipedia
%NIST disclaimer
%\footnote{The mention of specific products, trademarks, or brand names is for purposes of identification only. Such mention is not to be interpreted as an endorsement or certification of such products or brands by the National Institute of Standards and Technology, nor does it imply that the products so identified are necessarily the best available for the purpose. All trademarks mentioned herein belong to their respective owners.}
%
with a variety of applications, such as Named Entity Recognition (NER), relationship extraction, entity summarization~\cite{DBLP:conf/amw/Rosales-MendezP18}.
In analogy, methods to link mathematical expressions in scientific documents to Wikipedia articles using their surrounding text have been developed~\cite{DBLP:conf/icadl/KristiantoTA16,DBLP:conf/wsdm/KristiantoA17}. One of the conclusions was that for the linking to be reliable, a balanced combination of textual and mathematical elements must be considered. As potential candidates for MathEL, Mathematical Objects of Interest (MOI) were defined to elaborate methods for their discovery~\cite{DBLP:conf/www/Greiner-PetterS20}.
MathEL is expected to enhance mathematical subject classification~\cite{DBLP:conf/jcdl/ScharpfSYHMG20,DBLP:conf/mkm/SchubotzSTKBG20}.

To implement our FCR methods, we employ Wikidata as the semantic grounding for Wikification (entity linking to Wiki web resources). Since Wikipedia is only semi-structured, Wikidata\footnote{\url{http://www.wikidata.org}} was launched to provide direct access to specific interlingual facts (RDF\footnote{\url{https://www.w3.org/RDF}} triples) and to retrieve information systematically.\\ Wikidata is a free and open semantic knowledge base that can be read and edited by humans and machines~\cite{DBLP:journals/cacm/VrandecicK14}. Wikidata stores items with statements and references. In the case of mathematical knowledge, this may include formulas. For example, one may describe the physics concept `pressure' (item ID Q39552) with a `defining formula' property (property ID P2534) $p = F/S$.
To scalably seed information into Wikidata, a primary sources tool (PST)\footnote{\url{https://www.wikidata.org/wiki/Wikidata:Primary\_sources\_tool}} was introduced. This tool allows active users to quickly browse through new claims and references in order to approve or reject their validity. Currently, Wikidata contains approximately 5,7K items with a `defining formula' property\footnote{\url{https://w.wiki/z8p}}.
%TODO (camera-ready version): update number

Besides Wikidata, other semantic databases exist that store mathematical formula knowledge. The NIST Digital Library of Mathematical Functions~\cite{NIST:DLMF} and NIST Digital Repository of Mathematical Formulae (DRMF)~\cite{DBLP:conf/mkm/CohlMSSW14} are two examples of maintained high-quality semantic datasets. Moreover, the benchmark \textsc{MathMLben}~\cite{DBLP:conf/jcdl/SchubotzGSMCG18} was created to evaluate tools for mathematical format conversion (from \LaTeX{} to \MathML{} to Computer Algebra Systems), containing almost 400 formulas from Wikipedia, the arXiv\footnote{\url{https://arxiv.org}}, and DLMF. These were augmented by Wikidata macros in~\cite{DBLP:conf/sigir/ScharpfSG18}.

Mathematical Information Retrieval (MathIR) systems address the information need of people working in STEM fields by retrieving, processing, and analyzing mathematical formulas~\cite{DBLP:conf/sigir/ScharpfSG18}. Up until now, various formula search engines have been developed. Furthermore, translations between different markups (\LaTeX{}, Presentation, and Content \MathML{}) and standards have been introduced~\cite{DBLP:journals/mics/GuidiC16}.
Schubotz et al. present a framework to translate \MathML{} into Computer Algebra System (CAS) syntax. Furthermore, standards like OpenMath\footnote{\url{http://openmath.org}} and OMCDoc\footnote{\url{https://mathweb.org}} provide extensible ways to represent the semantics of mathematical objects in mathematical documents~\cite{DBLP:books/sp/Kohlhase06}. They can be used to annotate formula expressions in definitions, theorems, and proofs. Given markup on object, statement, and theory level, the soundness of mathematical systems can be assessed~\cite{DBLP:conf/sigir/ScharpfSG18}. In addition, the PhysML variant accounts for the special characteristics of physics: observables, physical systems, and experiments~\cite{DBLP:conf/mkm/HilfKS06}.
Moreover, Mathematical Question Answering (MathQA) systems have been built~\cite{mathqajournal,DBLP:conf/jcdl/ScharpfSG22} to provide quick and concise formula answers to mathematical questions in natural language which are commonly asked on the web~\cite{DBLP:conf/clef/ScharpfSGOTG20}. MathQA systems can retrieve answers from unstructured text passages or structured knowledge bases. In the latter case, MathEL needs to be employed to assign natural language concept names to mathematical formulas. While classical math search engines typically map a mathematical language query (formula string) to a collection of web resources that include the natural language name of the Formula Concept~\cite{DBLP:conf/aisc/KohlhaseS06}, MathQA systems perform the reverse transformation from natural to mathematical language. Another application of the mapping from mathematical to natural language using MathEL is question generation~\cite{Scharpf2022b}.

For some Mathematical Language Processing (MLP) applications, the formula constituents (operators, identifiers, numbers) have to be annotated using Mathematical Markup Language (\MathML{}). There are several tools available to convert \LaTeX{} into \MathML{}, most prominently the \LaTeXML{} converter\footnote{\url{https://dlmf.nist.gov/LaTeXML}}. Furthermore, the occurring symbols (variables, constants) need to be disambiguated, i.e., their meaning inferred from the context by unsupervised retrieval or supervised annotation. There have been previous attempts to automatically retrieve the semantics of identifiers from the surrounding text~\cite{DBLP:conf/sigir/SchubotzGLCMGYM16,GreinerPetter2022a}. However, it was found that not all identifier names could be extracted from the text. To address this, Schubotz et al. cluster identifier namespaces to enable a fallback retrieval from the definition cluster. While Wikipedia articles commonly contain variable definitions in the text, many paper articles often omit them, assuming expert reader domain knowledge. To build machine-interpretable datasets, manual annotation is thus inevitable. Since this is very time-consuming, formula and identifier annotation recommender systems, such as `AnnoMathTeX'~\cite{Scharpf2019b,Scharpf2021} are built to speed up the process.

To create labeled formula data benchmarks, we need open access corpora of STEM documents. For research experiment reproducibility, snapshots must be defined. The arXiv.org e-Print archive~\cite{mckiernan2000arxiv} makes available free preprints for an extensive collection of publications from physics, mathematics, computer science, economics, and other fields. On the arXiv, many authors provide their \LaTeX{} source code. Both Wikipedia and arXiv articles were extracted as part of the NTCIR MathIR Task~\cite{DBLP:conf/ntcir/AizawaKOS14}. We employ the NTCIR arXiv dataset for our research in this paper.
In 2017, the Special Interest Group for Math Linguistics (SIGMathLing)\footnote{\url{https://sigmathling.kwarc.info}} was initiated as a forum and resource cooperative for the linguistics of mathematical or technical documents.

\section{Formula Concept Discovery} \label{sec:fcd}

In this section, we attempt to formally define a \textit{Formula Concept} and set up \textit{Formula Concept Retrieval Tasks}.

\subsection{Formula Concept Retrieval Tasks}

\paragraph{Definition.}

Following%
%Scharpf et al.
~\cite{DBLP:conf/sigir/ScharpfSG18}, we define the \textit{formula content} as the sets of operators, identifiers\footnote{\url{https://www.w3.org/TR/MathML3/chapter4.html\#contm.ci}}, and numbers that a formula contains.
Furthermore, we define a \textit{Formula Concept} as a collection of equivalent formulas with different representations featuring the same formula content (operators, identifiers, and numbers).
Consider the Klein--Gordon equation representations in Figure \ref{fig:Klein-Gordon-equation} as an example of a Formula Concept. Obviously, the formula content may vary as the occurring operators, identifiers, and numbers change from instance to instance.
Operators such as partial derivatives can be represented in several ways ($\partial^2 u / \partial t^2$ vs.~$u_{tt}$ vs.~$\ddot{u}$), identifiers can be subsumed into others (e.g., $\alpha = m c / \hbar$), and physical constants can be transformed to different unit systems (e.g., natural units with $\hbar = c = 1$).
The Formula Concept Discovery challenges will be discussed in more detail in Section \ref{sec:challengeidentification}.
This motivates our study to find out what equivalent representations can occur and how to handle them.

\paragraph{Tasks.}

Our goal is to map diverse representations of a formula to one unique Formula Concept ID\footnote{The Formula Concept ID (here Wikidata QID) for the whole formula must not be confused with a formula identifier, which is a constituent of the formula with no fixed value.}, e.g., linking all occurrences of the Klein--Gordon equation shown in Figure \ref{fig:Klein-Gordon-equation} to the Wikidata item Q868967\footnote{\url{https://www.wikidata.org/wiki/Q868967}}.
We define two subtasks of the \textit{Formula Concept Retrieval Task}:
\begin{itemize}
\item \textit{Formula Concept Discovery} is a method to find common equivalent representations and a name candidate for a given formula, and
\item \textit{Formula Concept Recognition} is an approach for recognizing formulas in documents as being instances of a previously defined Formula Concept.
\end{itemize}

In the following, we present our implementation and evaluation results for Formula Concept Discovery and Formula Concept Recognition. These results are based on analytical examinations, machine learning, fuzzy string matching, and Wikipedia article heuristics.

% OVERVIEW of FCD methods (mindmap)

%\begin{figure}
%    \centering
%    \includegraphics[width=\textwidth]{FCD_overview}
%    \caption{Overview of Formula Concept Retrieval Tasks}
%    \label{fig:FCD_overview}
%\end{figure}

For the discovery of Formula Concepts, we define the following four tasks:
\begin{enumerate}[{Task }1{:}]
    \item \label{tsc:exampleretrieval} %Task:~
    Retrieval of Formula Concept examples,
    \item \label{tsc:exampleanalysis} %Task:~
    Analysis of Formula Concept examples,
    \item \label{tsc:challengeidentification} 
    %Task:~
    Identification of Formula Concept Discovery challenges,
    \item \label{tsc:requirementderivation} 
    %Task:~
    Derivation of Formula Concept Retrieval system requirements.
\end{enumerate}

In Task \ref{tsc:exampleretrieval}, we employ three methods to retrieve examples of Formula Concepts, which are suitable for discussing and identifying challenges of Formula Concept Discovery and Formula Concept Recognition.
In Task \ref{tsc:exampleanalysis}, we analyze and discuss three selected Formula Concept examples.
We choose three sets of differential equations from physics: the Klein--Gordon equation (KGE), Einstein's field equations (EFE), and Maxwell's equations (ME).
The examples are retrieved from search engine results for the Formula Concept name yielding publications (sources as in Figure \ref{fig:Klein-Gordon-equation}), as well as from Wikipedia article content\footref{foot:efelink}, and a textbook~\cite{fliessbach1990allgemeine}.
Given our background in theoretical physics and applied mathematics, we choose examples from this domain. Since we are domain experts on the topics, we can judge the Formula Concept semantics.
The formula annotation is achieved in a two-step process: 1) the retrieval by the concept name in the selected sources determines the annotation or assignment of the whole formula; 2) the domain expert subsequently semantically analyzes the formula and retrieves the semantic annotations of the formula constituents by considering the context and descriptions or explanations from the respective sources (text surrounding the formula).
In Task \ref{tsc:challengeidentification}, we identify and summarize the Formula Concept Discovery challenges, which we observe in the discussion of the three Formula Concept examples.
These challenges determine the requirements for technical implementations of FCD and FCR.
In Task \ref{tsc:requirementderivation}, we address the identified challenges by deriving requirements for a Formula Concept Retrieval system and proposing methods to tackle the challenges.

The developed algorithms, the dataset, and full result tables are available at \url{https://github.com/ag-gipp/formula-concept-retrieval}.
%TODO: add "under a Apache-2.0 license"?

\subsection{Task \ref{tsc:exampleretrieval}:~Retrieval of Formula Concept Examples}

For the retrieval of example Formula Concepts, we employ the following three methods:
\begin{enumerate}[{Method }1{:}]
    \item \label{mth:searchbyconceptname} %Method:~
    Search by Formula Concept Name,
    \item \label{mth:knearestneighbor} %Method:~
    $k$-Nearest-Neighbors ($k$NN) in Formula Vector Space,
    \item \label{mth:firstformula} %Method:~
    Wikipedia article First Formula Multi-Language Heuristic.
\end{enumerate}

In Method \ref{mth:searchbyconceptname}, we perform searches by the Formula Concept name in a corpus of publications, a Wikipedia article, and a textbook, respectively.
In Method \ref{mth:knearestneighbor}, we employ machine learning to retrieve equivalent representations of formulas~\cite{Scharpf2019b}, which occur most often (duplicates) in a selected corpus containing astrophysics papers from the NTCIR arXiv dataset~\cite{DBLP:conf/ntcir/AizawaKOS14}. For an introduction of the dataset, see the paragraph `Data selection' in Section \ref{parag:knn}.
In Method \ref{mth:firstformula}, we make use of a simple heuristic~\cite{schubotz2018introducing,DBLP:conf/icms/Halbach20}.
We take the tentative Formula Concept names of the examples retrieved using Method \ref{mth:knearestneighbor}.
We then extract the corresponding Wikipedia articles.
For each Formula Concept article, we retrieve the first five versions in different languages.
We then assess how many different representations of the individual Formula Concepts are among these articles.

\subsubsection{Method \ref{mth:searchbyconceptname}:~Search by Formula Concept Name}

For our first example, the Klein--Gordon equation, we perform a web search to retrieve ten representations from publications~\cite{arbab2010derivation,detweiler1980klein,haroun2017derivation,kaloyerou1989evolution,morawetz1968time,pecher1984nonlinear,strauss1978numerical,tiwari1988derivation,tretyakov2010derivation}.
Each publication contains the Formula Concept name as a keyword or in the full text.
For our second example, Einstein's field equations, we retrieve representations from the corresponding Wikipedia article\footnote{\label{foot:efelink}Available at \url{https://en.wikipedia.org/wiki/Einstein_field_equations}.}.
For our third example, Maxwell's field equations, we take derivations from a textbook on General Relativity~\cite{fliessbach1990allgemeine}.

\subsubsection{Method \ref{mth:knearestneighbor}:~$k$-Nearest-Neighbors in Formula Vector Space}

This subsection is based on our previous publication~\cite{Scharpf2019b}, in which we presented Formula Concept Discovery using $k$-Nearest-Neighbors for the first time.
Since it might be impossible to formally define all equivalence transformations exhaustively, we test approaching a Formula Concept in machine learning terms as a collection of approved formula vectors (comparing encodings) within a specified similarity range (comparing metrics).
We illustrate the formula space (formula content space in Figure \ref{fig:formcontspace} and formula semantic space in Figure \ref{fig:formsemspace}) in Experiment \ref{exp:FCclassclust} of FCR in Section \ref{subsec:FCclassclust}.
It represents formulas as encoded vectors.
Then, a Formula Concept can be defined as all vectors around a central vector within a specified distance (cutoff).

\paragraph{Method.}

We approach the discovery of Formula Concepts by retrieving equivalent formulations with different representations using machine learning (see Figure \ref{fig:ConcClust}).
The retrieved instances are augmented with name candidates from the surrounding text.
The initial step is to identify formula candidates that occur most often within a given dataset.
We assume that they are potential seeds of popular Formula Concepts.
We first tried formula clustering~\cite{adeel2012efficient,ma2010feature}.
However, we discovered that this was not a suitable method for FCD since the number of clusters is a priory unclear%
\footnote{However, in Experiment \ref{exp:FCclassclust} (Section \ref{subsec:FCclassclust}), we employ $k$-means clustering with a known number ($k=3$) of clusters.}.
The tested algorithms are not able to group equivalent formulas.
Subsequently, we decided to start with a ranking of formula duplicates (with the same \LaTeX{} string).
In contrast to the clustering, this yields valuable results for the selected Formula Concept examples.

\begin{figure}[tb]
\centering
\renewcommand{\arraystretch}{1.5}
\begin{tabular}{|c|c|}
\hline
\wdl{\tt{Hubble's law}}{\tt Q179916} & \wdl{\tt{Equation of state}}{\tt Q214967} \\  \hline
{{$\! \begin{aligned}
%  \nonumner \\
  p &= \omega \rho \nonumber \\
  p &= \kappa \rho  \nonumber\\
  \omega &= p / \rho  \nonumber\\
  p_d &= \omega \rho_d  \nonumber\\
%    \nonumner \\
\end{aligned}$}} &  {$\! \begin{aligned}
%  \nonumner \\
   \dot{a} &= a H  \nonumber\\
   H_i &= \dot{R} / R  \nonumber\\
   H &= \dot{a} / a  \nonumber\\
   H(t) &= \dot{a} / a  \nonumber\\
%     \nonumner\\
 \end{aligned}$} \\\hline
\end{tabular}
\caption{Clustering equivalent representations of formulas in the semantic space as named Formula Concept Wikidata items.}
\label{fig:ConcClust}
\end{figure}

\paragraph{Data Selection.}
\label{parag:knn}

We employ the NTCIR arXiv dataset~\cite{DBLP:conf/ntcir/AizawaKOS14}, which comprises 105,120 document sections containing over 60 million formulas. The formulas are enclosed in \verb|<math>| tag environments. The documents were converted from \LaTeX{}  to an XHTML format (\url{https://tei-c.org}). The disk size of the dataset is about 174GB uncompressed.
We confine our computations to the subject class of astrophysics (680 {\tt astro-ph} documents), employing a domain expert to evaluate the results semantically.
To get the most popular formulas in the dataset as potential candidates for important Formula Concepts, we first identify duplicates where the exact formula string reoccurs in multiple documents. We subsequently rank the results by their occurrence frequency, i.e., the number of duplicates $d$ (see the respective column in Table \ref{tab:FCD}).
From the duplicate ranking, we select a formula length range between 10 and 30 characters\footnote{Expressions with less than ten characters are often not equations, and identical formulas with more than 30 characters are rare.}
and restrict our selection to duplicates occurring in at least two documents $D \geq 2$.
This selection criteria processing results in 3,495 formulas.
We then manually select all equations (for now, we confine the Formula Concept definition to include equations only).
We discard all stubs without a right-hand side, as well as simple variable dependence definitions, such as $x = x\left(t\right)$ and $x=y$ or $x = \mathrm{const}$.
The algorithms for the data selection pipeline can be found in the source repository.

\paragraph{Evaluation.}

For the first 50 samples from the duplicate ranking, we retrieve the operators and identifiers from the provided \MathML{} \verb|<mo>| and \verb|<mi>| tags, as well as the surrounding text (words within a window of $\pm500$ characters around the formula).
We encode both tag contents using the \textit{TfidfVectorizer} from the Python package \textit{Scikit-learn}~\cite{DBLP:journals/jmlr/PedregosaVGMTGBPWDVPCBPD11} and {\tt Doc2Vec} model~\cite{DBLP:conf/icml/LeM14} from the Python package \textit{Gensim}~\cite{vrehuuvrek2011scalability}.
We then assess the performance of a $k$-Nearest-Neighbors classifier~\cite{shakhnarovish2005nearest} to retrieve equivalent formula representations.
For a given instance of a Formula Concept, we compute the $k$-Nearest-Neighbors formulas as candidates for variations of that Formula Concept.
Subsequently, we use our domain knowledge to judge whether these candidates are indeed equivalent representations of the given Formula Concept.
We test the effectiveness of our approach on four different formula vector encodings:
\begin{itemize}
    \item {\tt{math2vec}} encoding the formula constituents using the {\tt{Doc2Vec}} model as proposed in~\cite{DBLP:conf/mkm/YoussefM18};
    \item {\tt{math tf-idf}} encoding the formula constituents using the {\tt{TfidfVectorizer}};
    %for the formulas
    \item {\tt{semantics2vec}} encoding the surrounding text (containing tentative formula semantics) using  the {\tt{Doc2Vec}} model; and
    \item {\tt{semantics tf-idf}} encoding the surrounding text using the {\tt{TfidfVectorizer}}.
    %for the surrounding text
\end{itemize}

The computation of the {\tt Doc2Vec} formula vector encodings is more time-expensive than TF-IDF, due to the iterative learning process of the neural model.

\paragraph{Results.}

Table \ref{tab:FCD} shows the results of our approach for discovering Formula Concepts as published before~\cite{DBLP:conf/sigir/ScharpfSCG19}.
We rank the extracted formulas by the number of duplicates $d$ and list the number of documents $D$, in which they appear. Note that the likelihood of retrieving non-duplicate equivalent representations increases for higher values of distinct documents. If the Formula Concept representations are found in different documents, there are more than if they appear in the same document. This means that there are fewer variations within the same document.
We can see that only for the first 18 Formula Concept examples are there more than two duplicates from distinct documents, i.e., formulas appearing twice or more within the corpus. We evaluate the first 50 examples. The primary investigation was to compare the performance of four different formula vector encodings in terms of the retrieved number of equivalent representations.
In total, we can retrieve 163 equivalent Formula Concept representations for our 50 samples.
On average, this corresponds to more than three (163/50 = 3.3) per formula (from 3 different documents) or around one (163/50/4 = 0.8) per source per formula. Some of the retrieved formulas even contain different identifier symbols or varying indices (e.g., \verb|a| is replaced by \verb|R| in Example \ref{ex:KGE}, see the first line of Table \ref{tab:FCD}).
Increasing the number of formula neighbors parameter $k$ from 1 in integer steps, we can not find additional matching representations above $k=9$.
We define the retrieval success $s$ of an individual encoding as the percentage of retrieved representations compared to all other formula vector encodings. Calculating the overall success distribution, we discover that the \textit{math2vec} ($e_m$) encoding distinctively outperforms the others by yielding 71\% of the retrieved instances, followed by {\tt{semantics tf-idf}}, ($\hat e_s$) with 15\%, \textit{semantics2vec} ($e_s$) with 11\%, and {\tt{math tf-idf}} ($\hat e_m$) with 4\%.
Overall, for 34 of the investigated 50 sample formulas, i.e., 34/50 = 68\%, we are able to retrieve equivalent representations.
We conclude that while the math2vec encoding retrieves the most equivalent formula matches as candidates for a Formula Concept, it is most effective to employ all formula vector encodings simultaneously to maximize the retrieval.
Note that we can only determine false positives and compute precision but not the number of false negatives to compute recall. This is because we do not know a priori how many different equivalent representations, semantically close to the examined concept, still exist. We can neither determine this in general (how many notational variations are possible in principle) nor for the given corpus (how many do occur).
Finally, we list the top five name candidates from the surrounding text. The word window size is chosen to be $\pm$500 characters. Decreasing the window size in steps of 100, the top 50 coverage performance drops from 100\% to 17\% to 11\% for $\text{ws} = \{500, 400, 300\}$ to $\text{ws} = 200$ to $\text{ws} = 100$ respectively.
We evaluate whether they contain a suitable name for the Formula Concept to be seeded as a Wikidata item.
For our 50 Formula Concept examples, we achieve a recall of 36/50 = 72\% for the formula name.
Furthermore, for 41/50 = 82\% of the retrieved name candidates, a Wikidata QID is available to tag the Formula Concept.

% macros for following table

\newcommand*\rot{\rotatebox{0}}
\newcommand{\rotp}[1]{\rot{\parbox{1.5cm}{#1}}}
\newcommand{\rotpI}[1]{#1}
\newcommand{\sample}[2]{#2: #1\hfill}
\newcommand{\NA}[0]{N/A}
\newcommand{\sampleN}[0]{\NA}
\newcommand{\cand}[1]{#1}
\newcommand{\docs}[2]{#1 / #2}
\sisetup{round-integer-to-decimal,round-mode=places,round-precision=1 }
\newcommand{\round}[1]{\num{#1}} %SIunitx
\newcommand{\score}[4]{\round{#1}, \round{#2}, \round{#3}, \round{#4} }

\newcommand{\mathVec}[0]{\ensuremath{{e_m}}}
\newcommand{\mathTfidf}[0]{\ensuremath{{\hat e_m}}}
\newcommand{\semVec}[0]{\ensuremath{{e_s}}}
\newcommand{\semTfidf}[0]{\ensuremath{{\hat e_s}}}

%\vspace{10pt}

%Restore last column (name candidates) for Diss!

\begin{table*}[ht]
%\vspace{20pt}
\caption{Formula Concept Discovery~\cite{DBLP:conf/sigir/ScharpfSCG19}. Top-50 results of a cross-document duplicate search in the subject class {\tt astro-ph} of the NTCIR arXiv dataset. Equivalent formulas are retrieved to bundle mathematical concept candidates using a $k$-Nearest-Neighbors ($k$NN) recommendation, while comparing the relative success $s$ of different formula vector encodings (\textit{math2vec}:~$e_m$, {\tt{math tf-idf}}:~$\hat e_m$, \textit{semantics2vec}:~$e_s$, {\tt{semantics tf-idf}}:~$\hat e_s$). The number of duplicates $d$ and originating distinct documents $D$ are shown as well as a retrieved sample formula. Furthermore, it is evaluated whether the first five words of the surrounding text are candidates for the formula's name, and whether a Wikidata QID is available.}
\label{tab:FCD}
\resizebox{\textwidth}{!}{%
\def\arraystretch{1.2}% we could make this slightly smaller
\begin{tabular}{|l|l|l|l|l|l|l|l|l|l|}
\hline
\textbf{Nr.} & \textbf{Formula} &  \textbf{Name (QID)} & $d$ / $D$ &  $ s_{\mathVec}, s_{\mathTfidf}, s_{\semVec}, s_{\semTfidf}$ & Encoding:~sample \\ \hline
1 & $H=\dot{a}/a$ & \wdl{\tt{Hubble parameter}}{\tt Q179916} & \docs{32}{32} & \score{0}{0.1}{0}{ 0.9} & 
    \sample{$H_{i}=\dot{R}/R$}{ \semTfidf }  \\ \hline
2 & $p=\omega\rho$ & \wdl{\tt{Equation of state}}{\tt Q214967} & \docs{6}{5} & \score{0.29}{0}{0.14}{ 0.57} & 
    \sample{$p_{d}=w\rho_{d}$}{ \semVec }  \\ \hline
3 & $\omega=p/\rho$ & \wdl{\tt{Accelerating universe}}{\tt Q1049613} & \docs{4}{3} & \score{0.67}{0}{0}{ 0.33} & 
    \sample{$p=\omega\rho$}{ \mathVec }  \\ \hline
4 & $p=-A/\rho^{\alpha}$ & \wdl{\tt{Dark fluid}}{\tt Q5223514} & \docs{4}{4} & \score{0.67}{0}{0.33}{ 0} & 
    \sample{$p=-\frac{A}{\rho^{\alpha}}$}{ \mathVec } \\ \hline
5 & $p_{d}=w\rho_{d}$ & \wdl{\tt{Dark energy}}{\tt Q18343} & \docs{4}{3} & \score{0.33}{0}{0.33}{ 0.33} &
    \sample{$p_{X}=\omega_{X}\rho_{X}$}{ \semVec }  \\ \hline
6 & $H={\dot{a}}/a$ & \wdl{\tt{Hubble's law}}{\tt Q179916} & \docs{4}{4} & \score{0.4}{0.1}{0.2}{ 0.3} &
    \sample{${\mathcal H}=a^{\prime}/a$}{ \mathTfidf }  \\ \hline
7 & $k=|{\bf k}|$ & \wdl{\tt{Wavenumber}}{\tt Q192510} & \docs{3}{3} & \score{0.83}{0}{0.17}{ 0} &
    \sample{$k=|\vec{k}|$}{ \mathVec }  \\ \hline
8 & $f=e^{-\phi}R$ & \NA & \docs{3}{2} & \score{1}{0}{0}{ 0} &
    \sample{$f(\phi)=e^{-\phi}R$}{ \mathVec }  \\ \hline
9 & $p=\kappa\rho$ & \wdl{\tt{Equation of state}}{\tt Q214967} & \docs{3}{2} & \score{0.33}{0}{0.67}{ 0} &
    \sample{$p_{D}=w(z)\rho_{D}$}{ \semVec }  \\ \hline
10 & $w=p_{X}/\rho_{X}$ & \wdl{\tt{Equation of state}}{\tt Q214967} & \docs{3}{3} & \score{0.62}{0}{0.12}{ 0.25} &
    \sample{$p_{X}=w_{X}\rho_{X}$}{ \mathVec }  \\ \hline
11 & $\mu=m_{p}/m_{e}$ & \wdl{\tt{Proton-to-electron mass ratio}}{\tt Q2912520} & \docs{3}{3} & \score{1}{0}{0}{ 0} &
    \sample{$m_{i}=\mu m_{p}$}{ \mathVec } \\ \hline
12 & $\phi_{c}=M/g$ & \wdl{\tt{Critical value}}{\tt Q2189464} & \docs{3}{3} & \score{0}{0}{0}{ 0} &
    \sampleN  \\ \hline
13 & $p=-\frac{A}{\rho^{\alpha}}$ & \wdl{\tt{Chaplygin gas}}{\tt Q5073250} & \docs{3}{3} & \score{0.8}{0}{0}{ 0.2} &
    \sample{$p=-A\rho^{-\alpha}$}{ \mathVec }  \\ \hline
14 & $p=\alpha\rho$ & \wdl{\tt{Polytropic gas}}{\tt Q831024} & \docs{3}{2} & \score{0.67}{0}{0.17}{ 0.17} &
    \sample{$w_{\alpha}=p_{\alpha}/\rho_{\alpha}$}{ \semTfidf }  \\ \hline
15 & $M=\widetilde{M}/\Gamma$ & \wdl{\tt{Connected manifold}}{\tt Q2721559} & \docs{3}{3} & \score{0}{0}{0}{ 0} &
    \sampleN  \\ \hline
16 & $g(a)=\bigtriangleup(a)/a$ & \wdl{\tt{Dark energy}}{\tt Q18343} & \docs{3}{2} & \score{1}{0}{0}{ 0} &
    \sample{$g(a)=\Delta(a)/a$}{ \mathVec } \\ \hline
17 & $\alpha=dn_{s}/d\ln{k}$ & \wdl{\tt{Wavenumber}}{\tt Q192510} & \docs{3}{3} & \score{1}{0}{0}{ 0} &
    \sample{$dn_{s}/d\ln k=\alpha_{s}$}{ \mathVec } \\ \hline
18 & $\psi=-i\theta$ & \NA & \docs{3}{2} & \score{0}{0}{0}{ 0} &
    \sampleN \\ \hline
19 & $dt=a(\eta)d\eta$ & \wdl{\tt{Time}}{\tt Q11471} & \docs{2}{2} & \score{0.5}{0}{0.25}{ 0.25} &
    \sample{$t=\int a(\eta)d\eta$}{ \semTfidf } \\ \hline
20 & $\Delta x_{min}=\sqrt{\beta}$ & \wdl{\tt{Lower bound}}{\tt Q21067468} & \docs{2}{2} & \score{1}{0}{0}{ 0} &
    \sample{$\Delta x_{\rm min}=\hbar\sqrt{\beta}$}{ \mathVec } \\ \hline
21 & $k^{i}=ap^{i}$ & Modes (\NA) & \docs{2}{2} & \score{0}{0}{0}{ 0} &
    \sampleN \\ \hline
22 & $\varphi=\delta A_{\mu}$ & \wdl{\tt{Perturbations}}{\tt Q911364} & \docs{2}{2} & \score{0}{0}{0}{ 0} &
    \sampleN \\ \hline
23 & $h_{ab}=g_{ab}-n_{a}n_{b}$ & \wdl{\tt{Metric}}{\tt Q865746} & \docs{2}{2} & \score{0}{0}{0}{ 0} &
    \sampleN \\ \hline
24 & $K=K_{ab}h^{ab}$ & \wdl{\tt{Brane}}{\tt Q385601} & \docs{2}{2} & \score{1}{0}{0}{ 0} &
    \sample{$K=K_{\alpha\beta}h^{\alpha\beta}$}{ \mathVec } \\ \hline
25 & $v=\sqrt{|dp/d\rho|}$ & \wdl{\tt{Equation of state}}{\tt Q214967} & \docs{2}{2} & \score{1}{0}{0}{ 0} &
    \sample{$v_{c}=\sqrt{dp_{c}/d\rho_{c}}$}{ \mathVec } \\ \hline
26 & $Q=\sqrt{G}M$ & \wdl{\tt{Limit}}{\tt Q246639} & \docs{2}{2} & \score{0}{0}{0}{ 0} &
    \sampleN \\ \hline
27 & $\zeta=H\delta\phi/\dot{\phi}$ & \wdl{\tt{Perturbation theory}}{\tt Q10886678} & \docs{2}{2} & \score{1}{0}{0}{ 0} &
    \sample{${\mathcal R}=(H/\dot{\phi})\delta\phi_{\psi}$}{ \mathVec } \\ \hline
28 & $m_{\gamma}=e/\sqrt{\pi}$ & \wdl{\tt{Photon mass}}{\tt Q3198} & \docs{2}{2} & \score{0}{0}{0}{ 0} &
    \sampleN \\ \hline
29 & $d\eta=dt/a(t)$ & \wdl{\tt{Conformal time}}{\tt Q2482717} & \docs{2}{2} & \score{0.56}{0}{0.11}{ 0.33} &
    \sample{$t=\int a(\eta)d\eta$}{ \semTfidf } \\ \hline
30 & $T_{g}=H_{o}t_{g}$ & \wdl{\tt{Dimensionless quantity}}{\tt Q126818} & \docs{2}{2} & \score{0}{0}{0}{ 0} &
    \sampleN \\ \hline
31 & ${\mathcal H}=a^{\prime}/a$ & \wdl{\tt{Hubble's law}}{\tt Q179916} & \docs{2}{2} & \score{0.7}{0}{0.1}{ 0.2} &
    \sample{$H={\dot{a}}/a$}{ \semTfidf } \\ \hline
32 & $\theta=A\exp(-\zeta t)$ & \wdl{\tt{Exponential decrease}}{\tt Q574576} & \docs{2}{2} & \score{0}{1}{0}{ 0} &
    \sample{$\psi(t,r)=\psi(r)\exp(-i\omega t)$}{ \mathTfidf } \\ \hline
33 & $p_{i}=\omega_{i}\rho_{i}$ & \NA & \docs{2}{2} & \score{0.71}{0}{0.14}{ 0.14} &
    \sample{$w_{\rm X}=p_{\rm X}/\rho_{\rm X}$}{ \semVec } \\ \hline
34 & $i\partial_{t}\Phi=H\Phi$ & \wdl{\tt{Schr{\"o}dinger evolution}}{\tt Q165498} & \docs{2}{2} & \score{0}{0}{0}{ 0} &
    \sampleN \\ \hline
35 & $H(t)=\dot{a}/a$ & \wdl{\tt{Hubble's law}}{\tt Q179916} & \docs{2}{2} & \score{0.75}{0.12}{0}{ 0.12} &
    \sample{$\dot{a}=aH$}{ \mathVec } \\ \hline
36 & $p_{\Lambda}=-\rho_{\Lambda}$ & \wdl{\tt{Dark energy}}{\tt Q18343} & \docs{2}{2} & \score{1}{0}{0}{ 0} &
    \sample{$p_{D}=-\rho_{D}$}{ \mathVec } \\ \hline
37 & $P_{M}=w\rho_{M}$ & \wdl{\tt{Equation of state}}{\tt Q214967} & \docs{2}{2} & \score{0.57}{0}{0.29}{ 0.14} &
    \sample{$p_{x}=w\rho_{x}$}{ \semVec } \\ \hline
38 & $f_{\nu}=\rho_{\nu}/\rho_{d}$ & \wdl{\tt{Neutrino}}{\tt Q2126} & \docs{2}{2} & \score{0}{0}{0}{ 0} &
    \sampleN \\ \hline
39 & $A_{t}=rA_{s}$ & \wdl{\tt{fluctuation}}{\tt Q5462624} & \docs{2}{2} & \score{0}{0}{0}{ 0} &
    \sampleN \\ \hline
40 & $p_{m}=\gamma\rho_{m}$ & \wdl{\tt{Nonrelativistic matter}}{\tt Q55921784} & \docs{2}{2} & \score{1}{0}{0}{ 0} &
    \sample{$\gamma=p/\rho$}{ \mathVec } \\ \hline
41 & $\Omega_{i}=\rho_{i}/\rho_{c}$ & Expansion rate (\NA) & \docs{2}{2} & \score{1}{0}{0}{ 0} &
    \sample{$\Omega=\rho/\rho_{\rm crit}$}{ \mathVec } \\ \hline
42 & $P(k)=Ak^{n}$ & \wdl{\tt{Inflation}}{\tt Q273508} & \docs{2}{2} & \score{0}{0}{0}{ 0} &
    \sampleN \\ \hline
43 & $L_{I}=M(\tau)\phi[x(\tau)]$ & \NA & \docs{2}{2} & \score{0}{0}{0}{ 0} &
    \sampleN \\ \hline
44 & $L=\kappa h_{ab}T^{ab}$ & \NA & \docs{2}{2} & \score{0}{0}{0}{ 0} &
    \sampleN \\ \hline
45 & $w_{i}=P_{i}/\rho_{i}$ & \wdl{\tt{Equation of state}}{\tt Q214967} & \docs{2}{2} & \score{0.67}{0}{0.22}{ 0.11} &
    \sample{$w_{\alpha}=p_{\alpha}/\rho_{\alpha}$}{ \semTfidf } \\ \hline
46 & ${\bar{M}}={B}/{C}$ & \NA & \docs{2}{2} & \score{0.33}{0}{0.33}{ 0.33} &
    \sample{${\bar{M}}=\frac{B}{C}$}{ \semVec } \\ \hline
47 & $\Psi=\Psi_{\ell}+\Psi_{s}$ & \NA & \docs{2}{2} & \score{0}{0}{0}{ 0} &
    \sampleN \\ \hline
48 & $z=a\dot{\phi}/H$ & \wdl{\tt{Equation}}{\tt Q11345} & \docs{2}{2} & \score{0.67}{0}{0}{ 0.33} &
    \sample{$z_{q}=a\dot{\phi}/H$}{ \semTfidf } \\ \hline
49 & $u^{\mu}=dx^{\mu}/d\tau$ & \wdl{\tt{Comoving fluid}}{\tt Q5462744} & \docs{2}{2} & \score{1}{0}{0}{ 0} &
    \sample{$k^{\mu}=dx^{\mu}/dv$}{ \mathVec } \\ \hline
50 & $\dot{\phi}=-W_{\phi}$ & \wdl{\tt{Firstorder differential equation}}{\tt Q11214} & \docs{2}{2} & \score{1}{0}{0}{ 0} &
    \sample{$\dot{\chi}=-W_{\chi}$}{ \mathVec } \\ \hline
\end{tabular}}
\end{table*}

%Einstein's field equations (arXiv) by kNN vector search

\subsubsection{Method \ref{mth:firstformula}:~Wikipedia Article First Formula Multi-Language Heuristic}

Table \ref{tab:1stWikipedia} shows another approach to discover Formula Concepts.
We employ the tentative mathematical concept name candidate and retrieve the corresponding English Wikipedia article.
For each Formula Concept article, we retrieve the first five versions in different languages.
We then assess how many of these contain a first formula that is a different representation of the Formula Concept.
As an example, for formula number 1, the `Hubble parameter', the English article's first formula is \verb|v = H_0 D|, while in the German it is \verb|H(t) = \frac{\dot a(t)}{a(t)}|.
We show the success score $s$ in the last column.
It is the fraction of different representations within the first five language versions.
On average, a Formula Concept appears in two different representations. 
In our evaluation, we leave out all formulas, for which no concept name is available (N/A), to search for Wikipedia articles (-).
For the 32 formulas, for which we can select a Formula Concept name from the surrounding text candidates, we find 155 Wikipedia articles (for some names, there are less than five language versions available).
In total, 53/155 = 34\% of the individual versions contain Formula Concept variations.
This corresponds to 19/32 = 59\% of the formulas.
The results indicate that it is in principle possible to retrieve Formula Concept representations via Wikipedia article first formula multi-language heuristic. However, this does not work for a significant part of the sample. Our finding aligns with previous results in the literature~\cite{DBLP:conf/icms/Halbach20}, which report that considering multiple Wikipedia languages decreases both precision and recall compared to using only English Wikipedia.

%formulaeFCD_50samples_1stWikipedia.csv
\begin{table}[ht]
\caption{Formula Concept Discovery via Wikipedia article first formula multi-language heuristic. We assess whether the first formulas in different language versions of the Wikipedia article are different representations of a chosen Formula Concept. The success score $s$ is shown in the last column as the fraction of different representations within the first five language versions. On average, a Formula Concept appears in two different representations and 34\% of the individual versions contain Formula Concept variations. Formulas for which no mathematical concept name is available (N/A) are omitted (-) in the evaluation.}
\label{tab:1stWikipedia}
\resizebox{\textwidth}{!}{%
\def\arraystretch{1.0}% we could make this slightly smaller
\begin{tabular}{|l|l|l|l|l|}
\hline
\textbf{Nr.} & \textbf{Formula}                                                                                                   & \textbf{Formula name candidate}           & \textbf{Wikidata QID} & \textbf{s} \\ \hline
1   & \tt{H=\textbackslash{}dot\{a\}/a}                                                                              & \tt{Hubble parameter}                 & {\tt Q179916}      & 3/5                           \\ \hline
2   & \tt{p=\textbackslash{}omega\textbackslash{}rho}                                                                & \tt{Equation of state}                & {\tt Q214967}      & 4/5                           \\ \hline
3   & \tt{\textbackslash{}omega=p/\textbackslash{}rho}                                                               & \tt{Accelerating universe}            & {\tt Q1049613}     & 0/5                           \\ \hline
4   & \tt{p=-A/\textbackslash{}rho\textasciicircum{}\{\textbackslash{}alpha\}}                                       & \tt{Dark fluid    }                   & {\tt Q5223514}     & 0/5                           \\ \hline
5   & \tt{p\_\{d\}=w\textbackslash{}rho\_\{d\}}                                                                      & \tt{Dark energy}                      & {\tt Q18343}       & 0/5                           \\ \hline
6   & \tt{H=\{\textbackslash{}dot\{a\}\}/a}                                                                          & {\tt N/A}                              & {\tt Q179916}      & -                             \\ \hline
7   & \tt{k=|\{\textbackslash{}bf k\}|}                                                                              & \tt{Wavenumber}                       & {\tt Q192510}      & 2/5                           \\ \hline
8   & \tt{f=e\textasciicircum{}\{-\textbackslash{}phi\}R}                                                            & {\tt N/A}                              & {\tt N/A}          & -                             \\ \hline
9   & \tt{p=\textbackslash{}kappa\textbackslash{}rho}                                                                & \tt{Equation of state}                & {\tt Q214967}      & 4/5                           \\ \hline
10  & \tt{w=p\_\{X\}/\textbackslash{}rho\_\{X\}}                                                                     & \tt{Equation of state}                & {\tt Q214967}      & 4/5                           \\ \hline
11  & \tt{\textbackslash{}mu=m\_\{p\}/m\_\{e\}}                                                                      & \tt{Proton-to-electron mass ratio}    & {\tt Q2912520}     & 1/5                           \\ \hline
12  & \tt{\textbackslash{}phi\_\{c\}=M/g}                                                                            & \tt{Critical value}                   & {\tt Q2189464}     & 0/5                           \\ \hline
13  & \tt{p=-\textbackslash{}frac\{A\}\{\textbackslash{}rho\textasciicircum{}\{\textbackslash{}alpha\}\}}            & \tt{Chaplygin gas}                    & {\tt Q5073250}     & 1/5                           \\ \hline
14  & \tt{p=\textbackslash{}alpha\textbackslash{}rho}                                                                & \tt{Polytropic gas}                   & {\tt Q831024}      & 4/5                           \\ \hline
15  & \tt{M=\textbackslash{}widetilde\{M\}/\textbackslash{}Gamma}                                                    & \tt{Connected manifold}               & {\tt Q2721559}     & 0/5                           \\ \hline
16  & \tt{g(a)=\textbackslash{}bigtriangleup(a)/a}                                                                   & \tt{Dark energy}                      & {\tt Q18343}       & 0/5                           \\ \hline
17  & \tt{\textbackslash{}alpha=dn\_\{s\}/d\textbackslash{}ln\{k\}}                                                  & {\tt N/A}                              & {\tt Q192510}      & -                             \\ \hline
18  & \tt{\textbackslash{}psi=-i\textbackslash{}theta}                                                               & {\tt N/A}                              & {\tt N/A}          & -                             \\ \hline
19  & \tt{dt=a(\textbackslash{}eta)d\textbackslash{}eta}                                                             & {\tt N/A}                              & {\tt Q11471}       & -                             \\ \hline
20  & \tt{\textbackslash{}Delta x\_\{min\}=\textbackslash{}sqrt\{\textbackslash{}beta\}}                             & \tt{Lower bound}                      & {\tt Q21067468}    & 0/5                           \\ \hline
21  & \tt{k\textasciicircum{}\{i\}=ap\textasciicircum{}\{i\}}                                                        & \tt{Modes}                            & {\tt N/A}          & 0/5                           \\ \hline
22  & \tt{\textbackslash{}varphi=\textbackslash{}delta A\_\{\textbackslash{}mu\}}                                    & \tt{Perturbations}                    & {\tt Q911364}      & 0/5                           \\ \hline
23  & \tt{h\_\{ab\}=g\_\{ab\}-n\_\{a\}n\_\{b\}}                                                                      & \tt{Metric}                           & {\tt Q865746}      & 1/5                           \\ \hline
24  & \tt{K=K\_\{ab\}h\textasciicircum{}\{ab\}}                                                                      & \tt{Brane}                            & {\tt Q385601}     & 1/5                           \\ \hline
25  & \tt{v=\textbackslash{}sqrt\{|dp/d\textbackslash{}rho|\}}                                                       & \tt{Equation of state}                & {\tt Q214967}      & 4/5                           \\ \hline
26  & \tt{Q=\textbackslash{}sqrt\{G\}M}                                                                              & \tt{Limit}                            & {\tt Q246639}      & 4/5                           \\ \hline
27  & \tt{\textbackslash{}zeta=H\textbackslash{}delta\textbackslash{}phi/\textbackslash{}dot\{\textbackslash{}phi\}} & {\tt N/A}                              & {\tt N/A}    & -                             \\ \hline
28  & \tt{m\_\{\textbackslash{}gamma\}=e/\textbackslash{}sqrt\{\textbackslash{}pi\}}                                 & \tt{Photon mass}                      & {\tt Q3198}        & 0/5                           \\ \hline
29  & \tt{d\textbackslash{}eta=dt/a(t)}                                                                              & \tt{Conformal time}                   & {\tt Q2482717}     & 2/5                           \\ \hline
30  & \tt{T\_\{g\}=H\_\{o\}t\_\{g\}}                                                                                 & {\tt N/A}                              & {\tt N/A}      & -                             \\ \hline
31  & \tt{\{\textbackslash{}cal} H\}=a\textasciicircum{}\{\textbackslash{}prime\}/a                                  & {\tt N/A}                              & {\tt N/A}      & -                             \\ \hline
32  & \tt{\textbackslash{}theta=A\textbackslash{}exp(-\textbackslash{}zeta t)}                                       & \tt{Exponential decrease}             & {\tt Q574576}      & 3/5                           \\ \hline
33  & \tt{p\_\{i\}=\textbackslash{}omega\_\{i\}\textbackslash{}rho\_\{i\}}                                           & {\tt N/A}                              & {\tt N/A}          & -                             \\ \hline
34  & \tt{i\textbackslash{}partial\_\{t\}\textbackslash{}Phi=H\textbackslash{}Phi}                                   & \tt{Schrödinger evolution}            & {\tt Q165498}      & 2/5                           \\ \hline
35  & \tt{H(t)=\textbackslash{}dot\{a\}/a}                                                                           & {\tt N/A}                              & {\tt N/A}      & -                             \\ \hline
36  & \tt{p\_\{\textbackslash{}Lambda\}=-\textbackslash{}rho\_\{\textbackslash{}Lambda\}}                            & \tt{Dark energy}                      & {\tt Q18343}       & 0/5                           \\ \hline
37  & \tt{P\_\{M\}=w\textbackslash{}rho\_\{M\}}                                                                      & \tt{Equation of state}                & {\tt Q214967}      & 4/5                           \\ \hline
38  & \tt{f\_\{\textbackslash{}nu\}=\textbackslash{}rho\_\{\textbackslash{}nu\}/\textbackslash{}rho\_\{d\}}          & \tt{Neutrino}                         & {\tt Q2126}        & -                             \\ \hline
39  & \tt{A\_\{t\}=rA\_\{s\}}                                                                                        & \tt{Fluctuation}                      & {\tt Q5462624}     & -                             \\ \hline
40  & \tt{p\_\{m\}=\textbackslash{}gamma\textbackslash{}rho\_\{m\}}                                                  & \tt{Nonrelativistic matter}           & {\tt Q55921784}    & -                             \\ \hline
41  & \tt{\textbackslash{}Omega\_\{i\}=\textbackslash{}rho\_\{i\}/\textbackslash{}rho\_\{c\}}                        & \tt{Expansion rate}                   & {\tt N/A}          & 2/5                           \\ \hline
42  & \tt{P(k)=Ak\textasciicircum{}\{n\}}                                                                            & \tt{Inflation}                        & {\tt Q273508}      & -                             \\ \hline
43  & \tt{L\_\{I\}=M(\textbackslash{}tau)\textbackslash{}phi{[}x(\textbackslash{}tau){]}}                            & {\tt N/A}                              & {\tt N/A}          & -                             \\ \hline
44  & \tt{L=\textbackslash{}kappa h\_\{ab\}T\textasciicircum{}\{ab\}}                                                & {\tt N/A}                              & {\tt N/A}          & -                             \\ \hline
45  & \tt{w\_\{i\}=P\_\{i\}/\textbackslash{}rho\_\{i\}}                                                              & \tt{Equation of state}                & {\tt Q214967}      & 4/5                             \\ \hline
46  & \tt{\{\textbackslash{}bar\{M\}\}=\{B\}/\{C\}}                                                                  & {\tt N/A}                              & {\tt N/A}          & -                             \\ \hline
47  & \tt{\textbackslash{}Psi=\textbackslash{}Psi\_\{\textbackslash{}ell\}+\textbackslash{}Psi\_\{s\}}               & {\tt N/A}                              & {\tt N/A}          & -                             \\ \hline
48  & \tt{z=a\textbackslash{}dot\{\textbackslash{}phi\}/H}                                                           & \tt{Equation}                         & {\tt Q11345}    & 0/5                             \\ \hline
49  & \tt{u\textasciicircum{}\{\textbackslash{}mu\}=dx\textasciicircum{}\{\textbackslash{}mu\}/d\textbackslash{}tau} & \tt{Comoving fluid}                   & {\tt Q5462744}     & 0/5                             \\ \hline
50  & \tt{\textbackslash{}dot\{\textbackslash{}phi\}=-W\_\{\textbackslash{}phi\}}                                    & \tt{First order differential equation} & {\tt Q11214}       & 3/5                           \\ \hline
\end{tabular}}
\end{table}

\subsection{Task \ref{tsc:exampleanalysis}:~Analysis of Formula Concept Examples}
\label{sec:task2}

In the following, we do step-by-step examinations of three differential equations from physics:
\begin{enumerate}[{Example }1{:}]
    \item \label{ex:KGE} 
    %Example:~
    Klein--Gordon Equation,
    \item \label{ex:EFE} 
    %Example:~
    Einstein's Field Equations,
    \item \label{ex:ME} 
    %Example:~
    Maxwell's Equations.
\end{enumerate}

The presented representations are not exhaustive. Only some of the most interesting representations are selected and presented to discuss important aspects and derive a list of challenges for Formula Concept Retrieval.

The challenge analysis framework is the following: The domain expert thoroughly examines the formula at hand to understand its specific particularities. Performing a `semantic analysis' means that constraints, notation (see, for example, \url{https://dlmf.nist.gov/not}), substitutions, and equivalences are carefully considered.

%Differential equations from (astro)physics.
%Shortly describe their physical and mathematical meaning first

% -> derive Formula Concept definition and requirements for FCR/FCDB/AnnoMathTeX from examples [paper pdfs (google), arXiv (kNN), Wikipedia (1st formulas), textbook (Fliessbach),...]

%equivalence class?

\subsubsection{Example \ref{ex:KGE}:~Klein--Gordon Equation.}

The Klein--Gordon equation is a relativistic wave equation.
It describes the behavior of particles (modeled as waves) at high energies and velocities comparable to the speed of light (relativistic).
Being a partial differential equation containing second partial derivatives in both time $\partial^2 / \partial t^2$ and space $\partial^2 / \partial x_k^2$ it can be employed to compute the evolution of a quantum wave function $\psi$ in time $t$ and space $\vec x$~\cite{gross2008relativistic}.
Apart from the terms containing the derivatives of the wave function, there is an additional term with the undifferentiated wave function.
Depending on the notation, some terms are additionally multiplied by some factors of constants (not changing in time and space).
The signs of the terms depend on the metric signature, a notational convention of how to combine time and space~\cite{fliessbach1990allgemeine}.

In the first retrieved representation
\begin{align}
    \frac{1}{c^2} \frac{\partial^2 \psi}{\partial t^2} - \nabla^2 \psi + \left( \frac{m_0 c}{\hbar} \right)^2 \psi = 0,
    \label{eq:KGE1}
%\tag{a}
\end{align}
the term pre-factors are $1/c^2$ and $(m_0 c/\hbar)^2$. The spatial derivatives with respect to the coordinates $\vec x = (x,y,z)$ are encapsulated in the Laplace operator
\begin{align*}
    \nabla^2 = {\nabla \cdot \nabla =  (\partial_x, \partial_y, \partial_z)\cdot (\partial_x, \partial_y, \partial_z)}.
\end{align*}
In the second representation
\begin{align}
    u_{tt} + A u + f(u) = 0,
    \label{eq:KGE2}
\end{align}
the wave function is denoted $u$ instead of $\psi$. Additionally, the second derivative with respect to time is denoted using subscripts
%\begin{align*}
$
    u_{tt} = \frac{\partial^2 u }{\partial t^2}$.
%\end{align*}
The space derivative is operated using a matrix multiplication $A \cdot u$ corresponding to $\nabla^2 u$, and the metric signature is chosen such that the term has a positive sign.
Finally, the constant factors are absorbed in the function $f(u)$, which is proportional to $(m_0 c/\hbar)^2 u$. 
In both the previous and following representations, the multiplication is always implicit, i.e., the multiplication sign "$\cdot$" is omitted.
The equation representation allows any function of $u$, $f(u)$ linear or nonlinear to be added. For it to be the Klein--Gordon equation, $f(u)$ has to equal a non-zero constant times $u$.
In this case, the parameters are set to
\begin{align*}
    A:=-\Delta+m^2, m\ne 0, \quad f(u):=\lambda|u|^{\rho-1}u, \lambda\in{\mathbb R},
\end{align*}
such that the equation contains the second space derivatives in the Laplace operator $\Delta$ and is linear in $u$, e.g., $f(u) = \lambda u$ for $\rho = 1$.
The need to automatically retrieve this additional constraint information is a major challenge for FCR.
In the third representation
\begin{align}
    \partial^2_{ct} h_n (z,t) - \partial^2_z h_n (z, t) + \nu_n^2 h_n (z,t) = 0,
    \label{eq:KGE3}
\end{align}
the time derivatives includes the factor $c$ (speed of light) and is again denoted using subscripts, such that
\begin{align*}
    \partial^2_{ct} =
    \frac{ \partial^2} { \partial (ct)^2}
    =  \frac{1}{c^2}
    \partial^2_t.
\end{align*}
This is equivalent to the absorption of the factor $1/c^2$ from the first representation \eqref{eq:KGE1}.
The wave function is here denoted $h(z,t)$, explicitly emphasizing the dependence on space $z$ and time $t$.
Here, only one dimension is considered---the coordinate $z$, such that the spacial derivative is reduced to $\partial^2_z = \partial^2 / \partial z^2$.
The metric signature is the same as in \eqref{eq:KGE1} with a minus sign in front of the second term.
In the fourth representation
\begin{align}
    \nabla^a \nabla_a \psi = \mu^2 \psi,
    \label{eq:KGE4}
\end{align}
the wave function is again denoted $\psi$ as in \eqref{eq:KGE1}.
The constants are absorbed in the factor $\mu^2$, such that the linear term containing the undifferentiated wave function is now shifted from the left-hand to the right-hand side of the equation.
Both the space and time derivatives are combined into one single term by using Einstein's notation of summation convention~\cite{einstein1916foundation}.
It states implicit summation over double indices.
In our case, $a$, the summation index, denotes the dimension coordinates of time $t$ and space $x, y, z$. Without additional remarks, it is now clear whether all coordinates are considered or some omitted.
It could possibly be a time-independent ($\partial^2 \psi / \partial t^2$ = 0) or one-dimensional form ($\psi(\vec x) = \psi(z)$), as in \eqref{eq:KGE3}.
In the fifth representation\footnote{Labeled by the authors of the source article as `evolution time Klein--Gordon equation'.}
\begin{align}
    \frac{\hbar^2}{c^2} \frac{\partial^2 \psi}{\partial t^2} - \frac{\hbar^2 \partial^2 \psi}{\partial x^2} = - 2 i \hbar \frac{\partial \psi}{\partial \tau},
    \label{eq:KGE5}
\end{align}
there is an additional term containing a first derivative with respect to proper time $\tau$, which is proportional to time $t$ for constant speed.
The term is imaginary, denoted by the imaginary unit $i$.
Physically, it introduces an exponential decay of the wave function (damping).
The sixth representation
\begin{align}
    -\hbar^2 \frac{\partial^2 \Psi}{\partial t^2} + c^2 \hbar^2 \nabla^2 \Psi = m_0^2 c^4 \Psi,
    \label{eq:KGE6}
\end{align}
has a different signature (the term signs differ from the previous representations).
However, the term without derivative appears positive on the right-hand side as in \eqref{eq:KGE4}.
Moreover, the pre-factors containing the constants---Planck's constant $\hbar$, the speed of light $c$, and the rest mass $m_0$---are distributed differently.
In the seventh representation
\begin{align}
    \nabla^2 \phi - \frac{1}{c^2} \frac{\partial^2 \phi}{\partial t^2} - \frac{2\alpha + a}{c^2} \frac{\partial \phi}{\partial t} - \frac{\alpha^2 + a \alpha}{c^2} \phi = 0,
    \label{eq:KGE7}
\end{align}
the wave function is denoted $\phi$.
The second space derivatives appear again using the Laplace operator $\nabla^2$ as in \eqref{eq:KGE1}.
Here, some additional constants $\alpha$ and $a$ are introduced, and a term containing a first partial time derivative $\partial \phi / \partial t$, similar to \eqref{eq:KGE5}.
By setting $a = -2\alpha$ in the publication, this term vanishes, and the equation becomes the Klein--Gordon equation.
The eight representation
\begin{align}
    u_{tt} - \Delta u + m^2 u + G'(u) = 0,
    \label{eq:KGE8}
\end{align}
uses the same variable $u$ and time derivative $u_{tt}$ as in \eqref{eq:KGE2}.
The Laplace operator performing the second spatial derivatives is denoted as $\Delta = \nabla^2$.
The constants are absorbed in the factor $m^2$, and there is an additional term, the function $G'(u)$ of the wave function. This $G(u)$ must be equal to a non-zero constant times $u$ in order for $'G(u) = 0$ and the representation to be the Klein--Gordon equation.
The ninth representation
\begin{align}
    \left( \eta^{\mu \nu} \frac{\partial}{x^\mu} \frac{\partial}{x^\nu} - \left(\frac{mc}{\hbar} \right)^2 \right) \varphi = 0,
    \label{eq:KGE9}
\end{align}
again uses Einstein notation as in \eqref{eq:KGE4} for the partial (time and space) derivatives.
For the signature (the term signs), the Minkowski metric $\eta_{\mu \nu}$ is employed.
The wave function $\phi$ can then be factored out.
The tenth representation
\begin{align}
    \left(-\frac{1}{c^2} \frac{\partial^2}{\partial t^2} \sum_{i=1}^p \frac{\partial}{x^i} \frac{\partial}{x^i} - \left(\frac{mc}{\hbar} \right)^2 \right) \varphi = 0
    \label{eq:KGE10}
\end{align}
is similar to \eqref{eq:KGE9}.
However, it explicitly displays the summation using the sign $\sum$ and limits the considered dimensions to $p$.
Lastly, the eleventh representation
\begin{align}
    u_{tt} - \Delta u + m u + P'(u) = 0
    \label{eq:KGE11},
\end{align}
is almost identical to \eqref{eq:KGE8}---the only difference being that the constant $m^2$ is replaced by $m$ and the function $G$ by $P$. This again means that to be the Klein--Gordon equation, the function derivative $P'(u)$ must vanish.

To summarize, in the different representations of the Klein--Gordon equation extracted from the 11 publications, there are several different symbols used to denote the wave function:~$\psi$, $u$, $h$, $\Psi$, and $\phi$.
The constant factors $m_0$, $c$, $\hbar$, etc., appear at different places in different terms of the equation or are omitted entirely in particular unit systems.
The derivative notation varies significantly, e.g., from $\partial^2 \psi / \partial t^2$ to $\partial^2_{ct}$ to $u_{tt}$ for the time derivative of the wave function.
In \eqref{eq:KGE4} and \eqref{eq:KGE9}, Einstein's summation notation is used to compactify the derivatives, while omitting summation signs.
The signs of the terms differ with the metric signature that is used.
Additional terms and functions are introduced (e.g., the damping term in \eqref{eq:KGE5} and $G'(u)$ and $P'(u)$ in equations \eqref{eq:KGE8} and \eqref{eq:KGE11}).
Note that there are potentially more representation variations, which were not considered due to their absence in the examples. For instance, there are forms of the KGE, in which the D'Alembert operator
\begin{align*}
    \Box = \frac{1}{c^2}\frac{\partial^2}{\partial t^2}-\sum_{i=1}^{d-1}\frac{\partial^2}{\partial x_i{}^2}
\end{align*}
takes care of the time and space derivatives.

\subsubsection{Example \ref{ex:EFE}:~Einstein's Field Equations.}

%Einstein field eqn.s as example, retrieval in astro-ph dataset
%explore possible equiv. trafos for FC definition and augmentation generator

%Wikipedia:~11 different representations
%\url{https://en.wikipedia.org/wiki/Einstein_field_equations} (06/02/19)

%arXiv NTCIR:~77 matches reducing to xx different representations retrieved by matching the set of identifiers

Einstein's field equations are the fundamental differential equations in Einstein's theory of general relativity.
They relate the curved geometry of spacetime (space and time are united in the framework) to the distribution of matter, which generates a gravitational field~\cite{einstein1916foundation}.
Mathematically, the EFEs form a system of ten coupled nonlinear partial differential equations~\cite{rendall2005theorems}.
As in the previously discussed representations \eqref{eq:KGE4} and \eqref{eq:KGE9} of the Klein--Gordon equation, four-dimensional indices are used.

The first representation
\begin{align}
    G_{\mu\nu}+\Lambda g_{\mu\nu}=\kappa 
    T_{\mu\nu}
    \label{eq:EFE1}
\end{align}
is a very compact form.
The Einstein tensor
\begin{align*}  
    G_{\mu \nu} = R_{\mu \nu} - \tfrac{1}{2} R g_{\mu \nu}
\end{align*}
subsumes the spacetime curvature Ricci tensor $R_{\mu \nu}$ and scalar curvature $R$, and metric tensor $g_{\mu \nu}$, which describes the gravitational field.
The stress-energy tensor $T_{\mu \nu}$ describes the density and flux of energy and momentum in spacetime.
A tensor is a generalization of a matrix and a vector in higher dimensions. The two-dimensional tensors with two indices $\mu$ and $\nu$ can also be written as a matrix (cf.~field tensor in Example \ref{ex:ME}), where the indices correspond to the column and row numbers.
In equation \eqref{eq:EFE1}, The cosmological constant $\Lambda$ quantifies the contribution of dark energy to the expansion of the universe.
Furthermore, there is another constant
\begin{align*}
    \kappa = 8 \pi G / c^4,
\end{align*}
containing the gravitational constant $G$ and the constant which represents the speed of light $c$.
The second representation
\begin{align}
    G_{\mu \nu} + \Lambda g_{\mu \nu} = 8 \pi T_{\mu \nu}~(G=c=1),
    \label{eq:EFE2}
\end{align}
explicitly states that geometric units are used with the constants $G=c=1$, which sets the pre-factor on the right-hand side to $\kappa = 8 \pi$.
The third representation
\begin{align}
    R_{\mu\nu}-\frac{1}{2}g_{\mu\nu}R-\Lambda g_{\mu\nu}=(8\pi G_{N})T_{\mu\nu},
\end{align}
writes out the definition of $G_{\mu \nu}$ on the left-handy side, and uses $c=1$ but $G=G_N$ with an additional index $N$.
The fourth representation
\begin{align}
    G_{\mu\nu}=-\Lambda g_{\mu\nu}+\kappa^{2}T^{\rm tot}_{\mu\nu}
    \label{eq:EFE4}
\end{align}
shows the term with the cosmological constant $\Lambda$ moved to the right-hand side, $\kappa^2$ listed instead of $\kappa$, and $T^\text{tot}_{\mu \nu}$ is listed with an additional superscript.
The fifth representation
\begin{align}
    G_{\mu\nu}=R_{\mu\nu}-g_{\mu\nu}R/2=\kappa T^{\mu\nu}-\Lambda g_{\mu\nu}
\end{align}
is a combination of \eqref{eq:EFE1} and \eqref{eq:EFE4}.
The sixth representation
\begin{align}
    R_{\mu\nu}-\frac{1}{2}Rg_{\mu\nu}=\kappa_{r}(T)T_{\mu\nu}+\Lambda(T)g_{\mu\nu}
    \label{eq:EFE5}
\end{align}
has the sign of the $\Lambda$-term changed again, while showing its dependence of $T$. Furthermore, $\kappa$ here has the index $r$ and its dependence of $T$ is shown as well. 
In the seventh representation
\begin{align}
    K_{\mu\nu}-Kg_{\mu\nu}=-\frac{\kappa^{2}}{2}T_{\mu\nu}+r_{c}G_{\mu\nu},
\end{align}
the units are chosen, such that the pre-factor of the $T_{\mu \nu}$-term is $-\kappa^2/2$, and $G_{\mu \nu}$ is multiplied by an additional factor $r_c$ (critical radius of the universe.
The eight representation
\begin{align}
    G_{AB}\equiv R_{AB}-{1\over 2}g_{AB}R=\kappa^{2}\,T_{AB}
\end{align}
uses the Latin letters $A$ and $B$ instead of the Greek letters $\mu$ and $\nu$.
The ninth representation
\begin{align}
    R_{\mu\nu}-\frac{1}{2}g_{\mu\nu}R+\Lambda g_{\mu\nu}=-8\pi GT_{\mu\nu}
f_{R}\,G_{\mu\nu}
\end{align}
introduces an additional function $f_R$ and an explicit occurrence of the Newtonian gravitational constant $G$.
The tenth and eleventh representations
\begin{align}
    R_{\mu\nu}-\frac{1}{2}g_{\mu\nu}R+\Lambda_{c}g_{\mu\nu}=8\pi GT_{\mu\nu}
\end{align}
and
\begin{align}
    R_{\mu\nu}-{1\over 2}Rg_{\mu\nu}+\Lambda_{eff}g_{\mu\nu}=8\pi GT_{\mu\nu}
\end{align}
highlight that the cosmological constant $\Lambda$ is critical ($c$) and effective ($eff$) using subscripts.
The twelfth representation
\begin{align}
    G_{\mu\nu}-g_{\mu\nu}\Lambda=\frac{8\pi G}{c_{0}^{4}\phi^{4}}T_{\mu\nu}
    \label{eq:EFE12}
\end{align}
displays an additional identifier $\phi$ within $\kappa$ and index of $c_0$.
The thirteenth representation
\begin{align}
    E^{\mu\nu}=-G^{\mu\nu}+\kappa T^{\mu\nu}-\Lambda g^{\mu\nu}
\end{align}
relates a fourth tensor $E_{\mu \nu}$ to the other three ($G_{\mu \nu}$, $g_{\mu \nu}$, and $T_{\mu \nu}$).
For $E_{\mu \nu} = 0$ it reduces to \eqref{eq:EFE1}.
In the fourteenth representation
\begin{align}
    R_{\mu\nu}-\frac{1}{2}g_{\mu\nu}R=8\pi G_{5}T_{\mu\nu}-\Lambda_{5}g_{\mu\nu},
\end{align}
another index $5$ is added to the constants $G$ and $\Lambda$.
In the fifteenth representation
\begin{align}
    R_{\mu\nu}-\frac{1}{2}Rg_{\mu\nu}=8\pi GT_{\mu\nu}-\Lambda g_{\mu\nu}
T^{\rm RG}_{\mu\nu},
    \label{eq:EFE15}
\end{align}
an additional superscript RG is displayed.
The sixteenth representation
\begin{align}
    R_{\mu \nu} - \frac{ \Lambda g_{\mu \nu}}{\frac{D}{2}-1} = \frac{8 \pi G}{c^4} \left(T_{\mu \nu} - \frac{1}{D-2}Tg_{\mu \nu}\right),
\end{align}
contains an additional constant $D$, which is the dimension of the spacetime.
Finally, the seventeenth representation
\begin{align}
    G_{\mu\nu}=\kappa_{4}^{2}T_{\mu\nu}-\Lambda g_{\mu\nu}+Q_{\mu\nu}\textbf{}
\end{align}
adds another subscript for $\kappa$ and the electromagnetic charge tensor $Q_{\mu \nu}$ (`Einstein-Maxwell equations').
Summarizing, Example \ref{ex:EFE} reiterates that the same Formula Concept can be represented using different unit systems, which modify the coefficients of the individual terms.
As in Example \ref{ex:KGE}, different names for identifiers and sub- or superscripts can occur.
Furthermore, sometimes a variable dependence is explicitly displayed as in \eqref{eq:EFE5}.

\subsubsection{Example \ref{ex:ME}:~Maxwell's Equations.}

Maxwell's equations are the foundation of classical electromagnetism and optics.
They describe how charges and electric currents generate electric and magnetic fields and model light as electromagnetic waves~\cite{jackson1999classical}.
Mathematically, they form a set of four coupled partial differential equations, which---like the Klein--Gordon equation (Example \ref{ex:KGE})---contain time and space derivatives.
While the Klein--Gordon equation is a scalar equation (wave function), Einstein's field equations relate tensors (curvature and mass-energy), Maxwell's equations are vector (electric and magnetic field) equations.

The first two equations are Gauß' law for electric and magnetic fields
\begin{align}
    \text{div} \ \vec{E} = 4 \pi \rho, \ \text{div} \ \vec{B} = 0.
\label{eq:Gausslaws}
\end{align}
They state that the source (given by the divergence) of the electric field ($\vec E$) is a charge (the density distribution $\rho$), while the magnetic field ($\vec B$) has no source distribution (equals zero).
The third and fourth of Maxwell's equations are Faraday's law of induction and Ampère's circuital law
\begin{align}
    \text{rot} \ \vec{E} = - \frac{1}{c} \frac{\partial\vec{B}}{\partial t}, \ \text{rot} \ \vec{B} = \frac{4 \pi}{c} \vec{j} + \frac{1}{c} \frac{\partial\vec{E}}{\partial t}.
\label{eq:FaradayandAmpere}
\end{align}
They state that electric fields (rot $\vec E$) (or curl) are generated by changing magnetic fields ($\partial\vec{B} / \partial t$) and magnetic fields (rot $\vec B$) are generated by changing electric fields ($\partial\vec{E} / \partial t$) and charge current density distributions ($\vec j$).
Both the existence of a non-zero curl (rot), i.e., vortex strength, and divergence (div), i.e., source strength of the electric and magnetic fields, are obtained using permutations of the field components.
While the second and third equations are homogeneous, the first and the fourth equations are inhomogeneous.
The latter two contain source terms (electric charge and current density distributions).

% integral forms from Wikipedia: https://en.wikipedia.org/wiki/Maxwell%27s_equations (27.12.19)

Equations \eqref{eq:Gausslaws} and \eqref{eq:FaradayandAmpere} are the differential forms of Maxwell's equations. However, it is also possible to represent them in their integral forms.
Gauß's law for the electric field then writes
\begin{align}
\oiint_{\partial \Omega} \mathbf{E} \cdot \mathrm{d} \mathbf{S} = \frac{1}{\varepsilon_0} \iiint_\Omega \rho \mathrm{d} V,
\label{eq:Gausselecint}
\end{align}
where $\oiint_{\partial \Omega}$ is a surface integral over the boundary surface $\partial \Omega$ (with the loop indicating that the surface is closed), and $\iiint_\Omega$ is a volume integral over the volume $\Omega$.
Gauß law for the magnetic field then becomes
\begin{align}
\oiint_{\partial \Omega} \mathbf{B} \cdot \mathrm{d} \mathbf{S} = 0.
\label{eq:Gaussmagnint}
\end{align}
Faraday's law of induction can be written as
\begin{align}
\oint_{\partial \Sigma} \mathbf{E} \cdot \mathrm{d}\boldsymbol{l} = -\operatorname{\frac{d}{dt}} \iint_{\Sigma} \mathbf{B} \cdot \mathrm{d}\mathbf{S},
\label{eq:Faradayint}
\end{align}
where $\oint_{\partial \Sigma}$ is a line integral integrating over the boundary curve $\partial \Sigma$ (with the loop indicating that the curve is closed), and $\iint_{\Sigma}$ is a surface integral over the surface $\Sigma$.
Finally, Ampère's law becomes
\begin{align}
\oint_{\partial \Sigma} & \mathbf{B} \cdot \mathrm{d} \boldsymbol{l} = \mu_0 \left(\iint_{\Sigma} \mathbf{j} \cdot \mathrm{d}\mathbf{S} + \varepsilon_0 \frac{\mathrm{d}}{\mathrm{d}t} \iint_{\Sigma} \mathbf{E} \cdot \mathrm{d} \mathbf{S} \right).
\label{eq:Ampereint}
\end{align}
Maxwell's equations can also be transformed into a four-vector notation, which includes tensors and Einstein's summation convention (as in Example \ref{ex:EFE}:~Einstein's field equations).
In this notation, the two inhomogeneous partial differential equations are reduced to
\begin{align}
    \partial_\alpha F^{\alpha \beta} = \frac{4 \pi}{c} j^\beta,
\label{eq:Maxwellinhom}
\end{align}
and the homogeneous partial differential equation is reduced to
\begin{align}
    \varepsilon^{\alpha \beta \gamma \delta} \partial_\beta F_{\gamma \delta} = 0.
\label{eq:Maxwellhom}
\end{align}
The charge and current sources density distributions ($\rho$ and $\vec j$) are combined into one four-vector \begin{align*}
    (j^\beta) = (c \rho, j^i).
\end{align*}
The four-derivative of both space and time is defined as
%\begin{align*}
$\partial_\alpha = \frac{\partial}{\partial x^\alpha}$.
%\end{align*}
The permutations needed for the curl and the divergence of the electric and magnetic field are encapsulated in the Levi-Civita symbol
\begin{align*}
    \varepsilon^{\alpha \beta \gamma \delta} =
    \begin{cases}
    +1, & (\alpha, \beta, \gamma, \delta) = \text{even permutation of} \ (0,1,2,3)\\
    -1, & (\alpha, \beta, \gamma, \delta) = \text{odd permutation of} \ (0,1,2,3)\\
    0 & \text{otherwise}
    \end{cases}.
\end{align*}
The electromagnetic field tensor is then defined as
\begin{align*}
    (F^{\alpha \beta}) = 
    \left(\begin{matrix}
    0  &   -E_x/c &  -E_y/c &  -E_z/c \\
    E_x/c &   0  &  -B_z & B_y \\
    E_y/c & B_z &   0  &  -B_x \\
    E_z/c &  -B_y & B_x &   0  \\
    \end{matrix}\right),
\end{align*}
containing all six components of both the electric and magnetic fields in three dimensions.

Summarizing, Example \ref{ex:ME} shows how unification into a single physics framework (Maxwell's equation of electromagnetism) combines multiple Formula Concepts:~Gauß' law of electric and magnetic fields; Faraday's law of induction; and Ampère's circuital law.
Equation \eqref{eq:Maxwellinhom} could either be labeled `Gauß' electric law' and `Ampère's law' or `Maxwell's inhomogeneous equations'.
Analogously, equation \eqref{eq:Maxwellhom} could either be labeled `Gauß' magnetic law' and `Faraday's law' or `Maxwell's homogeneous equations'.
By transforming to the more compact notation, tensors and indices are introduced.
Notably, the electromagnetic field tensor $F^{\alpha \beta}$ subsumes multiple components of two vectors.

%\subsubsection{(Example 4: Hubble's Law)}

%\subsubsection{(Example 5: Conformal Time)}
%or another from retrieval method 3

\subsection{Task \ref{tsc:challengeidentification}:~Identification of Challenges}\label{sec:challengeidentification}

In the following, we identify the challenges for Formula Concept Discovery and Recognition.
They are derived from the discussion of the three Formula Concept examples. The challenges provide an impression of the peculiarities that need to be considered by FCD and FCR approaches.
%
%PPP: check how this table can be shifted from the appendix to this position
\begin{table}[h]
\caption{Challenges for Formula Concept Discovery and Recognition, derived from the discussion of three Formula Concept examples (differential equations presented in Section \ref{sec:task2}).}
\label{tab:challenges}
\centering
\begin{tabularx}{\textwidth}{|l|l|X|}
\hline
\textbf{Challenge}                              & \textbf{Type}           & \textbf{Description}                                                                                                                                                                                                                                                                                                                                                                                                  \\ \hline \hline
1                                               & Symbols                 & Different symbols for constants or variables (cf. Klein--Gordon equation \eqref{eq:KGE1} and \eqref{eq:KGE2}) are used.                                                                                                                                                                                                                                                        \\ \hline
\newtag{2}{ch:substitutions} & Symbols                 & Substitutions, i.e., identifiers are subsumed into others and then appear implicitly (e.g., $\kappa = 8 \pi G / c^4$ linking representation \eqref{eq:EFE1} and \eqref{eq:EFE12} of Einstein's field equations).                                                                                                                                                                    \\ \hline
\newtag{3}{ch:index}         & Symbols                 & Additional (index or semantic) sub- or superscripts (cf. equation \eqref{eq:KGE10} and \eqref{eq:EFE15}) are introduced.                                                                                                                                                                                                                                                           \\ \hline
4                                               & Symbols                 & Sometimes, a variable dependence is explicitly displayed, as in equation \eqref{eq:EFE5}.                                                                                                                                                                                                                                                                                                            \\ \hline \hline
5                                               & Terms                   & Constants appear in different terms (cf. Klein--Gordon equation \eqref{eq:KGE1} and \eqref{eq:KGE6}).                                                                                                                                                                                                                                                                          \\ \hline
6                                               & Terms                   & Additional terms and functions are introduced (e.g., the damping term in Klein--Gordon equation \eqref{eq:KGE5} and $G'(u)$ and $P'(u)$ in equations \eqref{eq:KGE8} and \eqref{eq:KGE11}).                                                                                                                                                                        \\ \hline
7                                               & Terms                   & Signs of the terms differ with the metric signature that is used (cf. again Klein--Gordon equation \eqref{eq:KGE1} and \eqref{eq:KGE6}).                                                                                                                                                                                                                                       \\ \hline
8                                              & Terms                   & Einstein's summation notation can be used to compactify terms (e.g., the derivatives in equation \eqref{eq:KGE4}) while omitting summation signs.                                                                                                                                                                                                                                                    \\ \hline \hline
9                                               & Differential / Integral & Varying derivative notation is used, e.g., from $\partial^2 \psi / \partial t^2$ to $\partial^2_{ct}$ to $u_{tt}$ for the time derivative of the wave function in Example \ref{ex:KGE}. Another commonly used notation would be the double-dot $\ddot{u}$, where each dot represents a time derivative.                                                                                              \\ \hline
\newtag{10}{ch:diffint}      & Differential / Integral & Differential and integral forms are employed interchangeably. Maxwell's equations can be written using derivatives (equations \eqref{eq:Gausslaws} and \eqref{eq:FaradayandAmpere}) or integrals (equations \eqref{eq:Gausselecint}, \eqref{eq:Gaussmagnint}, \eqref{eq:Faradayint}, and \eqref{eq:Ampereint}). \\ \hline \hline
\newtag{11}{ch:framework}    & Compactification        & Unification into a single (physics) framework is applied. Maxwell's equations of electromagnetism combine multiple Formula Concepts: Gauß' law of electric and magnetic fields, Faraday's law of induction, and Ampère's circuital law.                                                                                                                                                                          \\ \hline
12                                              & Compactification        & Tensor notation is used. Transforming to the more compact forms \eqref{eq:Maxwellinhom} and \eqref{eq:Maxwellhom}, tensors and indices are introduced. The electromagnetic field tensor $F^{\alpha \beta}$ subsumes multiple components of two field vectors $\vec E$ and $\vec B$.                                                                                                 \\ \hline \hline
\newtag{13}{ch:units}         & Units                   & Different unit systems are applied. Constant factors or numbers can be transformed into different unit systems (e.g., natural units $G=c=1$ in equation \eqref{eq:EFE2}).                                                                                                                                                                                                                            \\ \hline
\end{tabularx}
\end{table}

Table \ref{tab:challenges} contains the results of our evaluation.
Most of them are notation issues.
Different names for symbols (constants or variables) are used.
Different notation systems are applied for signatures and units.
Different forms for derivatives, summations, and tensors are employed.
For some challenges, e.g., Challenge \ref{ch:index} there is an overlap between the different Formula Concept examples. For others, e.g., Challenges \ref{ch:diffint} and \ref{ch:framework}, the issues only apply to the specific example. We can note an average of four challenges per example.
It remains an open question whether this number increases or decreases with additional examples. There can potentially be more or less overlap of challenges shared by examples.
If the same challenges do not reoccur frequently and the number of challenges significantly increases with new examples, Formula Concept retrieval methods are faced with additional difficulties.

\subsection{Task \ref{tsc:requirementderivation}:~Derivation of Formula Concept Retrieval System Requirements}

In the following, we address the identified Formula Concept Discovery challenges by deriving requirements for a Formula Concept Retrieval system.
%
%\paragraph{Formula Concept Database.}
%
%create Wikidata related / attached Formula Concept Database / describe planned project
%"Formula augmentation": computationally generate equivalent forms, substitute identifiers by x_1, x_2, ..., x_n for standardization and ambiguity elimination
%
%TODO (camera-ready version): update number
Since currently, less than 6,000 formulas are seeded into Wikidata\footnote{To get the current number, run \url{https://w.wiki/3bL6}} and storing multiple representations as `defining formula' of the same Formula Concept item is not endorsed, we argue for the creation of a specific Wikidata-attached \textit{Formula Concept Database}~\cite{schubotz2018introducing}.
It should include formalized \textit{augmentation} to generate equivalent forms using, e.g., commutations, additional sub- and superscripts, unit and reference frame variations, etc.
Most importantly, a method for inferring substitutions or implicit terms needs to be developed.

%\paragraph{Equivalence Transformations and Computer Algebra Systems.}

We propose to formalize the augmentation of a Formula Concept as translation between its different representations.
One could use equivalence generations made by Computer Algebra Systems to train, e.g., a Siamese Network,~\cite{DBLP:conf/nips/BromleyGLSS93}, to assess whether two formulas are representations of the same Formula Concept.
For this, the choice of a suitable formula encoding needs to be explored.
A hypothesis we have to examine beforehand is whether Formula Concept Recognition relies on identifying equivalent representations or only requires the semantic annotations of formula identifiers.
We will discuss this further in future work, as well the exploration of practical implications of the interpretation of a Formula Concept as a mathematical `word' that can be translated between different representations (analogous to `languages').

%\paragraph{Formula Concept Discovery by Recognition (FCD by FCR).}

Apart from distinguishing FCD and FCR as separate methods, one could also combine them to discover Formula Concepts by recognizing (tagging) an increasing amount of formulas per mathematical concept over time.
Therefore, we propose an Active Learning system that shows randomly selected formulas to a user.
The system then has to figure out whether, for a shown formula, there is already a mathematical concept identifier available.
If missing, it should create one and match the following occurrences to it.
Unfortunately, CAS cannot generate all notation transformations (e.g., from vector to tensor, see Formula Concept Example \ref{ex:ME}).

%\paragraph{Semantic Formula Encoding.}

%-> semantification of identifiers necessary
%forget syntax, functionality, numbers, and operators -> we just need semantic identifiers
%-> how to prove that statement?
%database with combinations of identifiers for FCR

% example Klein-Gordon equation again

\begin{figure}
    \centering
    \includegraphics[width=\textwidth]{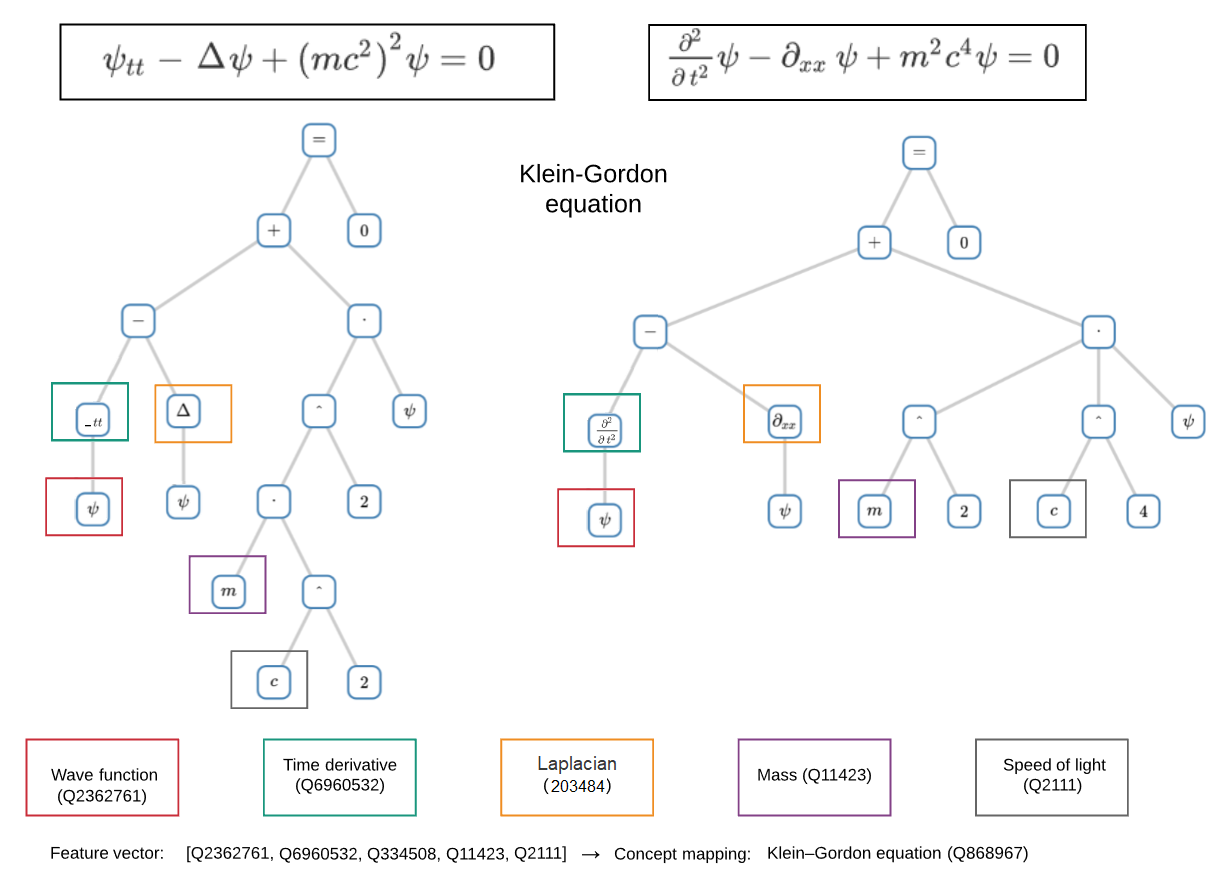}
    \caption{Comparison of two representations of the Klein--Gordon equation (left and right). Different constituents of the expression trees are marked as semantic entities that have a unique Wikidata ID (Qxxx).}
    \label{fig:KGeq_comparison}
\end{figure}

Figure \ref{fig:KGeq_comparison} shows the expression trees of two representations of the Klein--Gordon equation (left and right) in comparison.
Different constituents of the equations are marked in the trees as semantic entities. They can be matched to unique IDs in a semantic database, e.g., Wikidata. For example, the identifier $c$ representing the `speed of light' is assigned the Query ID (QID) `Q2111'.
Since both trees contain the same semantic entities, they can be matched as representing the same Formula Concept.

Summarizing, we derive the following Formula Concept Retrieval system requirements from the identified challenges for FCD:
\begin{enumerate}
    \item Set up a Formula Concept Database (FCDB);
    \item Employ equivalence transformations and Computer Algebra Systems;
    \item Enable Formula Concept Discovery by Recognition (FCD by FCR); and
    \item Integrate formula matching via semantic formula encoding.
\end{enumerate}

\subsection{Conclusion (FCD)}
%Länger / Diskussion

%COMPARISON OF APPROACHES

We compare the effectiveness of retrieving different Formula Concept representations of Method \ref{mth:knearestneighbor} ($k$-Nearest-Neighbors in formula vector space) with Method \ref{mth:firstformula} (Wikipedia article first formula multi-language heuristic). While Method \ref{mth:firstformula} achieves a precision of 34\% for retrieving Formula Concept representations from multilingual Wikipedia articles, Method \ref{mth:knearestneighbor} outperforms this with a precision of 68\% using machine learning. The $k$NN approach is not only performing well, but it also has the advantage of being easily usable and transferable to other corpora. Method \ref{mth:searchbyconceptname} can not be compared to the other two because it is a priori unclear where (at which number of webpages or textbooks) to stop the search. Therefore, we only concentrate on our three Formula Concept examples (KGE, EFE, and ME), for which we can retrieve a total of more than 30 representations, searching in publications, Wikipedia, and a textbook. We conclude that for Formula Concept Discovery to achieve the best results (retrieval of a large number of equivalent formula representations per concept), it is beneficial to combine the different methods optimally.

%=====================================================================================
% Formula Concept Recognition
%=====================================================================================

\section{Formula Concept Recognition} \label{sec:fcr}

In this section, we introduce methods for Formula Concept Recognition (FCR). Recall that the goal of FCR is to recognize formulas in documents as being instances of a previously defined Formula Concept.

The presented FCR methods were not introduced or published before. Prior work only included the first FCD experiments and results. Currently, to the best of our knowledge, no other FCR methods have been published so far. However, to establish comparability and replicability, we evaluate the performance of our approaches against that of open source and commercial formula search engines in Experiment \ref{exp:FCsearch} as presented in Section \ref{subsec:FCsearch}.

In the following, we describe and evaluate several different approaches for FCR.
To assess the feasibility and performance of the proposed methods, we set up the following three experiments:

\bigskip
\begin{enumerate}[{Experiment }1{:}]\interlinepenalty10000 
    %Experiment 1:~
    \item \label{exp:FCsearch}
    Formula Concept Search;
    %Experiment 2:~
    \item \label{exp:FCclassclust}
    Formula Concept Classification and Clustering;
    %Experiment 3:~
    \item \label{exp:FCsim}
    Formula Concept Similarity.
\end{enumerate}

In Experiment \ref{exp:FCsearch}, we investigate how well Formula Concepts can be retrieved by search queries using the formula \LaTeX{} string or the formula constituents. Therefore, we employ several sources, such as Wikidata items, as well as Wikipedia articles and arXiv documents from the NTCIR dataset. The results from Wikidata can be associated with a unique semantic ID (the Wikidata QID). We compare the performance of the open source retrieval to selected competitor (formula) search engines.
In Experiment \ref{exp:FCclassclust}, we assess how well a manually labeled balanced dataset of 100 Formula Concept examples from 10 classes can be automatically recognized by machine learning classification and clustering to separate the Formula Concepts in several vector encoding spaces.
In Experiment \ref{exp:FCsim}, we test how well formula (encoding) similarities can indicate that different formulas are representations of the same Formula Concept.
Therefore, we compute a similarity map matrix of pairwise formula or class similarities.
The developed algorithms, the dataset, and full result tables are available at \url{https://github.com/ag-gipp/formula-concept-retrieval}.

\subsection{Experiment \ref{exp:FCsearch}:~Formula Concept Search}\label{subsec:FCsearch}

We first approach the recognition of Formula Concepts (FCR) as a search ranking problem, in contrast to classification and clustering, examined in the subsequent experiment. To evaluate finding, i.e., recognizing FCs in large corpora of mathematical content, we employ three open data sources (Wikidata, Wikipedia, arXiv) and two methods (retrieval using formula \LaTeX{} string or constituents).
Furthermore, we compare the performance of our methods to two formula search engines, one open source (Approach0\footnote{\label{foot:approach0}\url{https://www.approach0.xyz}}), and one commercial (Google\footnote{\label{foot:google}\url{https://www.google.com}}).

\begin{table}[]
\caption{Ten classes of our test set with 100 Formula Concept differential equation examples, including a linked Wikidata QID and concept name with Wikipedia article source link (above), as well as an example equation \LaTeX~string.}
\label{tab:100eqns}
\centering
%\resizebox{\textwidth}{!}{
\begin{tabular}{|l|l|l|l|}
\hline
\textbf{Nr.} & \textbf{QID} & \textbf{Label} & \textbf{Name}                    \\ \hline
1            & \href{https://wikidata.org/wiki/Q868967}{Q868967}      & KGE            & \href{https://en.wikipedia.org/wiki/Klein\%E2\%80\%93Gordon_equation}{Klein--Gordon equation}            \\ \hline
2            & \href{https://wikidata.org/wiki/Q273711}{Q273711}      & EFE            & \href{https://en.wikipedia.org/wiki/Einstein_field_equations}{Einstein’s field equations}       \\ \hline
3            & \href{https://wikidata.org/wiki/Q51501}{Q51501}       & ME             & \href{https://en.wikipedia.org/wiki/Maxwell\%27s_equations}{Maxwell’s equations}              \\ \hline
4            & \href{https://wikidata.org/wiki/Q165498}{Q165498}      & SE             & \href{https://en.wikipedia.org/wiki/Schrödinger_equation}{Schrödinger equation}            \\ \hline
5            & \href{https://wikidata.org/wiki/Q860615}{Q860615}      & HE             & \href{https://en.wikipedia.org/wiki/Helmholtz_equation}{Helmholtz equation}               \\ \hline
6            & \href{https://wikidata.org/wiki/Q859808}{Q859808}      & BE             & \href{https://en.wikipedia.org/wiki/Biharmonic_equation}{Biharmonic equation}              \\ \hline
7            & \href{https://wikidata.org/wiki/Q104212301}{Q104212301}   & NSL            & \href{https://en.wikipedia.org/wiki/Newton\%27s_laws_of_motion#Newton's_second_law}{Newton’s second law of motion}    \\ \hline
8            & \href{https://wikidata.org/wiki/Q44746}{Q44746}        & HUP            & \href{https://en.wikipedia.org/wiki/Uncertainty_principle}{Heisenberg uncertainty principle} \\ \hline
9            & \href{https://wikidata.org/wiki/Q177045}{Q177045}      & SLT            & \href{https://en.wikipedia.org/wiki/Second_law_of_thermodynamics}{Second law of thermodynamics}     \\ \hline
10           & \href{https://wikidata.org/wiki/Q83152}{Q83152}       & CL             & \href{https://en.wikipedia.org/wiki/Coulomb\%27s_law}{Coulomb’s law}                    \\ \hline
\end{tabular}%}
\\ \vspace{0.3cm}
\centering
%\resizebox{\textwidth}{!}{
\begin{tabular}{|l|l|}
\hline
\textbf{Label} & \textbf{Example Equation}                                                                                                          \\ \hline
KGE            & \tt{u\_\{tt\} + A u + f(u) = 0}                                                                                                     \\ \hline
EFE            & \tt{G\_\{\textbackslash{}mu \textbackslash{}nu\} = \textbackslash{}kappa T\_\{\textbackslash{}mu \textbackslash{}nu\}}              \\ \hline
ME             & \tt{\textbackslash{}text\{div\} \textbackslash{}vec\{E\} = 4 \textbackslash{}pi \textbackslash{}rho}                                \\ \hline
SE             & \tt{\textbackslash{}hat H\}|\textbackslash{}Psi\textbackslash{}rangle = E |\textbackslash{}Psi\textbackslash{}rangle}               \\ \hline
HE             & \tt{(\textbackslash{}nabla\textasciicircum{}2 - k\textasciicircum{}2) A = -f}                                                       \\ \hline
BE             & \tt{\textbackslash{}nabla\textasciicircum{}4\textbackslash{}varphi=0}                                                               \\ \hline
NSL            & \tt{\textbackslash{}vec\{F\} = \textbackslash{}frac\{d\textbackslash{}vec\{p\}\}\{dt\}}                                             \\ \hline
HUP            & \tt{\textbackslash{}sigma\_\{x\}\textbackslash{}sigma\_\{p\} \textbackslash{}geq \textbackslash{}frac\{\textbackslash{}hbar\}\{2\}} \\ \hline
SLT            & \tt{\textbackslash{}oint \textbackslash{}frac\{\textbackslash{}delta Q\}\{T\} = 0}                                                    \\ \hline
CL             & \tt{|F| = \textbackslash{}\textbackslash{}frac\{|q\_1 \textbackslash{}\textbackslash{}times q\_2|\}\{r\textasciicircum{}2\}}        \\ \hline
\end{tabular}%}
\end{table}

For this and all subsequent FCR experiments, we collect a test set with 100 Formula Concept example differential equations from 10 classes. Table \ref{tab:100eqns} shows the concept class names and labels, together with the corresponding Wikidata QID (above) and example \LaTeX~string (below). The linked Wikipedia article is the source of the respective equations, which we collected for each class. A full list of all 100 collected equations can be found in the appendix. The selection extends the three classes discussed in Section \ref{sec:task2} by additional 7 classes with 10 examples each. Each class corresponds to a Wikipedia article (as indicated in Table \ref{tab:100eqns}). This means that we here apply the definition of a Formula Concept as a set of equation representations collected from the same Wikipedia article.
%def. of Formula Concept: Wikipedia article equivalence sets (domain expert judgement)

For each of our 100 example formulas, we evaluate the performance of 8 selected Formula Concept search retrieval sources: arXiv \LaTeX, arXiv constituents, Wikidata \LaTeX, Wikidata constituents, Wikipedia \LaTeX, Wikipedia constituents, Approach0, and Google. The first 6 represent our retrieval methods over open corpora, while the last 2 employ search engines. The method label \LaTeX~indicates that the formulae are compared by their \LaTeX~strings, whereas `constituents' means that the formula parts are aligned (set intersections of operators and identifiers).

We generated the top 10 results for each of the 8 sources on our 100 examples and manually assessed the ranking of the correct result for the resulting $10 \times 8 \times 100 = 8,000$ formulae. As ranking measures, we used `Top10 Recall' and `Top1 Recall' as well as `Mean Rank' (MR) and `Mean Reciprocal Rank' (MRR), which is defined as~\cite{DBLP:conf/trec/Voorhees99}
% from \url{https://en.wikipedia.org/wiki/Mean_reciprocal_rank}
\begin{align*}
\text{MRR} = 1/\text{MR} = \frac{1}{|Q|} \sum_{i=1}^{|Q|} \frac{1}{\text{rank}_i},
\end{align*}
summing over all query results Q = 10.
%The '''mean reciprocal rank''' is a [[statistic]] measure for evaluating any process that produces a list of possible responses to a sample of queries, ordered by the probability of correctness. The reciprocal rank of a query response is the [[multiplicative inverse]] of the rank of the first correct answer: 1 for first place, 1/2 for second place, 1/3 for third place, and so on. The mean reciprocal rank is the average of the reciprocal ranks of results for a sample of queries Q~
In this formula, $\text{rank}_i$ refers to the rank position of the first relevant document for the i-th query.
The reciprocal value of the mean reciprocal rank represents the harmonic mean of the ranks.

\begin{table}[]
\caption{FCR as Formula Concept Search problem. Several open corpus sources (Wikidata and NTCIR Wikipedia, arXiv) are employed to retrieve formulas from a test set of 100 differential equations either using their \LaTeX{} string or constituents. The performance is compared in several ranking metrics (MRR, etc.) to competitors, an open source (Approach0), and a commercial (Google).}
\label{tab:FCRsearch}
\centering
%\resizebox{\textwidth}{!}{
\begin{tabular}{lllll}
\hline
\multicolumn{1}{|l|}{\textbf{Source / Metric}}                          & \multicolumn{1}{l|}{\textit{MRR}}  & \multicolumn{1}{l|}{\textit{MR}}   & \multicolumn{1}{l|}{\textit{Top10 Recall}} & \multicolumn{1}{l|}{\textit{Top1 Recall}} \\ \hline
                                                                        &                                    &                                    &                                            &                                           \\
\multicolumn{1}{l}{\textit{Formula Concept Retrieval methods (FCRs)}} & \multicolumn{1}{l}{}              & \multicolumn{1}{l}{}              & \multicolumn{1}{l}{}                      & \multicolumn{1}{l}{}                     \\ \hline
\multicolumn{1}{|l|}{arXiv \LaTeX{}}                                     & \multicolumn{1}{l|}{0.70}          & \multicolumn{1}{l|}{2.38}          & \multicolumn{1}{l|}{0.48}                  & \multicolumn{1}{l|}{0.27}                 \\ \hline
\multicolumn{1}{|l|}{arXiv constituents}                                & \multicolumn{1}{l|}{0.71}          & \multicolumn{1}{l|}{2.91}          & \multicolumn{1}{l|}{0.11}                  & \multicolumn{1}{l|}{0.07}                 \\ \hline \hline
\multicolumn{1}{|l|}{Wikidata \LaTeX{}}                                  & \multicolumn{1}{l|}{0.75}          & \multicolumn{1}{l|}{2.28}          & \multicolumn{1}{l|}{0.68}                  & \multicolumn{1}{l|}{0.44}                 \\ \hline
\multicolumn{1}{|l|}{Wikidata constituents}                             & \multicolumn{1}{l|}{0.54}          & \multicolumn{1}{l|}{2.65}          & \multicolumn{1}{l|}{0.17}                  & \multicolumn{1}{l|}{0.05}                 \\ \hline \hline
\multicolumn{1}{|l|}{Wikipedia \LaTeX{}}                                 & \multicolumn{1}{l|}{\textbf{0.78}} & \multicolumn{1}{l|}{\textbf{1.78}} & \multicolumn{1}{l|}{\textbf{0.74}}         & \multicolumn{1}{l|}{\textbf{0.48}}        \\ \hline
\multicolumn{1}{|l|}{Wikipedia constituents}                            & \multicolumn{1}{l|}{0.66}          & \multicolumn{1}{l|}{2.70}          & \multicolumn{1}{l|}{0.40}                  & \multicolumn{1}{l|}{0.21}                 \\ \hline
                                                                        &                                    &                                    &                                            &                                           \\
\multicolumn{1}{l}{\textit{Search Engines (SEs)}} & \multicolumn{1}{l}{}              & \multicolumn{1}{l}{}              & \multicolumn{1}{l}{}                      & \multicolumn{1}{l}{}                     \\ \hline
\multicolumn{1}{|l|}{Approach0}                                         & \multicolumn{1}{l|}{0.64}          & \multicolumn{1}{l|}{2.59}          & \multicolumn{1}{l|}{0.44}                  & \multicolumn{1}{l|}{0.21}                 \\ \hline
\multicolumn{1}{|l|}{Google}                                            & \multicolumn{1}{l|}{0.63}          & \multicolumn{1}{l|}{2.85}          & \multicolumn{1}{l|}{0.55}                  & \multicolumn{1}{l|}{0.26}                 \\ \hline
                                                                        &                                    &                                    &                                            &                                           \\
\multicolumn{1}{l}{\textit{FCRs vs.~SEs}} & \multicolumn{1}{l}{}              & \multicolumn{1}{l}{}              & \multicolumn{1}{l}{}                      & \multicolumn{1}{l}{}                     \\ \hline
\multicolumn{1}{|l|}{Mean (FCRs)}                                       & \multicolumn{1}{l|}{\textbf{0.69}} & \multicolumn{1}{l|}{\textbf{2.45}} & \multicolumn{1}{l|}{0.43}                  & \multicolumn{1}{l|}{\textbf{0.25}}        \\ \hline
\multicolumn{1}{|l|}{Mean (SEs)}                                        & \multicolumn{1}{l|}{0.63}          & \multicolumn{1}{l|}{2.72}          & \multicolumn{1}{l|}{\textbf{0.50}}         & \multicolumn{1}{l|}{0.24}                 \\ \hline
\end{tabular}%}
\end{table}

Table \ref{tab:FCRsearch} shows the results of the Formula Concept search evaluation. The performance of different FCR methods is compared to state-of-the-art (formula) search engine competitors (Approach0\footref{foot:approach0} and Google\footref{foot:google}). We also tested other formula search engines, such as MathWebSearch\footnote{\url{https://search.mathweb.org}}, \footnote{\url{https://www.searchonmath.com}}, zbMATH Open formulae\footnote{\url{https://zbmath.org/formulae}}, and Wolfram Alpha\footnote{\url{https://www.wolframalpha.com}} but they were either not working, access-restricted or too low performing to be included in the result table. The best results (lowest Mean Rank MR, highest Mean Reciprocal Rank MRR, and Recall) are marked in bold. The results exhibit that the FCR method source `Wikipedia \LaTeX' outperformed all other method sources in all metrics. This can be explained by the fact that our FCR examples were extracted from Wikipedia articles. However, not all equations were present in the NTCIR Wikipedia dataset. We find that the formula \LaTeX~string retrieval outperformed the retrieval using formula constituents. Furthermore, we compare our retrieval methods (FCRs) to the selected search engines (SEs). Our methods outperform the search engines in all metrics except `Top10 Recall' (it is very close in the `Top1 Recall' metrics). Summarizing, we compare the performance of different retrieval methods and sources in several ranking measures to demonstrate that it is possible to recognize Formula Concepts using search with a Mean Rank of down to 1.78, Mean Reciprocal Rank up to 0.78, and Recall up to 0.74. Our FCR methods outperform state-of-the-art search engines.

\subsection{Experiment \ref{exp:FCclassclust}:~Formula Concept Classification and Clustering}
\label{subsec:FCclassclust}

To assess how well the computer could separate our 100 Formula Concept examples into classes, we examine their joint formula (content or semantic) space.
Recall that the formula content was defined following~\cite{DBLP:conf/sigir/ScharpfSG18} as the sets of operators, identifiers, and numbers that a formula contains.
Because of Challenge \ref{ch:substitutions} (substitutions) and Challenge \ref{ch:units} (different unit systems), we decided to neglect the set of numbers.
Compared to the operators and identifiers, there are significantly fewer numbers, and they heavily depend on substitutions and unit systems (e.g., the number $8$ in the factor $8 \pi$ or the exponents $4$ in \eqref{eq:EFE12}).

Since formulas in mathematics can be similar to each other syntactically, yet address completely different concepts semantically or vice versa, we analyze the relationship between syntactic and semantic encodings.
There are two challenging cases: (1) syntactically similar but semantically different formulas (syntactic inter-class coherence but semantic inter-class separability) and (2) syntactically different but semantically coherent formulas (syntactic inner-class separability but semantic inner-class coherence).
An example for (1) from our selected classes can be:
\begin{align*}
    a \ \Psi_t + b \ \nabla^2 \Psi + c \ \Psi = 0 \
    \text{(class KGE) vs.} \
    a \ \Psi_{tt} + b \ \nabla^2 \Psi + c \ \Psi = 0 \
    \text{(class SE)}
\end{align*}
or
\begin{align*}
    -\partial \Psi / \partial t^2 + \nabla^2 \Psi - m^2 \Psi = 0 \
    \text{(KGE) vs.} \
    i \partial \Psi / \partial t + 1/2m \nabla^2 \Psi - V \Psi = 0 \
    \text{(SE)}.
\end{align*}
An example for (2) can be: $F = m a$ vs. $F = p / t$ (class NSL expressed using mass $m$ and acceleration $a$ vs. momentum $p$ and time $t$).

Encoding and classifying the syntactic or semantic formula content is indispensable, since the surrounding text is often noisy and the formula concepts are not explicitly named or described. Some authors of mathematical content implicitly assume the reader’s profound background knowledge. This limits the use of text-based encoding and classification methods.
In the following, we describe and discuss our tests of the content vs.~semantic coherence of Formula Concepts in terms of separability (classification accuracy and cluster centroid distance and purity).

%\paragraph{Data Preparation.}

%We took the \LaTeX{} strings of the 100 formulas and cleaned them. This means, we discarded styling macros, such as \verb|\,| and \verb|\textbf{}|. Furthermore, we neglected elementary operators, such as \verb|+,-,*,/,^| and splitted combinations of operators and identifiers, e.g., \verb|\vec{E}| (operator \verb|\vec| acting on identifier $E$) to operator \verb|\vec| and identifier $E$.
%
For the machine learning experiments, we create four files with the equation labels, \LaTeX{} strings, content, as well as semantic annotations, including Wikidata QIDs.
Each of the files has 100 lines corresponding to the individual formulas, i.e., (10 Formula Concept examples from each of the 10 classes KGE
EFE, ME, etc. respectively, see Table \ref{tab:100eqns}).
As an example, consider the first formula \eqref{eq:EFE1}.
It belongs to the first class, so the line in the label file reads \verb|EFE|.
In the \LaTeX{} string file, the corresponding line reads\\
\vspace{-0.5cm}
\begin{flushleft}
    \verb|\frac{1}{c^2}\frac{\partial^2 \psi}{\partial t^2} - \nabla^2\psi|
    \verb|+ \left(\frac{m_0 c}{\hbar} \right)^2\psi = 0|.
\end{flushleft}

The content line, containing the set of parsed operators and identifiers, then reads
\begin{flushleft}
    \verb|c, \partial, \psi, t, \nabla, m, \hbar|.
\end{flushleft}

We encode their semantics as
\begin{flushleft}
    \verb|c: "speed of light" (Q2111),|
    \verb|\partial: "partial derivative" (Q186475),|
    \verb|\psi: "wave function" (Q2362761), t: "time" (Q11471),|
    \verb|\nabla: "del" (Q334508), m: "mass" (Q11423) ,|
    \verb|\hbar: "Planck constant" (Q122894)|
\end{flushleft}
where the ID in parenthesis is the unique \textit{QID} from the item, we find in the semantic knowledge base Wikidata.

Summarizing, the data pipeline is the following: We parse the formula \LaTeX~strings (`formula TeX') to formula constituents (`formula content') and annotate them (`formula semantics') to get Wikidata encodings (`formula qids').
% semantics_dict:
% Min length of annotations per formula part: 1
% Max length of annotations per formula part: 4
% Mean length of annotations per formula part: 1.5974025974025974 (1-2)
% Average information entropy of constituent annotation frequencies: 0.29059740314006505
The yields a dictionary of formula constituent meanings with an average of 2 different annotations per constituent.
% example of constituent annotations: 
%'R': {'"distance" (Q126017)', '"Ricci curvature" (Q1195879)', '"molar gas constant" (Q182333)'}
As an example, the identifier `R' appears as `distance (Q126017)' or `Ricci curvature' (Q1195879)'.

% Content vs. Semantics
% formula peculiarities, identified by annotation

%KGeq
%\frac{\hbar^2}{c^2} \frac{\partial^2 \psi}{\partial t^2} - \frac{\hbar^2 \partial^2 \psi}{\partial^2 x^2} = - 2 i \hbar \frac{\partial \psi}{\partial \tau}
%(\tau term)

%EFeq
%R_{\mu \nu} - \frac{\Lambda g_{\mu \nu}}{\frac{D}{2}-1} = \frac{8 \pi G}{c^4} \left(T_{\mu \nu} - \frac{1}{D-2} T g_{\mu \nu}\right)
%(D unusual)
%G_{\mu \nu} - g_{\mu \nu} \Lambda = \frac{8 \pi G}{c_{0}^{4} \phi^{4}} T_{\mu \nu}
%(\phi unusual)
%G_{\mu \nu} + \Lambda g_{\mu \nu} = 8 \pi T_{\mu \nu} ~ (G = c = 1)
%((G = c = 1) constraint)
%G_{AB} \equiv R_{AB} - {1\over 2} g_{AB} R = \kappa^{2} T_{AB}
%(AB unusual)
%R_{\mu \nu} - \frac{1}{2} R g_{\mu \nu} = 8 \pi G T_{\mu \nu} - \Lambda g_{\mu \nu} T^{\rm RG}_{\mu \nu}
%(RG unusual)
%E^{\mu \nu} = -G^{\mu \nu} + \kappa T^{\mu \nu} - \Lambda g^{\mu \nu}
%(E unusual)
%G_{\mu \nu} = \kappa_{4}^{2} T_{\mu \nu} - \Lambda g_{\mu \nu} + Q_{\mu \nu}
%(Q unusual)

%Meq
%https://en.wikipedia.org/wiki/Maxwell%27s_equations (27.12.19)

%\paragraph{Formula Encoding.}

In our experiment, we employ the following formula vector encodings of both operators and identifiers:

\begin{itemize}
    \item Formula content {\tt TF-IDF};
    \item Formula content {\tt Doc2Vec};
    \item Formula semantics {\tt TF-IDF}; and
    \item Formula semantics {\tt Doc2Vec}.
\end{itemize}

For the formula content encodings, the sets of the parsed operator and identifier \LaTeX{} strings from the content file are employed.
For the formula semantics encodings, we use the sets of Wikidata QIDs.
It is important to note that while the sequence of formula constituents does not matter for the TF-IDF encoding, it is considered by the Doc2Vec encoding. In our experiments, we focus on a relative evaluation, i.e., a comparison of different encodings, rather than optimizing the overall performance by tuning hyperparameters.

\paragraph{30 Examples}

\begin{figure}
    \centering
    \subfloat{\includegraphics[width=\textwidth]{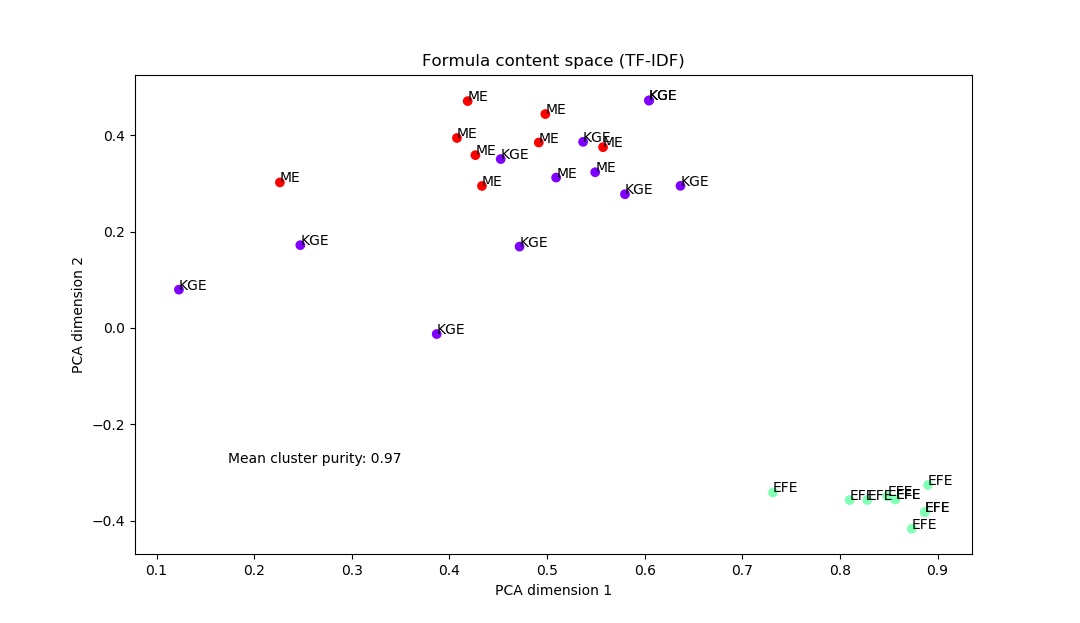}}\\ \vskip -3pt
    \subfloat{\includegraphics[width=\textwidth]{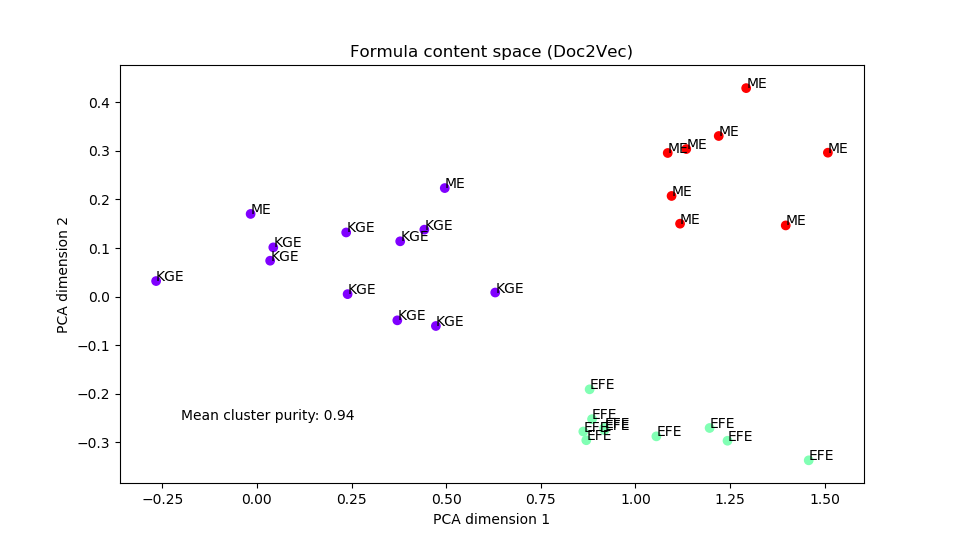}}
    \caption{Formula content space of three selected Formula Concepts (KGE, EFE, ME), using {\tt TF-IDF} or {\tt Doc2Vec} encodings, reduced by Principal Component Analysis (PCA) to two dimensions. The color code corresponds to the clusters computed by $k$-means ($k=3$) clustering. The three classes are best separated in the formula content {\tt Doc2Vec} encoding (second plot) with cluster mean centroid distance of $0.81$, purity of $0.94$, and classification accuracy of $0.90$.}
    \label{fig:formcontspace}
\end{figure}

\begin{figure}
    \centering
    \subfloat{\includegraphics[width=\textwidth]{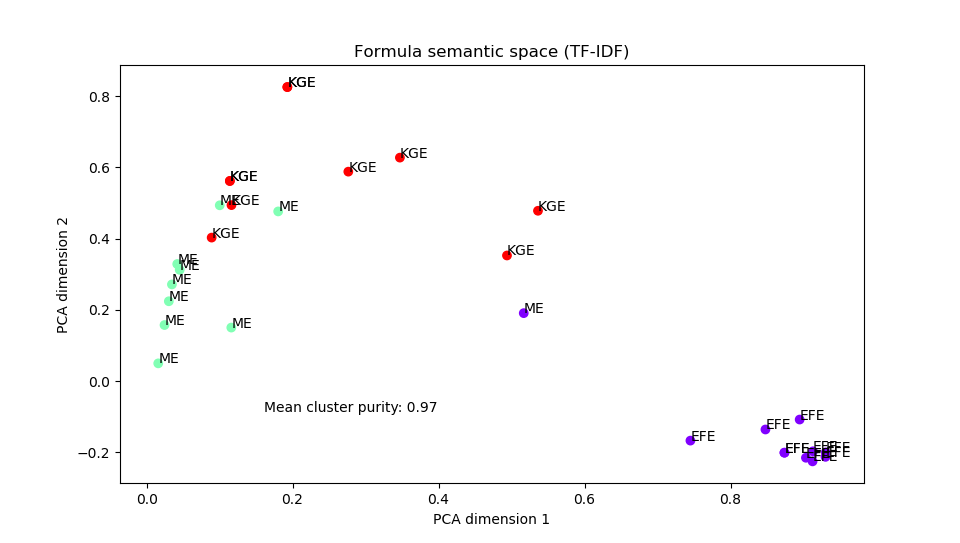}}\\ \vskip -3pt
    \subfloat{\includegraphics[width=\textwidth]{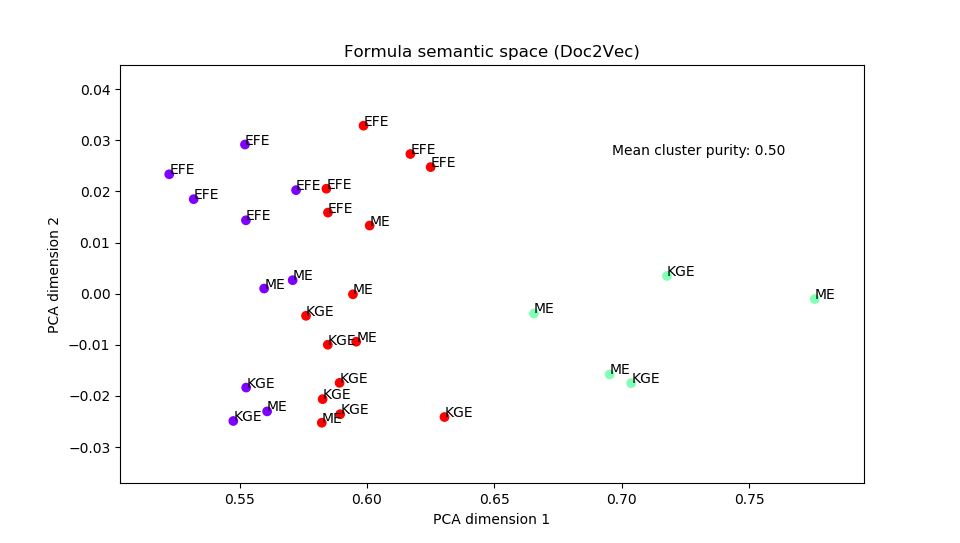}}
    \caption{Formula semantic space of three selected Formula Concepts (KGE, EFE, ME), using {\tt TF-IDF} or {\tt Doc2Vec} encodings, reduced by Principal Component Analysis (PCA) to two dimensions. The color code corresponds to the clusters computed by $k$-means ($k=3$) clustering.}
    \label{fig:formsemspace}
\end{figure}

%
%\paragraph{Formula Space.}
%
We first examine the separation of the three Formula Concepts by investigating the formula space in each of the four computed formula vector encodings.
Figures \ref{fig:formcontspace} and \ref{fig:formsemspace} show the resulting plots.
We reduce the dimensions via Principal Component Analysis (PCA) to two ($x$- and $y$-axes).
Furthermore, we color-code the results of our formula clustering experiment (see next paragraph), such that each datapoint color corresponds to a different cluster computed by $k$-means ($k=3$) clustering.
Apparently, in the formula content space with {\tt Doc2Vec} encodings (second plot), the three Formula Concept classes are separated best with the largest distances between the three cluster centroids (see Table \ref{tab:c_mean_dist}).
Only two Formula Concept examples of class ME are incorrectly located in the cluster, which primarily consists of class KGE.
We can identify these as being equation \eqref{eq:Maxwellinhom} and \eqref{eq:Maxwellhom}.
We suspect the partial derivative to be causing the mix-up of these ME, since they predominantly occur in the KGE.
\begin{table}[h]
\centering
\caption{Mean cluster centroid distance after employing PCA to reduce the number of datapoint dimensions to two (see the 2D plots in Figures  \ref{fig:formcontspace} and \ref{fig:formsemspace}).
The formula content {\tt Doc2Vec} encoding performs best (largest distance).}
\label{tab:c_mean_dist}
\begin{tabular}{|l|l|}
\hline
Encoding & Mean centroid distance \\
\hline
Formula content {\tt TF-IDF} & 0.57 \\ \hline
Formula content {\tt Doc2Vec} & 0.81 \\ \hline
Formula semantics {\tt TF-IDF} & 0.73 \\ \hline
Formula semantics {\tt Doc2Vec} & 0.11 \\ \hline
%ContTF: 0.5660515185639468
%Cont2V: 0.8057354425922362
%SemTF: 0.7274618077552994
%Sem2V: 0.10635234909585199
\end{tabular}
\end{table}

%\paragraph{Formula Clustering.}

As another measure for the separability of our three example Formula Concepts, we calculate the cluster purity as the number of datapoints of the class that makes up the largest fraction of a cluster divided by the cluster size, averaged over all clusters:
\vspace{-0.1cm}
\begin{align*}
    \text{purity} = \underset{\text{clusters}}{\text{mean}} \left[  \frac{1}{\text{cluster size}} \ \text{max}(\text{\#datapoints in cluster per class}) \right].
\end{align*}
\vspace{0.1cm}
\!\!Table \ref{tab:clust_pur} holds the cluster purities of a $k$-means clusterer on different formula vector encodings.
Apparently, the formula content {\tt Doc2Vec} encoding outperforms the others.
This is illustrated by comparing Figures \ref{fig:formcontspace} and \ref{fig:formsemspace}.
In the {\tt Doc2Vec} encoding, the smallest number of Formula Concept labels (only two) are mixed up.

\begin{table}[h]
\centering
\caption{Mean cluster purity of a $k$-means clusterer on different formula vector encodings.
The formula content {\tt Doc2Vec} encoding performs best (highest purity).}
\label{tab:clust_pur}
\begin{tabular}{|l|l|}
\hline
Encoding & Mean cluster purity \\
\hline
Formula content {\tt TF-IDF} & 0.97 \\ \hline
Formula content {\tt Doc2Vec} & 0.94 \\ \hline
Formula semantics {\tt TF-IDF} & 0.97 \\ \hline
Formula semantics {\tt Doc2Vec} & 0.50 \\ \hline
\end{tabular}
\end{table}

%\paragraph{Formula Classification.}

As the third measure for the separability of our three example Formula Concepts, we calculate the classification accuracy of a
%Logistic Regression (LogReg) and
Support Vector Machine (SVM) classifier on our four formula vector encodings.
%
%\begin{table}[h]
%\centering
%\caption{Classification accuracy (cross-validated) of a Logistic Regression (LogReg) classifier on different formula vector encodings.
%The formula semantic {\tt TF-IDF} encoding performs best (highest mean accuracy).}
%\label{tab:class_LogReg}
%\begin{tabular}{|l|l|}
%\hline
%Encoding & Classification accuracy (\textit{cv} = 3/10) \\
%\hline
%Formula content {\tt TF-IDF} & 0.93 / 1.00 \\ \hline
%Formula content {\tt Doc2Vec} & 0.90 / 0.90 \\ \hline
%Formula semantics {\tt TF-IDF} & 0.96 / 0.97 \\ \hline
%Formula semantics {\tt Doc2Vec} & 0.30 / 0.43 \\ \hline
%\end{tabular}
%\end{table}
%%
%\begin{table}[h]
%\centering
%\caption{Classification accuracy (cross-validated) of a Support Vector Machine (SVM) classifier on different formula vector encodings.
%The formula semantic {\tt TF-IDF} encoding performs best (highest mean accuracy).}
%\label{tab:class_SVM}
%\begin{tabular}{|l|l|}
%\hline
%Encoding & Classification accuracy (\textit{cv} = 3/10) \\
%\hline
%Formula content {\tt TF-IDF} & 0.93 / 1.00 \\ \hline
%Formula content {\tt Doc2Vec} & 0.93 / 0.93 \\ \hline
%Formula semantics {\tt TF-IDF} & 0.96 / 0.97 \\ \hline
%Formula semantics {\tt Doc2Vec} & 0.53 / 0.57 \\ \hline
%\end{tabular}
%\end{table}
%
%
%\paragraph{Conclusion (FCR using ML).}
%
Summarizing, we test FCR approaches for Formula Concept separation using machine learning techniques such as neural formula vector encodings ({\tt Doc2Vec}), dimensionality reduction (PCA), clustering ($k$-means), and classification (SVM).
Our three measures of separability are 1) mean cluster centroid distance, 2) mean cluster purity, and 3) classification accuracy (cross-validated).
While the formula semantic TF-IF encoding performs best (averaged over the two classifiers and cross-validation splittings), the formula content {\tt Doc2Vec} encodings outperform the others in both cluster centroid distance and purity.

We avoid data skewness by employing a balanced dataset of examples equally distributed over classes.
%For the classification and clustering to be effective, we need at least five to ten examples per Formula Concept.
%Currently, we are collecting more data (through the methods presented in \ref{sec:fut.work}) to run this experiment over a larger dataset with more individual Formula Concepts in the future.

The Formula Concept clustering using a $k$-means algorithm can assign 29/30 $\simeq$ 97\% correctly, while the fuzzy string matching performs\footnote{A formula is assigned to the Formula Concept class that achieves the highest sum of similarity values.} slightly worse with 28/30 $\simeq$ 93\%.
Random sampling only reaches 8/30 $\simeq$ 27\%.
So, the clustering outperforms the other methods.
However, this only works if the cluster number $k$ (number of Formula Concept classes in the dataset) is known a priori.

\paragraph{100 Examples}

In the next step, we extend our study to the full dataset of 100 examples FCs from 10 classes.

% Classification accuracy and cluster purity

% binomial choice distribution
% # scipy.special.binom(n,k) = scipy.special.binom(10,5) = 252.0
% # tot possibilities: sum(1,10,binom(n,k)) = 1275

% PLOTs

\begin{figure}
    \centering
    \subfloat{\includegraphics[width=0.5\textwidth]{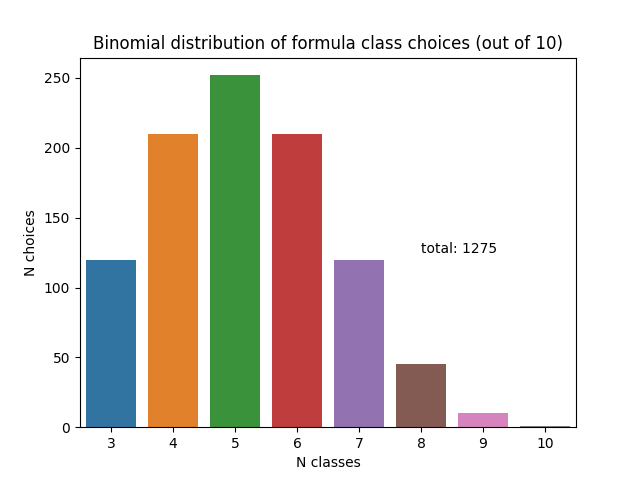}}\\
    \subfloat{\includegraphics[width=0.5\textwidth]{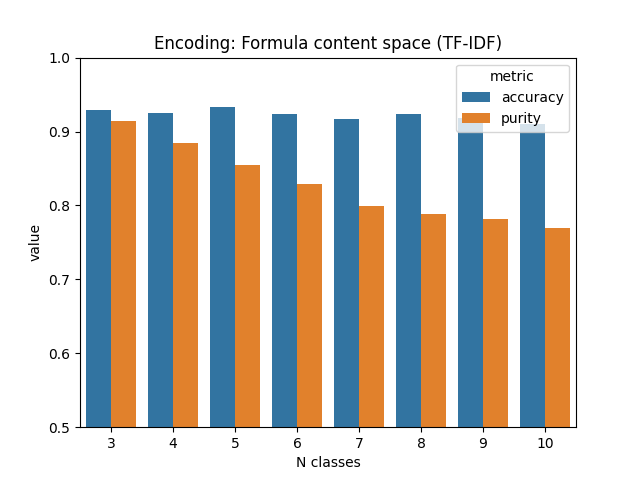}}
    \subfloat{\includegraphics[width=0.5\textwidth]{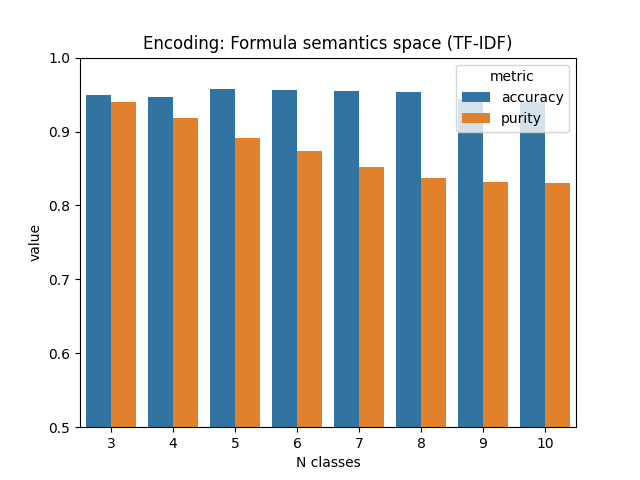}}\\
    \subfloat{\includegraphics[width=0.5\textwidth]{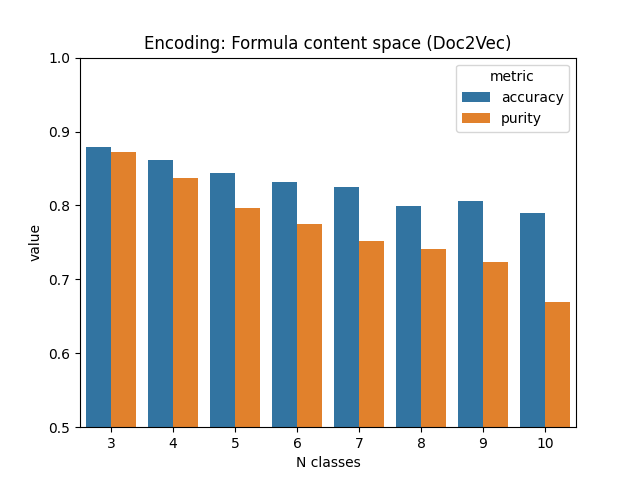}}
    \subfloat{\includegraphics[width=0.5\textwidth]{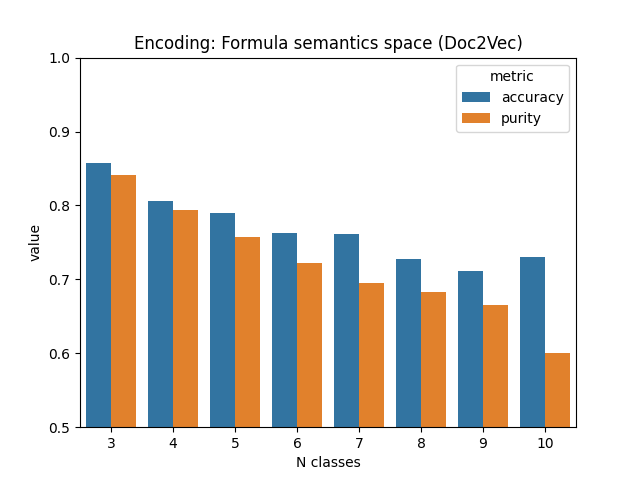}}
    \caption{Classification accuracies (cross-validated) and cluster purities (labeling-referenced) for a selection of 100 equations, semantically annotated (constituent QIDs) and sorted into 10 classes (formula QIDs). The binomial choice distribution for a selection of $N$ formulas out of the pool is shown above. Four different encodings (Content TF-IDF, Semantics TF-IDF, Content Doc2Vec, and Semantics Doc2Vec) are compared below.}
    \label{fig:class_clust_10eqn}
\end{figure}

% TABLE

\begin{table}[]
\caption{Classification accuracies (cross-validated) and cluster purities (labeling-referenced) for a selection of 100 equations, semantically annotated (constituent QIDs) and sorted into 10 classes (formula QIDs). The binomial choice distribution for selecting $N$ formulas out of the pool is featured in the first two columns. Four different encodings (Content TF-IDF, Semantics TF-IDF, Content Doc2Vec, and Semantics Doc2Vec) are compared.}
\label{tab:class_clust_10eqn}
\resizebox{\textwidth}{!}{
\begin{tabular}{|l|l|l|l|l|l|l|}
\hline
%\textbf{Nr. classes} & \textbf{Nr. choices} & \textbf{Metric / encoding} & \textbf{Content TFI-IDF} & \textbf{Content Doc2Vec} & \textbf{Semantics TF-IDF} & \textbf{Semantics Doc2Vec} \\ \hline
\textbf{Classes} & \textbf{Choices} & \textbf{Metric} & \textbf{Cont. TF.} & \textbf{Cont. D2V.} & \textbf{Sem. TF.} & \textbf{Sem. D2V.} \\ \hline
3                    & 120                  & accuracy                   & 0.93                     & 0.95                     & 0.88                      & 0.86                       \\ \hline
3                    & 120                  & purity                     & 0.91                     & 0.94                     & 0.87                      & 0.84                       \\ \hline
4                    & 210                  & accuracy                   & 0.93                     & 0.95                     & 0.86                      & 0.81                       \\ \hline
4                    & 210                  & purity                     & 0.88                     & 0.92                     & 0.84                      & 0.79                       \\ \hline
5                    & 252                  & accuracy                   & 0.93                     & 0.96                     & 0.84                      & 0.79                       \\ \hline
5                    & 252                  & purity                     & 0.86                     & 0.89                     & 0.80                       & 0.76                       \\ \hline
6                    & 210                  & accuracy                   & 0.92                     & 0.96                     & 0.83                      & 0.76                       \\ \hline
6                    & 210                  & purity                     & 0.83                     & 0.87                     & 0.77                      & 0.72                       \\ \hline
7                    & 120                  & accuracy                   & 0.92                     & 0.96                     & 0.82                      & 0.76                       \\ \hline
7                    & 120                  & purity                     & 0.80                      & 0.85                     & 0.75                      & 0.69                       \\ \hline
8                    & 45                   & accuracy                   & 0.92                     & 0.95                     & 0.80                       & 0.73                       \\ \hline
8                    & 45                   & purity                     & 0.79                     & 0.84                     & 0.74                      & 0.68                       \\ \hline
9                    & 10                   & accuracy                   & 0.92                     & 0.94                     & 0.81                      & 0.71                       \\ \hline
9                    & 10                   & purity                     & 0.78                     & 0.83                     & 0.72                      & 0.67                       \\ \hline
10                   & 1                    & accuracy                   & 0.91                     & 0.94                     & 0.79                      & 0.73                       \\ \hline
10                   & 1                    & purity                     & 0.77                     & 0.83                     & 0.67                      & 0.60                        \\ \hline
\hline
\textbf{Mean}        & /                    & accuracy                   & 0.92                     & \textbf{0.95}                     & 0.83                      & 0.77                       \\ \hline
\textbf{Mean}        & /                    & purity                     & 0.83                     & \textbf{0.87}                     & 0.77                      & 0.72                       \\ \hline
\end{tabular}
}
\end{table}

% Cont. D2V. performs best!

Figure \ref{fig:class_clust_10eqn} and Table \ref{tab:class_clust_10eqn} show the performance evaluation of classification (cross-validated) and clustering (labeling-referenced) of the labeled selection of 100 FC examples from 10 FC classes.
Classification accuracy (blue bars) and cluster purity (orange bars) is computed for each encoding (content or semantics in TF-IDF or Doc2Vec) in all 1275 combinatoric class choices individually (with $N$ ranging from 3 to 10, see the top plot for the binomial distribution). The displayed values ($y$-axis) are averaged over all respective combinations for a given number of class choices ($x$-axis). For each of the 4*1275 runs, we perform N-fold cross-validation retrieving the classification accuracy.

For the TF-IDF encoding (upper plots), the results are the following:
While the classification accuracy remains approximately stable with increasing N, the cluster purity decreases.
This means that in the supervised retrieval case (FCR), clustering is most appropriate for a small number of classes. However, it can still be helpful in the unsupervised case for discovering (FCD) and labeling unknown classes.
For the Doc2Vec encoding (lower plots), the results are the following:
The classification accuracy also decreases with increasing N, and the cluster purity more strongly. This means that it might be preferable to employ TF-IDF instead of Doc2Vec, which even has the additional advantage of being faster to compute.

We conclude that the classification is potentially more useful than the clustering for labeled FCR (if the formulas are already annotated). Yet, also for unlabeled formulas, the clustering might not be helpful because, as stated before, the cluster number of different concepts is not known a priori.
However, in the upcoming Experiment \ref{exp:FCsim}, we showed that a formula similarity map could be used instead as a means for both FCD and FCR.
%However, a formula similarity map as shown in Figure \ref{fig:unlabeled_vs_labeled}, \ref{fig:comparing_encodings_eqn}, and \ref{fig:comparing_encodings_cls} might be helpful for both FCD and FCR.

\subsection{Experiment \ref{exp:FCsim}:~Formula Concept Similarity}

In this experiment, we investigate the FC separability using FC similarity map matrices. We start with a preliminary analysis of the small set of 30 examples to be subsequently extended to all 100 examples.

\paragraph{30 Examples}

\begin{figure}
    \centering
    \includegraphics[width=\textwidth]{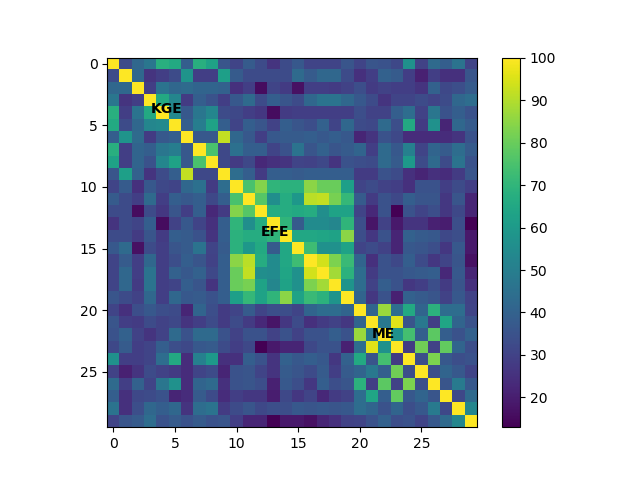}
    \caption{Matrix of the Formula Concept \LaTeX{} fuzzy string similarity percentages.
    On the $x$- and $y$-axes, the equation number is displayed such that each little square corresponds to one similarity value between one equation and another.}
    \label{fig:Fuzzy_matrix(content)}
\end{figure}

\begin{figure}
    \centering
    \includegraphics[width=\textwidth]{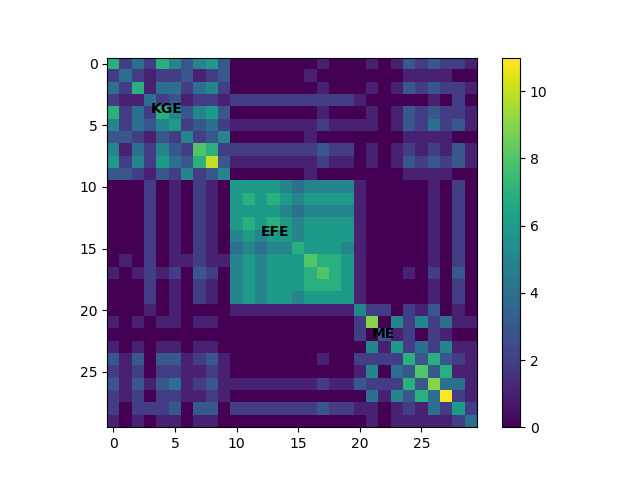}
    \caption{Matrix of the matching numbers of formula semantic QIDs.
    On the $x$- and $y$-axes, the equation number is displayed such that each
    %little
    square corresponds to one similarity value between one equation and another.}
    \label{fig:Matching_matrix(semantics)}
\end{figure}

Figure \ref{fig:Fuzzy_matrix(content)} shows the matrix of Formula Concept \LaTeX{} fuzzy string similarities for the small selection of 30 formulas discussed in Section \ref{sec:task2}.
We employ the \textit{fuzz.partial\_ratio} function of the Python package fuzzywuzzy\footnote{\url{https://github.com/seatgeek/fuzzywuzzy}}.
Each %little
square corresponds to the similarity percentage of the example equation with the number displayed on the $x$-axis to the example equation with the number displayed on the $y$-axis.
Since pairwise similarities are symmetric, the matrix is symmetric, and we can concentrate the investigations only on the part above or below the diagonal.
Apparently, the three Formula Concepts (KGE equation number 1-10, EFE number 10-20, ME 20-30) form three large squares (or triangles) aligned on the diagonal (containing the individual 100\% self-similarities).
Particularly striking is the EFE square in the center of the matrix with its high values and density.
This means that the Einstein Field Equations are the most similar, and the Formula Concept is highly coherent.
The considered representations of the other two Formula Concepts are much more diverse and more difficult to match or identify.
Figure \ref{fig:Matching_matrix(semantics)} shows the matrix of the Formula Concept semantic similarities.
The color code corresponds to the number of matching Wikidata QIDs of the corresponding Formula Concept examples (the $x$- and $y$-axes). 
The distribution is very similar to the fuzzy \LaTeX{} string content matching shown in Figure \ref{fig:Fuzzy_matrix(content)} (except the EFE square is slightly more distinct).
Thus, semantification has no significant advantage here.
However, in cases where the identifier symbols vary more, we expect an improvement.
%This thesis remains to be investigated in future work.

\paragraph{100 Examples}

% Unlabeled vs. labeled

\begin{figure}
    \centering
    \subfloat{\includegraphics[width=0.5\textwidth]{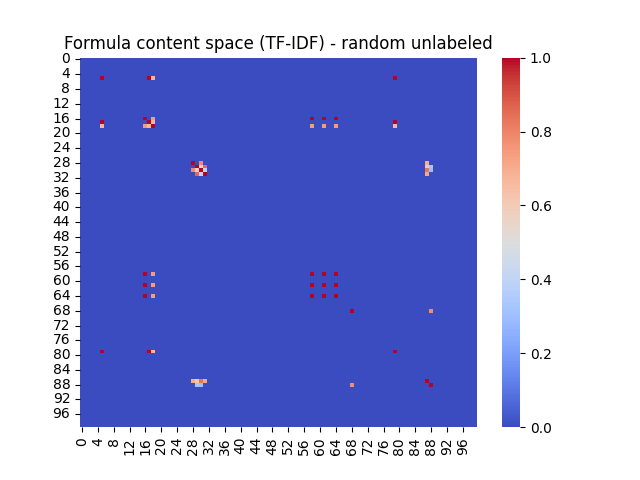}}
    \subfloat{\includegraphics[width=0.5\textwidth]{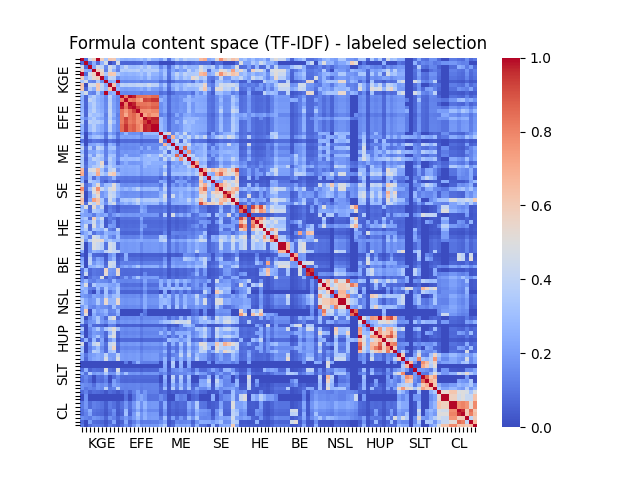}}\\
    \subfloat{\includegraphics[width=0.5\textwidth]{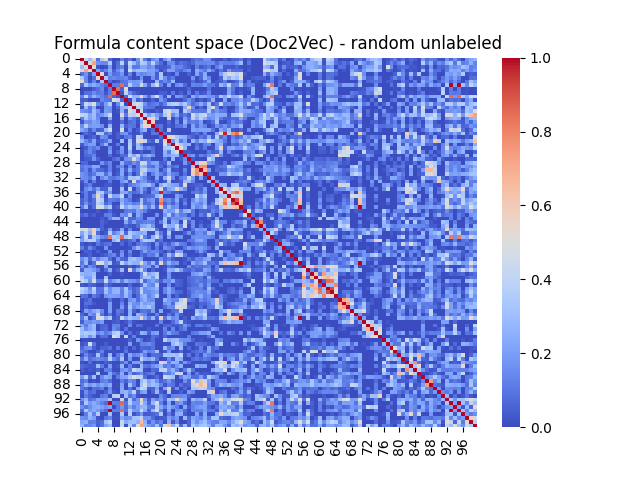}}
    \subfloat{\includegraphics[width=0.5\textwidth]{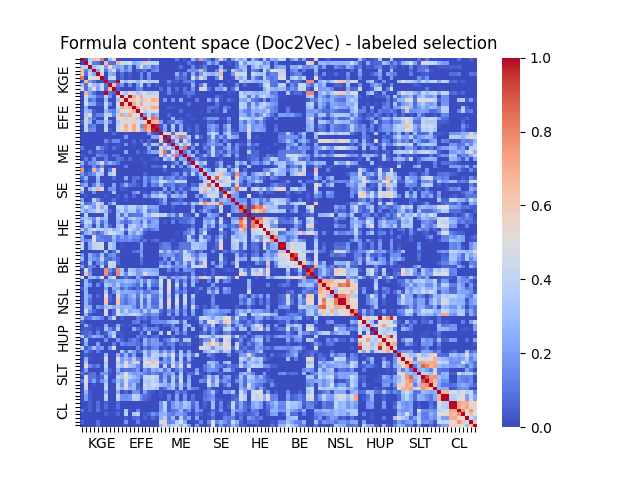}}\\
    \subfloat{\includegraphics[width=0.5\textwidth]{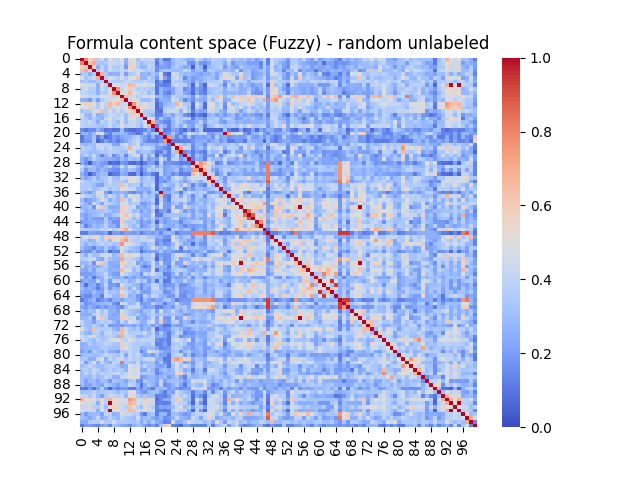}}
    \subfloat{\includegraphics[width=0.5\textwidth]{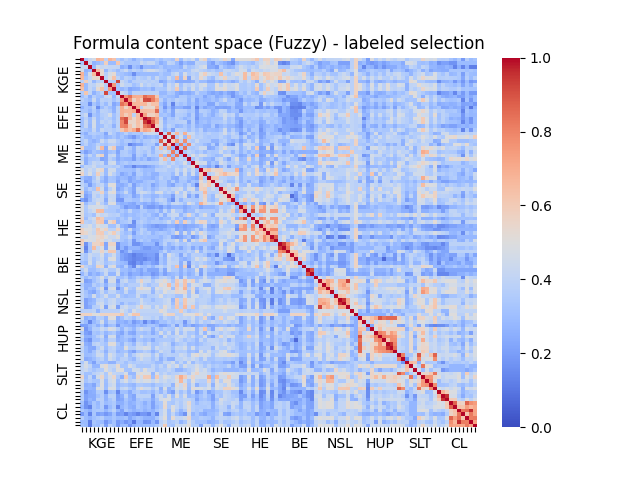}}
    \caption{Comparing unlabeled random equations (left) from the arXiv NTCIR dataset (astro-ph domain) to selected labeled equations (right) annotated by a human domain expert in different encodings (TF-IDF above, Doc2Vec middle, Fuzzy below, Content left, and Semantic right). Axes show random numbers or selected equation class labels. Very high TF-IDF, Doc2Vec cosine, or fuzzy string similarity between equations are marked in red. Figure best viewed in color.}
    \label{fig:unlabeled_vs_labeled}
\end{figure}

Figure \ref{fig:unlabeled_vs_labeled} shows a comparison of the formula similarities of random unlabeled to all of our 100 selected labeled example formulas.
While the random formulas are extracted from the arXiv NTCIR dataset, the labeled selection is taken from Wikipedia articles.

In Doc2Vec and Fuzzy encodings, the random unlabeled similarity map appears to be very similar to that of the labeled selection. This indicates that in both random sampling and labeled sampling, most of the formulas are not very similar to each other (blue background). However, for the labeled selection, there is an apparent self-coherence of the individual labeled FC classes (brighter red squares on the diagonal line).

We conclude that since the similarity map of labeled FCs is not weaker (less similarity) than that for random formulas, we can justify the classification and clustering as an appropriate tool or suitable means to recognize FCs. The lack of similarity or distinctness of the labeled classes does reflect the real-world situation for formulas in corpora, which is fortunate since it makes search and machine learning methods effective.

We can show that in the random sampling, the formula distinctness (low similarity) is equally low as for the labeled selection. This means that our machine learning experiments presented in Section \ref{exp:FCclassclust} are reasonable since they represent an information retrieval scenario that could occur.

% Comparing encodings

\begin{figure}
    \centering
    \subfloat{\includegraphics[width=0.5\textwidth]{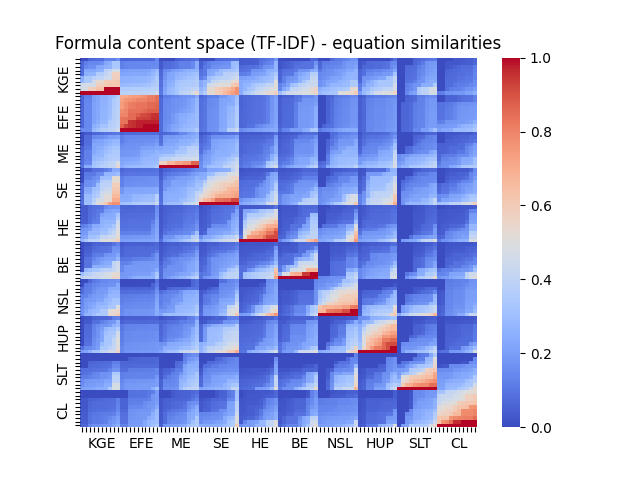}}
    \subfloat{\includegraphics[width=0.5\textwidth]{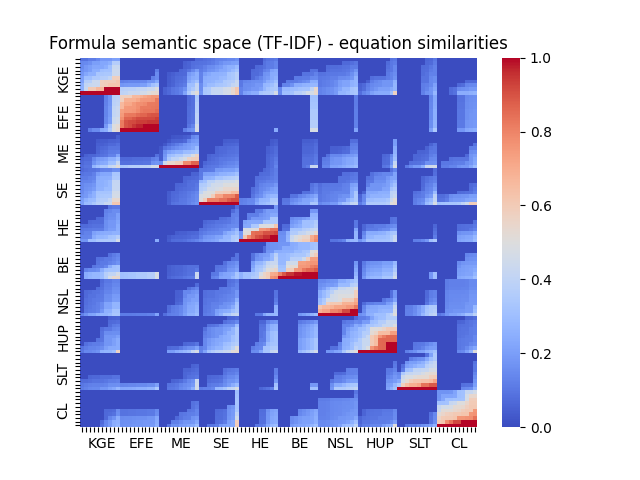}}\\
    \subfloat{\includegraphics[width=0.5\textwidth]{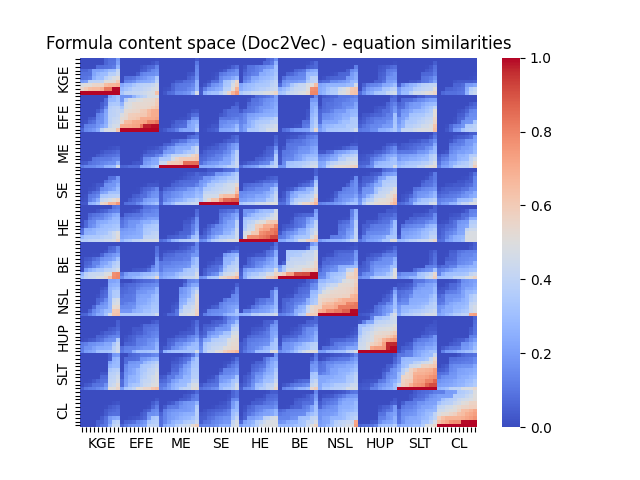}}
    \subfloat{\includegraphics[width=0.5\textwidth]{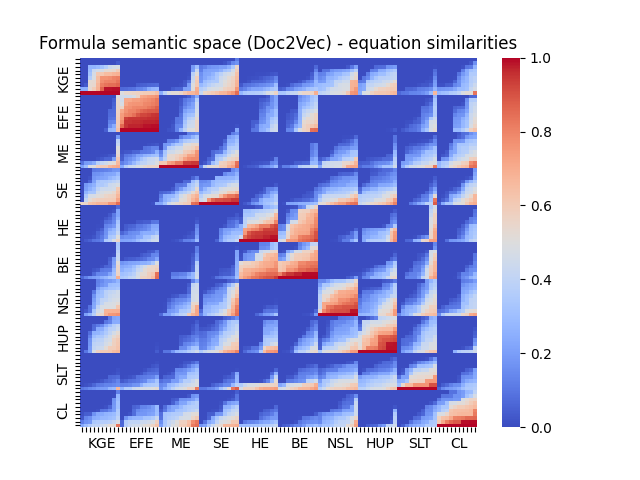}}\\
    \subfloat{\includegraphics[width=0.5\textwidth]{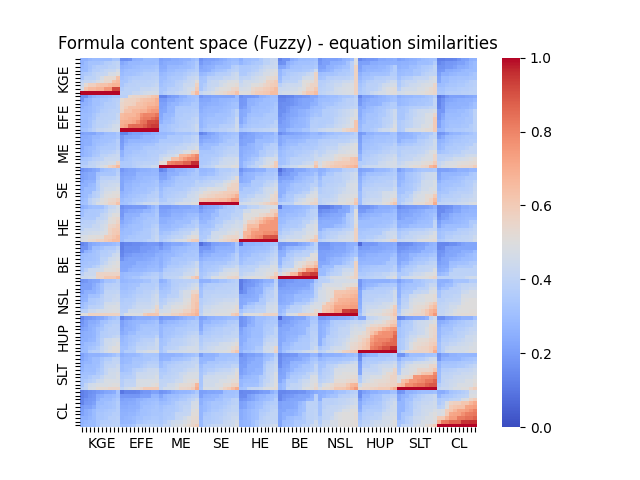}}
    \subfloat{\includegraphics[width=0.5\textwidth]{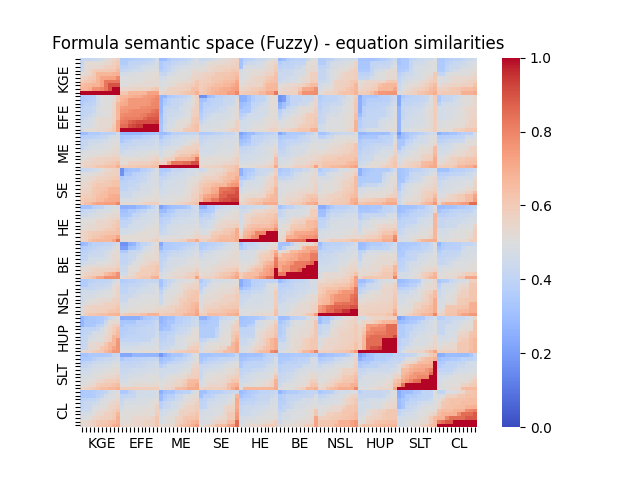}}
    \caption{Comparing labeled equation similarities for different encodings (TF-IDF above, Doc2Vec middle, Fuzzy below, Content left, and Semantic right). Axes show equation class labels. Very high TF-IDF, Doc2Vec cosine or fuzzy string similarity between equations are marked in red. Similarities are sorted within classes. Figure best viewed in color.}
    \label{fig:comparing_encodings_eqn}
\end{figure}

Figure \ref{fig:comparing_encodings_eqn} shows the formula similarities in different encodings (TF-IDF, Doc2Vec, Fuzzy) for all 100 examples, comparing the content space (formula constituent symbols encoded) to the semantic space (formula constituent QIDs encoded). Similarities are sorted within classes. The self-coherence of the labeled formula classes (labels on axes) is evident in all encodings. However, in the semantic space (Doc2Vec) encoding, additional inter-class / cross-class coherences are visible (some squares span several classes, e.g., `BE' and `HE' in the middle). This indicates latent semantic coherences that are less visible in the unsemantified content encoding.

\begin{figure}
    \centering
    \subfloat{\includegraphics[width=0.5\textwidth]{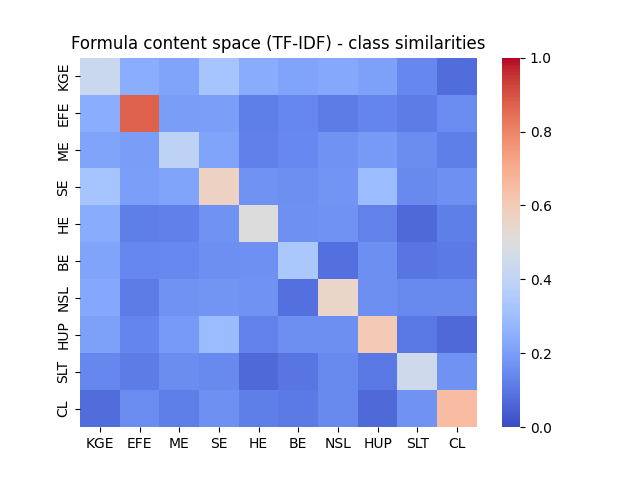}}
    \subfloat{\includegraphics[width=0.5\textwidth]{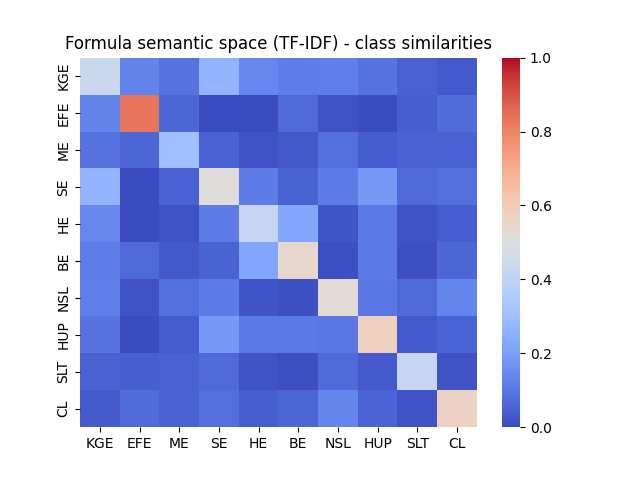}}\\
    \subfloat{\includegraphics[width=0.5\textwidth]{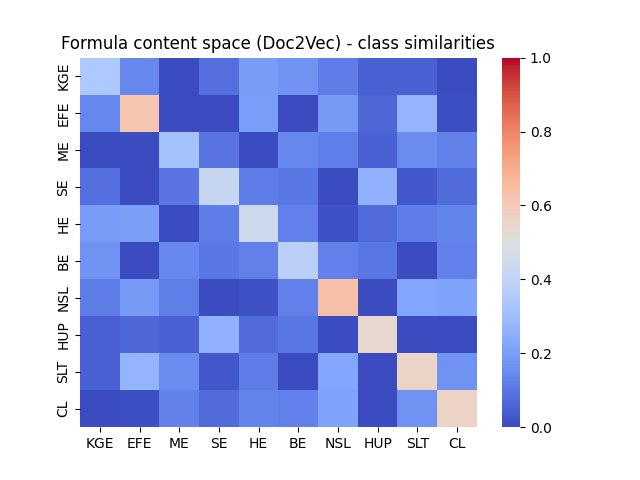}}
    \subfloat{\includegraphics[width=0.5\textwidth]{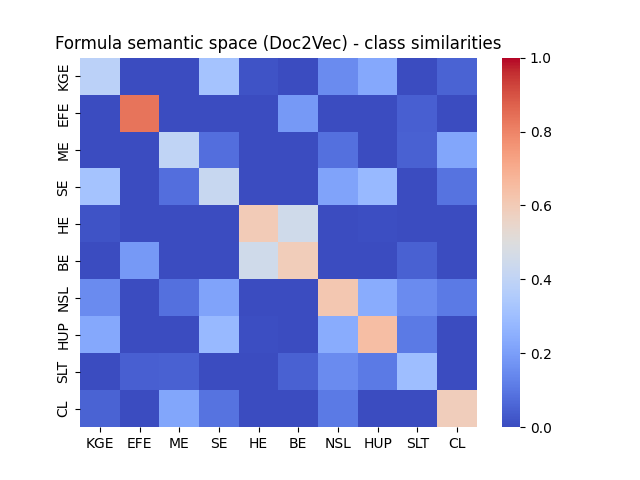}}\\
    \subfloat{\includegraphics[width=0.5\textwidth]{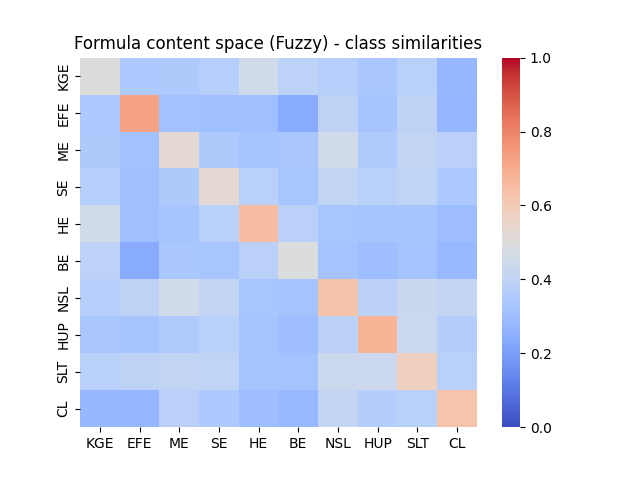}}
    \subfloat{\includegraphics[width=0.5\textwidth]{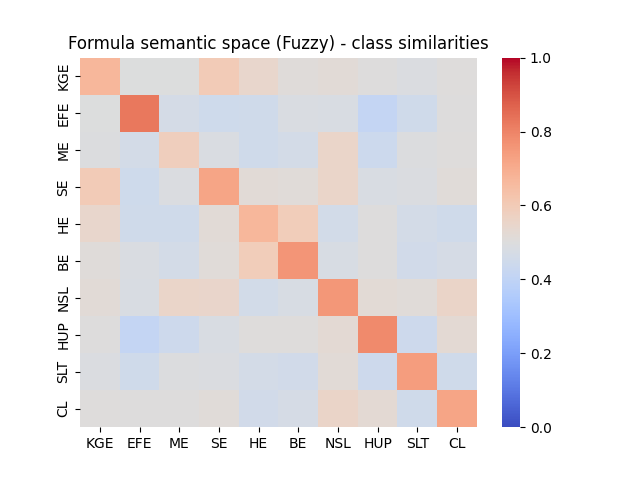}}
    \caption{Comparing labeled averaged class similarities for different encodings (TF-IDF above, Doc2Vec middle, Fuzzy below, Content left, and Semantic right). Axes show equation class labels. Very high TF-IDF, Doc2Vec cosine or fuzzy string similarity between equations are marked in red. Figure best viewed in color.}
    \label{fig:comparing_encodings_cls}
\end{figure}

Figure \ref{fig:comparing_encodings_cls} shows the formula similarities in different encodings and spaces averaged over classes (mean pooling). This view helps to better highlight the intra-class and inter-class coherences. On the top-left, the high intra-class coherence of the `EFE' formulas is illustrated by the prominent darker (more red intense) square. Moreover, the cross-class coherence mentioned in the description of Figure \ref{fig:comparing_encodings_eqn} is apparent again in the semantic space (Doc2Vec) encoding shown in the center of the middle right plot. Besides, other class similarities, such as that of the Klein--Gordon equations (`KGE') and Schrödinger equations (`SE'), can be identified as brighter squares.
Notice that the semantic (Fuzzy) space map (Figure \ref{fig:comparing_encodings_eqn} bottom right) shows that the inter-class similarity between KGE and SE in the semantic space is comparably high as the intra-class similarity of the ME class. This is reasonable, since they are indeed semantically very close. In the quantum physics framework, one equation can be derived from the other and vice versa. On the other hand, the intra-class similarity of the ME instances is high, since they are mutually semantically related.
The FC class similarity maps are also helpful for FCD, discovering FCs as coherent similarity areas to be subsequently analyzed and labeled.

\begin{figure}
    \centering
    \subfloat{\includegraphics[width=0.5\textwidth]{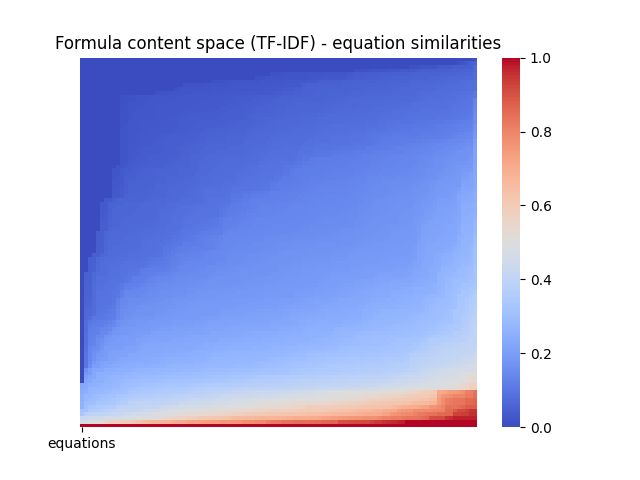}}
    \subfloat{\includegraphics[width=0.5\textwidth]{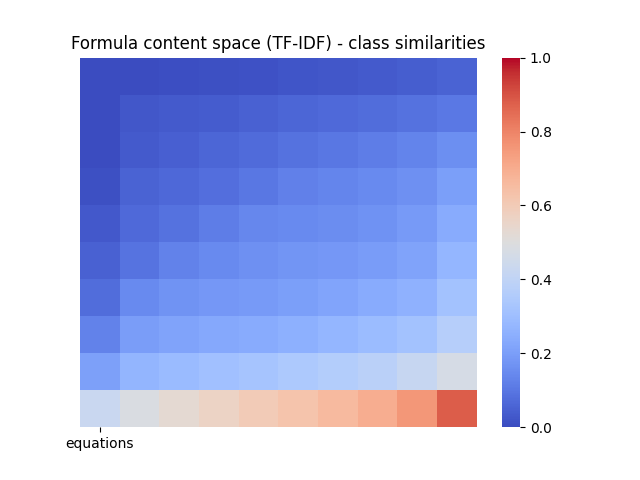}}
    \caption{Sorted similarity maps (Content TF-IDF encoding) for equations (left) and classes (right). The mean equation similarity is 0.2.}
    \label{fig:fully_sorted}
\end{figure}

Figure \ref{fig:fully_sorted} illustrates the overall dissimilarity of the equations in a sorted similarity map. The blue space (low similarity) significantly outweighs the red area (high similarity) at the bottom. The low mean equation similarity of 0.2 motivates FCR methods to exploit the separability.

\subsection{Conclusion (FCR)}

%COMPARISON OF APPROACHES

In three different experiments, we investigate the feasibility and effectiveness of methods to retrieve, separate, and recognize Formula Concepts (FCR).
For all experiments, we employ a manually labeled dataset of 100 Formula Concept examples from 10 classes retrieved from Wikipedia articles.

In Experiment \ref{exp:FCsearch} (Formula Concept search), we compare 8 different formula search methods on open corpora (Wikidata, Wikipedia, arXiv) and the web. 
We test how well Formula Concepts can be retrieved by search queries using either the formula \LaTeX{} string or the formula constituents, respectively.
The results show that using different retrieval methods and sources, it is possible to recognize Formula Concepts using search with a Mean Rank down to 1.78, Mean Reciprocal Rank up to 0.78, and Recall up to 0.74. Our FCR methods outperform the state-of-the-art search engines Approach0 and Google.

Experiment \ref{exp:FCclassclust} (Formula Concept classification and clustering), we assess Formula Concept separability by machine learning classification and clustering in selected formula encodings.
The results show while the cluster purity decreases with more FC classes, classification accuracy remains approximately stable around 0.9 when using TF-IDF formula encodings. This means that with stable accuracies, FC classification might be a more powerful means for FCR than FC clustering.

Experiment \ref{exp:FCsim} (Formula Concept similarity), we visualize formula (encoding) separability in similarity map matrices to illustrate coherence and overlap of Formula Concepts.
The results show that similarity maps are a valuable method for identifying both intra-class coherence and inter-class separability or overlap, which is useful for both FCD and FCR. Furthermore, the results motivate the employed machine learning methods since a comparison of our manual formula selection to randomly chosen formulas shows that in both cases, Formula Concepts are rather dissimilar and thus their classes separable from each other.

%
%\begin{figure}
%   \centering
%   \includegraphics[width=\textwidth]{FCR_assigment_comparison.png}
%    \caption{Comparison of FCR methods. The assignments of the Formula Concepts (KGE: 1, EFE: 2, ME:3) in the first column are examined. Formula Concept clustering using a $k$-means algorithm is able to assign 29/30 = 97\% correctly, while the fuzzy string matching performs slightly worse with 28/30 = 93\%. Random sampling only achieves 8/30 = 27\%.}
%    \label{fig:FCR_assignment_comparison}
%\end{figure}
%
We conclude that the search for specific formulas within a large dataset of STEM documents is a challenging problem.
Furthermore, we note that for FCR, there is an urgent need to augment semantic formula databases, for example, \textsc{MathMLben}\footnote{\url{https://mathmlben.wmflabs.org}} and Wikidata, such that they allow for multiple representations of a formula to be stored as a Formula Concept.
Having formulas tagged by Wikidata QIDs enables using them as markers in documents that can be cited (math citations).
Additionally, they can be employed to improve content-based recommender systems for academic literature, plagiarism detection systems, and ontology learning.

Note that our study’s aim is not a large-scale evaluation but rather a deductive conceptual work. The data, plots, and results we presented serve to illustrate the methodological concepts. We demonstrate the fundamental feasibility using examples and outline the potential for machine learning on labeled formula data. For a large-scale analysis using unlabeled formula data, we refer to the literature~\cite{DBLP:conf/jcdl/ScharpfSYHMG20,DBLP:conf/www/Greiner-PetterS20}.

%=====================================================================================
% Future Work
%=====================================================================================

\section{Future Work} \label{sec:fut.work}

This section outlines future endeavors and challenges, which we plan to address to further improve, evaluate, and apply FCD and FCR methods to additional use cases. These include exploring the practicability of a `Formula Rank', investigating a formula semantics sufficiency hypothesis, and developing methods for efficient semantic formula and triple annotation.

%\paragraph{Formula Concept Retrieval Task.}
%especially challenging: FCD - retrieving concept name, FCR in general

%\paragraph{Formula Encodings.}
%fut. work: encode sub/sup scripts differently

\paragraph{FormulaRank and Semantic Indexing.}

In analogy to Google's `PageRank'~\cite{DBLP:journals/cn/BrinP98}, and `TextRank'\\~\cite{mihalcea2004textrank}, we propose to employ a `FormulaRank' for Formula Concept popularity retrieval.
FormulaRank is supposed to rank formulas by the number of neighbors ($k$NN) or constituent intersections to estimate their importance. For this experiment, we first need to elaborate on interpretation standards and evaluation metrics for the results.
Secondly, we will develop and evaluate semantic indexing of the arXiv datasets containing formulas, their \LaTeX{} string, constituents retrieved from \MathML{} tags, surrounding text, and more.

\paragraph{Functional vs.~Semantic Recognition.}

Furthermore, we will investigate the following research question:~``Does the recognition of Formula Concepts require to take the functional relations of the formulas into account, or is it sufficient to only consider the semantics of the formula constituents?''. As an example, the Klein--Gordon equation
\begin{align*}
    \frac{1}{c^2} \frac{\partial^2 \psi}{\partial t^2} - \nabla^2 \psi + \left( \frac{m_0 c}{\hbar} \right)^2 \psi = 0,
\end{align*}
can be encoded as the semantic fingerprint of its constituents:
\begin{flushleft}
    \verb|c: "speed of light" (Q2111),|
    \verb|\partial: "partial derivative" (Q186475),|
    \verb|\psi: "wave function" (Q2362761), t: "time" (Q11471),|
    \verb|\nabla: "del" (Q334508), m: "mass" (Q11423) ,|
    \verb|\hbar: "Planck constant" (Q122894)|
\end{flushleft}
Alternatively, one could additionally take into account that the partial derivatives $\partial/\partial t$ and $\partial/\partial x$ act on the wave function $\psi$ and are applied with respect to both time $t$ and space $x$. Considering this circumstance would mean taking the functional relations of the formulas into account instead of merely considering the set of the semantics (fingerprint) of the formula constituents.

\paragraph{Semantic Annotations.}

To enable FCD by FCR, we are building a \LaTeX{} formula annotation recommender system~\cite{Scharpf2019b}, which helps and motivates authors from the STEM disciplines to make their papers semantically machine-interpretable by annotating formula and identifier names with Wikidata items (name and QID). We need labeled formula data for the semantic encodings and formula classification introduced in Section \ref{subsec:FCclassclust}.
Our long-term goal for this system is to directly integrate the annotation recommendation into both Wikipedia and Overleaf's editing or composing views. 
This would allow the Wikipedia and research communities to be more easily included in the semantification process of mathematical articles and research papers.
Employing extended AI-aided formula annotation enables scaling our approaches in further research projects on our infrastructure at Wikimedia, zbMATH Open, and the University of Göttingen.

\paragraph{RDF Triple Extraction.}

In the future, the semantic annotator will provide recommendations of RDF triples, both for natural language and mathematical statements.
A natural language statement can be, for example, the triple \{theory of relativity (Q43514), instance of (P31), scientific theory (Q3239681)\}.
For the mathematical statements, the Formula Concepts are represented as the triple \{Formula Concept item name, defining formula, formula \LaTeX{} string\}.

% Acknowledgements

%\begin{acknowledgements}
\section*{Acknowledgements}
This work was funded by the Deutsche Forschungsgemeinschaft (DFG, German Research Foundation) – grant 350192710 and 437179652 as well as the Lower Saxony Ministry of Science and Culture and the VW Foundation.
%\end{acknowledgements}

% could have direct or potential influence or impact bias on

\section*{Conflict of Interest}

The authors declare that they have no conflict of interest.

%\newpage

%\bibliographystyle{splncs04}
\bibliography{short}
%\bibliographystyle{apalike}
\bibliographystyle{unsrtnat}

\newpage

\section*{Appendix:~Formula Concept Examples}

\paragraph{Einstein Field Equations in Wikipedia (10 Results).}\footnote{Extracted from:~\url{https://en.wikipedia.org/wiki/Einstein_field_equations}.}

\begingroup
\allowdisplaybreaks
\begin{align*}
R_{\mu \nu} - \tfrac{1}{2}R \, g_{\mu \nu} + \Lambda g_{\mu \nu} = \frac{8 \pi G }{c^4} T_{\mu \nu},\\
G_{\mu \nu} = R_{\mu \nu} - \tfrac{1}{2} R g_{\mu \nu},\\
G_{\mu \nu} + \Lambda g_{\mu \nu} = \frac{8 \pi G}{c^4} T_{\mu \nu},\\
G_{\mu \nu} + \Lambda g_{\mu \nu} = 8 \pi T_{\mu \nu}~(G=c=1),\\
R_{\mu \nu} - \tfrac{1}{2} R g_{\mu \nu} - \Lambda g_{\mu \nu} = -\frac{8 \pi G}{c^4} T_{\mu \nu},\\
R - \frac{D}{2} R + D \Lambda = \frac{8 \pi G}{c^4} T,\\
-R + \frac{D \Lambda}{\frac{D}{2} -1} = \frac{8 \pi G}{c^4} \frac{T}{\frac{D}{2}-1},\\
R_{\mu \nu} - \frac{ \Lambda g_{\mu \nu}}{\frac{D}{2}-1} = \frac{8 \pi G}{c^4} \left(T_{\mu \nu} - \frac{1}{D-2}Tg_{\mu \nu}\right),\\
R_{\mu \nu} - \Lambda g_{\mu \nu} = \frac{8 \pi G}{c^4} \left(T_{\mu \nu} - \tfrac{1}{2}T\,g_{\mu \nu}\right),\\
R_{\mu \nu} - \tfrac{1}{2} R \, g_{\mu \nu} + \Lambda g_{\mu \nu} = \frac{8 \pi G}{c^4} T_{\mu \nu}~\text{(duplicate)}.
\end{align*}
\endgroup

\paragraph{Einstein Field Equations in arXiv NTCIR (77 Results).}\footnote{Dataset available at:~\url{http://research.nii.ac.jp/ntcir/ntcir-11/data.html}.}

\begingroup
\allowdisplaybreaks
\begin{align*}
G_{\mu\nu}-{1\over 2}Rg_{\mu\nu}=\kappa(T^{\varphi}_{\mu\nu}+T_{\mu\nu}),\\
R^{\mu\nu}-{1\over 2}g^{\mu\nu}(R-2\Lambda)=8\pi GT^{\mu\nu},\\
R_{\mu\nu}-\frac{1}{2}g_{\mu\nu}R+\Lambda_{c}g_{\mu\nu}=8\pi GT_{\mu\nu},\\
R_{\mu\nu}-{1\over 2}Rg_{\mu\nu}+\Lambda g_{\mu\nu}=8\pi GT_{\mu\nu},\\
G_{\mu\nu}=-\Lambda g_{\mu\nu}+\kappa^{2}T^{\rm tot}_{\mu\nu},\\
G_{AB}\equiv R_{AB}-{1\over 2}g_{AB}R=\kappa^{2}\,T_{AB},\\
G_{\mu\nu}+\Lambda g_{\mu\nu}=\kappa T_{\mu\nu},\\
G_{\mu\nu}-\Lambda g_{\mu\nu}=\kappa T_{\mu\nu},\\
G_{\mu\nu}-g_{\mu\nu}\Lambda=\frac{8\pi G}{c_{0}^{4}\phi^{4}}T_{\mu\nu},\\
R_{\mu\nu}-\frac{1}{2}Rg_{\mu\nu}+\Lambda g_{\mu\nu}=8\pi G\,T_{\mu\nu},\\
R_{\mu\nu}-{1\over 2}g_{\mu\nu}R=g_{\mu\nu}\Lambda-8\pi GT_{\mu\nu},\\
R_{\mu\nu}-\frac{1}{2}g_{\mu\nu}R+\Lambda g_{\mu\nu}=-8\pi GT_{\mu\nu},\\
R_{\mu\nu}-{1\over 2}g_{\mu\nu}R=\kappa T_{\mu\nu}-{\Lambda\over 2}g_{\mu\nu},\\
R_{\mu\nu}-{1\over 2}g_{\mu\nu}R=8\pi G[T_{\mu\nu}^{c}+T_{\mu\nu}^{q}],\\
R_{\mu\nu}-{1\over{2}}g_{\mu\nu}R=g_{\mu\nu}\Lambda-8\pi GT_{\mu\nu},\\
R_{\mu\nu}-{1\over 2}R g_{\mu\nu}={8\pi G\over c^{4}}T_{\mu\nu},\\
R_{\mu\nu}-\frac{1}{2}g_{\mu\nu}R={\frac{8\pi G}{c^{4}}}T_{\mu\nu},\\
G^{\mu\nu}+\Lambda g^{\mu\nu}=\kappa{T_{e}}^{\mu\nu},\\
G^{\mu\nu}-T^{\mu\nu}=\kappa{T_{g}}^{\mu\nu},\\
R_{\mu\nu}-\frac{1}{2}R g_{\mu\nu}+\Lambda g_{\mu\nu}=8\pi G\,T_{\mu\nu}~\text{(duplicate)},\\
R_{\mu\nu}-\frac{1}{2}g_{\mu\nu}R=8\pi G\,T_{\mu\nu}+\Lambda\,g_{\mu\nu},\\
G_{\mu\nu}+\Lambda g_{\mu\nu}=\kappa T_{\mu\nu},\\
R_{\mu\nu}-\frac{1}{2}g_{\mu\nu}R+\Lambda g_{\mu\nu}=8\pi GT_{\mu\nu},\\
G_{\mu\nu}=-\Lambda_{4}g_{\mu\nu}+\frac{1}{2\alpha_{0}}T_{\mu\nu}^{\rm c.c.},\\
R_{\mu\nu}-\frac{1}{2}g_{\mu\nu}R+\Lambda_{c}g_{\mu\nu}=8\pi GT_{\mu\nu},\\
R_{\mu\nu}-\frac{1}{2}~{}g_{\mu\nu}R=8\pi GT_{\mu\nu}-\Lambda g_{\mu\nu}~{},\\
{G}_{\mu\nu}+\alpha{H}_{\mu\nu}+\Lambda g_{\mu\nu}=\kappa_{n}^{2}{T}_{\mu\nu},\\
R_{\mu\nu}-{1\over 2}Rg_{\mu\nu}=-\Lambda g_{\mu\nu}+8\pi GT_{\mu\nu},\\
G_{\mu\nu}+\Lambda_{\mathrm{R}}g_{\mu\nu}=8\pi G\langle\widetilde{T}_{\mu\nu}\rangle,\\[0.1cm]
G_{\mu\nu}+\Phi_{\mu\nu}+\Lambda g_{\mu\nu}=\kappa T_{\mu\nu}~\text{(repeated 3 times)},\\
%G_{\mu\nu}+\Lambda g_{\mu\nu}=\kappa\,T_{\mu\nu},\\
%G_{\mu\nu}+\Lambda g_{\mu\nu}=\kappa\,T_{\mu\nu},\\
%G_{\mu\nu}+\Lambda g_{\mu\nu}=\kappa\,T_{\mu\nu},\\
R_{(\mu\nu)}-\frac{1}{2}Rg_{\mu\nu}+\Lambda g_{\mu\nu}=\kappa T_{\mu\nu},\\
R_{\mu\nu}-{1\over 2}g_{\mu\nu}R=8\pi GT_{\mu\nu}+\Lambda g_{\mu\nu},\\
R_{\mu\nu}-\frac{1}{2}Rg_{\mu\nu}=\kappa_{r}(T)T_{\mu\nu}+\Lambda(T)g_{\mu\nu},\\
R_{\mu\nu}-\frac{1}{2}Rg_{\mu\nu}=\kappa(T_{\mu\nu}^{m}+T_{\mu\nu}^{\Lambda}),\\
R_{\mu\nu}-\frac{1}{2}Rg_{\mu\nu}=\kappa T_{\mu\nu}+\Lambda(T)g_{\mu\nu},\\
R_{\mu\nu}-\frac{1}{2}Rg_{\mu\nu}=\kappa_{r}T_{\mu\nu}+\Lambda(T)g_{\mu\nu},\\
K_{\mu\nu}-Kg_{\mu\nu}=-\frac{\kappa^{2}}{2}T_{\mu\nu}+r_{c}G_{\mu\nu},\\
R_{\mu\nu}-\frac{1}{2}g_{\mu\nu}R+\Lambda_{c}g_{\mu\nu}=\kappa T_{\mu\nu},\\
R_{\mu\nu}-\frac{1}{2}g_{\mu\nu}R+\Lambda g_{\mu\nu}=8\pi GT_{\mu\nu},\\
R_{\mu\nu}-\Lambda g_{\mu\nu}=8\pi G(T_{\mu\nu}-{1\over 2}g_{\mu\nu}T),\\
R_{\mu\nu}-{1\over 2}g_{\mu\nu}R+\Lambda\,g_{\mu\nu}=8\pi G_{N}\,T_{\mu\nu},\\
R_{\mu\nu}-\frac{1}{2}g_{\mu\nu}R=\frac{8\pi G}{c^{4}}T_{\mu\nu},\\
G_{\mu\nu}=R_{\mu\nu}-\frac{1}{2}g_{\mu\nu}R=8\pi GT_{\mu\nu}-\Lambda g_{\mu\nu},\\
R_{\mu\nu}-\frac{1}{2}\,Rg_{\mu\nu}=\kappa\,T_{\mu\nu}-\Lambda\,g_{\mu\nu},\\
R_{\mu\nu}-\frac{1}{2}g_{\mu\nu}R-\Lambda g_{\mu\nu}=(8\pi G_{N})T_{\mu\nu},\\
R_{\mu\nu}-\frac{1}{2}g_{\mu\nu}R+\Lambda g_{\mu\nu}=-\kappa T_{\mu\nu},\\
G_{\mu\nu}=\kappa_{4}^{2}T_{\mu\nu}-\Lambda g_{\mu\nu}+Q_{\mu\nu},\\
R_{\mu\nu}-\frac{1}{2}g_{\mu\nu}R=\frac{8\pi G}{c^{4}}T_{\mu\nu},\\
R_{\mu\nu}-\frac{1}{2}g_{\mu\nu}R+\Lambda g_{\mu\nu}=-8\pi GT_{\mu\nu}
f_{R}\,G_{\mu\nu},\\
R_{\mu\nu}-\frac{1}{2}Rg_{\mu\nu}=8\pi GT_{\mu\nu}-\Lambda g_{\mu\nu}
T^{\rm RG}_{\mu\nu},\\
R_{\mu\nu}-\frac{1}{2}g_{\mu\nu}R+\Lambda g_{\mu\nu}=8\pi GT_{\mu\nu},\\
E^{\mu\nu}=-G^{\mu\nu}+\kappa T^{\mu\nu}-\Lambda g^{\mu\nu},\\
G_{\mu\nu}=R_{\mu\nu}-g_{\mu\nu}R/2=\kappa T^{\mu\nu}-\Lambda g_{\mu\nu},\\
R_{\mu\nu}-\frac{1}{2}g_{\mu\nu}R=\frac{8\pi G}{c^{4}}T_{\mu\nu},\\
R_{\mu\nu}-\frac{1}{2}g_{\mu\nu}R=8\pi G_{5}T_{\mu\nu}-\Lambda_{5}g_{\mu\nu},\\
R_{\mu\nu}-{1\over 2}Rg_{\mu\nu}+\Lambda_{eff}g_{\mu\nu}=8\pi GT_{\mu\nu},\\
R_{\mu\nu}-{1\over 2}Rg_{\mu\nu}+\Lambda g_{\mu\nu}=8\pi GT_{\mu\nu},\\
R_{\mu\nu}-{1\over 2}g_{\mu\nu}R={8\pi G\over c^{4}}T_{\mu\nu},\\
R_{\mu\nu}-\frac{1}{2}g_{\mu\nu}R-g_{\mu\nu}\Lambda=8\pi GT_{\mu\nu},\\
G_{\mu\nu}+\Lambda g_{\mu\nu}=\frac{\kappa}{e^{2}}T_{\mu\nu},\\
R_{\mu\nu}-\frac{1}{2}\,g_{\mu\nu}\,R=8\pi GT_{\mu\nu}+\Lambda,\\
G_{\mu\nu}\equiv R_{\mu\nu}-\frac{1}{2}Rg_{\mu\nu}=\kappa^{2}T_{\mu\nu},\\
G_{\mu\nu}=R_{\mu\nu}-\frac{1}{2}Rg_{\mu\nu}=\kappa T_{\mu\nu},\\
G^{\mu\nu}=-\Lambda(x)g^{\mu\nu}+\kappa T^{\mu\nu}_{\rm M},\\
R_{\mu\nu}-{g_{\mu\nu}\over 2}R={8\pi G\over c^{4}}T_{\mu\nu}
{1\over 2}{\rm Tr}H_{\chi}^{2},\\
R_{\mu\nu}-\frac{1}{2}g_{\mu\nu}R+\Lambda g_{\mu\nu}=\kappa T_{\mu\nu},\\
R_{\mu\nu}-\frac{1}{2}g_{\mu\nu}R+g_{\mu\nu}\Lambda=\kappa T_{\mu\nu},\\
G_{\mu\nu}-g_{\mu\nu}\Lambda={{8\pi G}\over c^{4}}T_{\mu\nu},\\
G_{\mu\nu}=R_{\mu\nu}-\frac{1}{2}g_{\mu\nu}R=\kappa^{2}T_{\mu\nu},\\
R^{\mu\nu}-\frac{1}{2}g^{\mu\nu}R=\frac{8\pi G}{c^{4}}T^{\prime\mu\nu},\\
R^{\mu\nu}-\frac{1}{2}g^{\mu\nu}R=\Lambda g^{\mu\nu}-8\pi GT^{\mu\nu}.
\end{align*}
\endgroup

\paragraph{Differential Equation Concept Class Examples (100 from 10 classes).}

% 10 EXAMPLES
%\begingroup
%\allowdisplaybreaks
%\begin{align*}
%\frac{1}{c^2} \frac{\partial^2 \psi}{\partial t^2} - \nabla^2 \psi + \left( \frac{m_0 c}{\hbar} \right)^2 \psi = 0,\\
%G_{\mu \nu} + \Lambda g_{\mu \nu} = \kappa T_{\mu \nu},\\
%\text{div} \vec{E} = 4 \pi \rho,\\
%i \hbar \frac{\partial}{\partial t} | \psi (t) \rangle = \hat{H} | \psi (t) \rangle,\\
%(\nabla^2 - k^2) A = -f,\\
%\nabla^4\varphi=0,\\
%\vec{F} = \frac{d\vec{p}}{dt},\\
%\sigma_{x}\sigma_{p} \geq \frac{\hbar}{2},\\
%\oint \frac{\delta Q}{T}=0,\\
%|F_1| = |F_2| = \frac{|q_1 \times q_2|}{r^2}.
%\end{align*}
%\endgroup

% ALL EQUATIONS
\begingroup
\allowdisplaybreaks
\begin{align*}
\text{\textbf{Klein--Gordon Equation (KGE)}}:\hspace{2.5cm}\\
\frac{1}{c^2} \frac{\partial^2 \psi}{\partial t^2} - \nabla^2 \psi + \left( \frac{m_0 c}{\hbar} \right)^2 \psi = 0,\hspace{2.5cm}\\
u_{tt} + A u + f(u) = 0,\hspace{2.5cm}\\
\partial^2_{ct} h_n (z,t) - \partial^2_z h_n (z, t) + \nu_n^2 h_n (z,t) = 0,\hspace{2.5cm}\\
\nabla^a \nabla_a \psi = \mu^2 \psi,\hspace{2.5cm}\\
-\hbar^2 \frac{\partial^2 \Psi}{\partial t^2} + c^2 \hbar^2 \nabla^2 \Psi = m_0^2 c^4 \Psi,\hspace{2.5cm}\\
\nabla^2 \phi - \frac{1}{c^2} \frac{\partial^2 \phi}{\partial t^2} - \frac{2\alpha + \alpha}{c^2} \frac{\partial \phi}{\partial t} - \frac{\alpha^2 + a \alpha}{c^2} \phi = 0,\hspace{2.5cm}\\
u_{tt} - \Delta u + m^2 u + G'(u) = 0,\hspace{2.5cm}\\
\left( \eta^{\mu \nu} \frac{\partial}{x^\mu} \frac{\partial}{x^\nu} - \left(\frac{mc}{\hbar} \right)^2 \right) \varphi = 0,\hspace{2.5cm}\\
\left(-\frac{1}{c^2} \frac{\partial^2}{\partial t^2} \sum_{i=1}^p \frac{\partial}{x^i} \frac{\partial}{x^i} - \left(\frac{mc}{\hbar} \right)^2 \right) \varphi = 0,\hspace{2.5cm}\\
u_{tt} - \Delta u + m u + P'(u) = 0.\hspace{2.5cm}\\
\text{\textbf{Einstein Field Equations (EFE)}}:\hspace{2.5cm}\\
G_{\mu \nu} + \Lambda g_{\mu \nu} = \kappa T_{\mu \nu},\hspace{2.5cm}\\
R_{\mu \nu} - \frac{1}{2} g_{\mu \nu} R - \Lambda g_{\mu \nu} = (8 \pi G_{N}) T_{\mu \nu},\hspace{2.5cm}\\
G_{\mu \nu} = -\Lambda g_{\mu \nu} + \kappa^{2} T^{\rm tot}_{\mu \nu},\hspace{2.5cm}\\
G_{\mu \nu} = R_{\mu \nu} - g_{\mu \nu} R/2 = \kappa T^{\mu \nu} - \Lambda g_{\mu \nu},\hspace{2.5cm}\\
R_{\mu \nu} - \frac{1}{2} R g_{\mu \nu} = \kappa_{r}(T) T_{\mu \nu} + \Lambda(T) g_{\mu \nu},\hspace{2.5cm}\\
K_{\mu \nu} - K g_{\mu \nu} = -\frac{\kappa^{2}}{2} T_{\mu \nu} + r_{c} G_{\mu \nu},\hspace{2.5cm}\\
R_{\mu \nu} - \frac{1}{2} g_{\mu \nu} R + \Lambda g_{\mu \nu} = -8 \pi G T_{\mu \nu} f_{R} G_{\mu \nu},\hspace{2.5cm}\\
R_{\mu \nu} - \frac{1}{2} g_{\mu \nu} R + \Lambda_{c} g_{\mu \nu} = 8 \pi G T_{\mu \nu},\hspace{2.5cm}\\
R_{\mu \nu} - {1\over 2} R g_{\mu \nu} + \Lambda_{eff} g_{\mu \nu} = 8 \pi G T_{\mu \nu},\hspace{2.5cm}\\
R_{\mu \nu} - \frac{1}{2} g_{\mu \nu} R = 8 \pi G_{5} T_{\mu \nu} - \Lambda_{5} g_{\mu \nu}.\hspace{2.5cm}\\
\text{\textbf{Maxwell's Equations (ME)}}:\hspace{2.5cm}\\
\text{div} \vec{E} = 4 \pi \rho,\hspace{2.5cm}\\
\oiint_{\partial \Omega} \mathbf{E} \cdot \mathrm{d} \mathbf{S} = \frac{1}{\varepsilon_0} \iiint_\Omega \rho \mathrm{d} V,\hspace{2.5cm}\\
\text{div} \vec{B} = 0,\hspace{2.5cm},\\
\oiint_{\partial \Omega} \mathbf{B} \cdot \mathrm{d} \mathbf{S} = 0,\hspace{2.5cm}\\
\text{rot} \vec{E} = - \frac{1}{c} \frac{\partial\vec{B}}{\partial t},\hspace{2.5cm}\\
\oint_{\partial \Sigma} \mathbf{E} \cdot \mathrm{d}\boldsymbol{l} = -\operatorname{\frac{d}{dt}} \iint_{\Sigma} \mathbf{B} \cdot \mathrm{d}\mathbf{S},\hspace{2.5cm}\\
\text{rot} \vec{B} = \frac{4 \pi}{c} \vec{j} + \frac{1}{c} \frac{\partial \vec{E}}{\partial t},\hspace{2.5cm}\\
\oint_{\partial \Sigma} \mathbf{B} \cdot \mathrm{d} \boldsymbol{l} = \mu_0 \left(\iint_{\Sigma} \mathbf{j} \cdot \mathrm{d}\mathbf{S} + \varepsilon_0 \frac{\mathrm{d}}{\mathrm{d}t} \iint_{\Sigma} \mathbf{E} \cdot \mathrm{d} \mathbf{S} \right),\hspace{2.5cm}\\
\partial_\alpha F^{\alpha \beta} = \frac{4 \pi}{c} j^\beta,\hspace{2.5cm}\\
\varepsilon^{\alpha \beta \gamma \delta} \partial_\beta F_{\gamma \delta} = 0.\hspace{2.5cm}\\
\text{\textbf{Schrödinger Equation (SE)}}:\hspace{2.5cm}\\
i \hbar \frac{\partial}{\partial t} | \psi (t) \rangle = \hat{H} | \psi (t) \rangle,\hspace{2.5cm}\\
i\hbar\frac{\partial}{\partial t} \Psi(x,t) = \left [ - \frac{\hbar^2}{2m}\frac{\partial^2}{\partial x^2} + V(x,t)\right ] \Psi(x,t),\hspace{2.5cm}\\
i \hbar \frac{d}{d t}\vert\Psi(t)\rangle = \hat H\vert\Psi(t)\rangle,\hspace{2.5cm}\\
\operatorname{\hat H}|\Psi\rangle = E |\Psi\rangle,\hspace{2.5cm}\\
i\hbar \frac{d}{dt}|\Psi(t)\rangle = \left(\frac{1}{2m}\hat{p}^2 + \hat{V}\right)|\Psi(t)\rangle,\hspace{2.5cm}\\
i\hbar\frac{\partial}{\partial t} \Psi(\mathbf{r},t) = - \frac{\hbar^2}{2m} \nabla^2 \Psi(\mathbf{r},t) + V(\mathbf{r}) \Psi(\mathbf{r},t),\hspace{2.5cm}\\
- \frac {\hbar ^2}{2m} \frac {d ^2 \psi}{dx^2} = E \psi,\hspace{2.5cm}\\
E\psi = -\frac{\hbar^2}{2m}\frac{d^2}{d x^2}\psi + \frac{1}{2} m\omega^2 x^2\psi,\hspace{2.5cm}\\
E \psi = -\frac{\hbar^2}{2\mu}\nabla^2\psi - \frac{q^2}{4\pi\varepsilon_0 r}\psi,\hspace{2.5cm}\\
i\hbar \frac{\partial}{\partial t} \Psi\left(\mathbf{r},t\right) = \hat{H} \Psi\left(\mathbf{r},t\right).\hspace{2.5cm}\\
\text{\textbf{Helmholtz Equation (HE)}}:\hspace{2.5cm}\\
(\nabla^2 - k^2) A = -f,\hspace{2.5cm}\\
\nabla^2 f = -k^2 f,\hspace{2.5cm}\\
\frac{\mathrm{d}^2 T}{\mathrm{d}t^2} + \omega^2T = \left( \frac{\mathrm{d}^2}{\mathrm{d}t^2} + \omega^2 \right) T = 0,\hspace{2.5cm}\\
\nabla^2 A = -k^2 A,\hspace{2.5cm}\\
\nabla_{\perp}^2 A + 2ik\frac{\partial A}{\partial z} = 0,\hspace{2.5cm}\\
\nabla^2 A(x) + k^2 A(x) = -f(x),\hspace{2.5cm}\\
\nabla^2 u + k^2 u = 0,\hspace{2.5cm}\\
\frac{\partial^2 u}{\partial x^2} + \frac{\partial^2 u}{\partial y^2} + \frac{\partial^2 u}{\partial z^2} + k^2 u(x,y,z) = 0,\hspace{2.5cm}\\
\nabla^2 u + k^2 u(\rho,\psi,z) = 0,\hspace{2.5cm}\\
\qquad \frac{1}{\rho} \,\frac{\partial}{\partial \rho} \left( \rho\,\frac{\partial u}{\partial \rho} \right) + \frac{1}{\rho^2} \,\frac{\partial^2 u}{\partial \phi^2} + \frac{\partial^2 u}{\partial z^2} + k^2 u = 0.\hspace{2.5cm}\\
\text{\textbf{Biharmonic Equation (BE)}}:\hspace{2.5cm}\\
\nabla^4\varphi=0,\hspace{2.5cm}\\
\nabla^2\nabla^2\varphi=0,\hspace{2.5cm}\\
\Delta^2\varphi=0,\hspace{2.5cm}\\
\sum_{i=1}^n\sum_{j=1}^n\partial_i\partial_i\partial_j\partial_j \varphi = 0,\hspace{2.5cm}\\
\left(\sum_{i=1}^n\partial_i\partial_i\right)\left(\sum_{j=1}^n \partial_j\partial_j\right) \varphi = 0,\hspace{2.5cm}\\
{\partial^4 \varphi\over \partial x^4 } + {\partial^4 \varphi\over \partial y^4 } + {\partial^4 \varphi\over \partial z^4 } + 2{\partial^4 \varphi\over \partial x^2\partial y^2} + 2{\partial^4 \varphi\over \partial y^2\partial z^2} + 2{\partial^4 \varphi\over \partial x^2\partial z^2} = 0,\\
\frac{1}{r} \frac{\partial}{\partial r} \left(r \frac{\partial}{\partial r} \left(\frac{1}{r} \frac{\partial}{\partial r} \left(r \frac{\partial \varphi}{\partial r}\right)\right)\right) + \frac{2}{r^2} \frac{\partial^4 \varphi}{\partial \theta^2 \partial r^2} + \frac{1}{r^4} \frac{\partial^4 \varphi}{\partial \theta^4} - \frac{2}{r^3} \frac{\partial^3 \varphi}{\partial \theta^2 \partial r} + \frac{4}{r^4} \frac{\partial^2 \varphi}{\partial \theta^2} = 0,\\
\Delta\Delta u(x,y)= 0,\hspace{2.5cm}\\
\Delta\Delta u(x,y)= f(x,y),\hspace{2.5cm}\\
\phi_{rrrr} + \frac{2}{r}\phi_{rrr} - \frac{1}{r^2}\phi_{rr} + \frac{1}{r^3}\phi_r = 0.\hspace{2.5cm}\\
\text{\textbf{Newton's Second Law (NSL)}}:\hspace{2.5cm}\\
\vec{F} = \frac{d\vec{p}}{dt},\hspace{2.5cm}\\
\vec{F} = m\vec{a},\hspace{2.5cm}\\
\vec{F} = m\frac{d^2}{dt^2}  \vec{s},\hspace{2.5cm}\\
\textbf{F} = \frac{d}{dt} (m\textbf{v}),\hspace{2.5cm}\\
\vec{F} \Delta p = \Delta p,\hspace{2.5cm}\\
\vec{F} = \frac{m \Delta \vec{v}}{\Delta t},\hspace{2.5cm}\\
\vec{F} = m \frac{\Delta \vec{v}}{\Delta t},\hspace{2.5cm}\\
\vec{F} = \frac{ \vec{p}}{t},\hspace{2.5cm}\\
F = \propto ma,\hspace{2.5cm}\\
F = kma.\hspace{2.5cm}\\
\text{\textbf{Heisenberg Uncertainty Principle (HUP)}}:\hspace{2.5cm}\\
\sigma_{x}\sigma_{p} \geq \frac{\hbar}{2},\hspace{2.5cm}\\
\sigma_E \frac{\sigma_B}{\left| \frac{d\langle \hat B \rangle}{dt}\right |} \ge \frac{\hbar}{2},\hspace{2.5cm}\\
\sigma_{J_x}^2+\sigma_{J_y}^2+\sigma_{J_z}^2\ge j,\hspace{2.5cm}\\
\Delta x \Delta p \geq \frac{\hbar}{2},\hspace{2.5cm}\\
\sigma_{x}^2\sigma_{p}^2 \geq ( \frac{1}{2i} \langle \lbrack \hat{\vec{x}}, \hat{\vec{p}} \rbrack \rangle )^2,\hspace{2.5cm}\\
\sigma_{x}^2\sigma_{p}^2 \geq \frac{\hbar}{2}^2,\hspace{2.5cm}\\
\sigma_{x}^2\sigma_{p}^2 \geq -\frac{1}{4} (\langle \lbrack \hat{A}, \hat{B} \rbrack \rangle )^2,\hspace{2.5cm}\\
\sigma_{x}\sigma_{p} \geq \frac{1}{2} \left| -i\hbar \int \Psi* \Psi dx \right|,\hspace{2.5cm}\\
\sigma_{x}\sigma_{p} \geq \frac{1}{2} \left| -i\hbar \right|,\hspace{2.5cm}\\
\sigma_{x}\sigma_{p} \geq \frac{1}{2} \left| \int \Psi*  \lbrack \hat{x}, \hat{p} \rbrack \Psi dx \right|.\hspace{2.5cm}\\
\text{\textbf{Second Law of Thermodynamics (SLT)}}:\hspace{2.5cm}\\
\oint \frac{\delta Q}{T}=0,\hspace{2.5cm}\\
\Delta S \ge \int \frac{\delta Q}{T_{surr}},\hspace{2.5cm}\\
dS_{\mathrm{tot}}= dS + dS_R \ge 0,\hspace{2.5cm}\\
dS_{tot} \ge 0,\hspace{2.5cm}\\
dE + \delta w_u \le 0,\hspace{2.5cm}\\
\int \frac{\delta Q}{T} = -N,\hspace{2.5cm}\\
\frac{dS}{dt} \ge 0,\hspace{2.5cm}\\
\frac{dS}{dt} = \dot S_{i},\hspace{2.5cm}\\
\frac{dS}{dt} = \frac{\dot Q}{T}+\dot S+\dot S_{i},\hspace{2.5cm}\\
dS =\frac{\delta Q}{T}.\hspace{2.5cm}\\
\text{\textbf{Coulomb's Law (CL)}}:\hspace{2.5cm}\\
|F_1| = |F_2| = \frac{|q_1 \times q_2|}{r^2},\hspace{2.5cm}\\
|F|=k_\text{e} \frac{|q_1||q_2|}{r^2},\hspace{2.5cm}\\
|\mathbf{F}|=k_\text{e} \frac{|q_1q_2|}{r^2},\hspace{2.5cm}\\
\mathbf{F}_1 = \frac{q_1q_2}{4\pi\varepsilon_0} \frac{\mathbf{r}_1-\mathbf{r}_2}{|\mathbf{r}_1 - \mathbf{r}_2|^3},\hspace{2.5cm}\\
\mathbf{F}_1 = \frac{q_1q_2}{4\pi\varepsilon_0} \frac{\mathbf{\hat{r}}_{12}}{|\mathbf{r}_{12}|^2},\hspace{2.5cm}\\
\mathbf{F}(\mathbf{r}) = {q\over4\pi\varepsilon_0} \sum_{i=1}^N q_i \frac{\mathbf{r} - \mathbf{r}_i}{|\mathbf{r}- \mathbf{r}_i|^3},\hspace{2.5cm}\\
\mathbf{F}(\mathbf{r}) = {q\over 4\pi\varepsilon_0} \sum_{i=1}^N q_i {\hat{\mathbf{R}}_i\over|\mathbf{R}_i|^2},\hspace{2.5cm}\\
\mathbf{F}(\mathbf{r}) = \frac{q}{4\pi\varepsilon_0}\int dq' \frac{\mathbf{r} - \mathbf{r'}}{|\mathbf{r} - \mathbf{r'}|^3},\hspace{2.5cm}\\
\mathbf{E}(\mathbf{r}) = {1\over4\pi\varepsilon_0} \sum_{i=1}^N q_i \frac{\mathbf{r}-\mathbf{r}_i}{|\mathbf{r}-\mathbf{r}_i|^3},\hspace{2.5cm}\\
\mathbf{E}(\mathbf{r}) = \frac{Q}{4\pi \varepsilon_0} \frac{\hat{\mathbf{r}}}{r^2}\hspace{2.5cm}.
\end{align*}
\endgroup

\end{document}